\documentclass[]{article}

\usepackage{underscore}
\usepackage[T1]{fontenc}
\usepackage{xspace}
\usepackage{stmaryrd, amsmath, amssymb, amsthm}
\usepackage{nicefrac}
\usepackage{tikz}
	\usetikzlibrary{trees,decorations,arrows,automata,positioning,plotmarks,
	shapes,backgrounds,petri}
\usepackage{wrapfig}
\usepackage{cancel}
\usepackage[left=2.25cm, right=2.25cm, top=3cm, bottom=3cm]{geometry}
\usepackage{hyperref}
\usepackage{fancyhdr}

\usepackage{macros}

\title{On the Expressiveness of Mixed Choice Sessions\\(Technical Report)}
\def\titlerunning{Distributability of Mobile Ambients}
\author{
	\begin{minipage}[t]{0.3\textwidth}
		\centering
		Kirstin Peters\\
		Universität Augsburg\\
		Germany
	\end{minipage}
	\hspace{2em}
	\begin{minipage}[t]{0.3\textwidth}
		\centering
		Nobuko Yoshida\\
		Imperial College London\\
		UK
	\end{minipage}
}
\def\titlerunning{On the Expressiveness of Mixed Choice Sessions}
\def\authorrunning{K. Peters and N. Yoshida}

\begin{document}

\maketitle


\begin{abstract}
Session types provide a flexible programming style for structuring
interaction, and are used to guarantee a safe and consistent
composition of distributed processes. Traditional session types 
include only one-directional input (external) and output (internal)
guarded choices. This prevents the session-processes 
to explore the full expressive power  
of the \piCal where the mixed choices are proved more expressive 
than the (non-mixed) guarded choices. To account this issue,   
recently Casal, Mordido, and Vasconcelos proposed 
the binary session types with mixed choices (\CMVmix).
This paper carries a surprising, unfortunate result on \CMVmix:
in spite of an inclusion of unrestricted channels with mixed choice, 
\CMVmix's mixed choice is rather separate and not 
mixed. We prove this negative result using 
two methodologies (using either the leader election problem or a synchronisation pattern as distinguishing feature), showing that there exists no good encoding 
from the \piCal into \CMVmix, preserving distribution.  
We then close their open problem on  
the encoding from \CMVmix into 
\CMV (without mixed choice), proving 
its soundness and thereby that the encoding is good up to coupled similarity.
This technical report extends a paper presented at the workshop EXPRESS/SOS'22.
\end{abstract}


\section{Introduction}
\label{sec:introduction}

Starting with the landmark result by Palamidessi in \cite{palamidessi97} and followed up by results such as \cite{nestmann00, palamidessi03, gorla10, peters12, petersNestmann12, breakingSymmetries16} it was shown that the key to the expressive power of the full \piCal in comparison to its sub-calculi such as \eg the asynchronous \piCal is \emph{mixed choice}.

\emph{Mixed choice} in the \piCal is a choice construct that allows to choose between inputs and outputs.
In contrast, \eg \emph{separate choices} are constructed from either only inputs or only outputs.
The additional expressive power of mixed choice relies on its ability to rule out alternative options of the opposite nature, \ie a term can rule out its possibility to perform an input by doing an output, whereas without mixed choice inputs can rule out alternative inputs only and outputs may rule out only alternative outputs.

To compare calculi with different variants of choice, we try to build an encoding or show that no such encoding exists \cite{boerPalamidessi1991, parrow08}.
The existence of an encoding that satisfies relevant criteria shows that the target language is expressive enough to emulate the behaviours in the source language.
Gorla \cite{gorla10} and others \cite{parrow08, petersNestmannGoltz13} introduced and
classified a set of general criteria
for encodability which are syntax-agnostic 
\cite{gorla10, petersNestmannGoltz13}: 
they are now commonly used for claiming expressiveness of a given
calculus, defining important features which a ``good encoding'' should
satisfy. These include \emph{compositionality} (homomorphism), 
\emph{name invariance} (bijectional renaming), 
sound and complete \emph{operational correspondence} (the source and
target can simulate each other), 
\emph{divergence reflection} (the target diverges only if the
source diverges), \emph{observability} (barb-sensitiveness), and \emph{distributability} preservation (the target has the same degree of distribution as the source). 
Conversely, a \emph{separation result}, \ie the proof of the absence of an encoding with certain criteria, shows that the source language can represent behaviours that can\emph{not} be expressed in the target. This paper gives a fresh look at
expressiveness of typed $\pi$-calculi,
focusing on choice constructs of \emph{session types}. 

Session types \cite{hondaVasconcelosKubo98,THK} specify and constrain the communication behaviour as a
protocol between components in a system. A session type system
excludes any non-conforming behaviour, statically preventing 
type and communication errors (i.e., mismatch of choice labels).
Several languages now have session-type support via libraries and
tools \cite{BETTYTOOLBOOK,DBLP:journals/ftpl/AnconaBB0CDGGGH16}. 
As the origin of session types is Linear Logic \cite{HondaK:typdyi}, 
traditional session types 
include only one-directional input (external) and output (internal)
guarded choices. To explore the full expressiveness of mixed choice
from the \piCal, 
recently Casal, Mordido, and Vasconcelos proposed 
the binary session types with mixed choices called \emph{mixed sessions}
\cite{casalMordidoVasconcelos22}.
We denote their calculus by \CMVmix. 
Mixed sessions include a mixture of branchings (labelled input choices) and 
selections (labelled output choices) 
at the same \emph{linear} channel or \emph{unrestricted} channel.
This extension gives us many useful and typable \emph{structured}
concurrent programming idioms which consist
of both unrestricted and linear non-deterministic choice behaviours.
We show that
in spite of its practical relevance,
 mixed sessions in \CMVmix 
are \emph{strictly less expressive} than
mixed choice in the \piCal
even with an unrestricted usage of choice channels.

This result surprised us.
We would have expected that using mixed choice with an unrestricted choice channel results into a choice construct comparable to choice in the \piCal.
But, as we show in the following, mixed choice in \CMVmix cannot express essential features of mixed choice in the \piCal.
First we observe that mixed sessions are not expressive enough to solve leader election in symmetric networks.
Remember that it was leader election in symmetric networks that was used to show that mixed choice is more expressive than separate choice in the \piCal (see \cite{palamidessi97}).
Second we observe that mixed sessions cannot express the synchronisation pattern \patternStar.
Synchronisation patterns were introduced in \cite{petersNestmannGoltz13} to capture the amount of synchronisation that can be expressed in distributed systems.
The synchronisation pattern \patternStar was identified in \cite{petersNestmannGoltz13} as capturing exactly the amount of synchronisation introduced with mixed choice in the \piCal.
Finally, we have a closer look at the encoding from \CMVmix into \CMV presented in \cite{casalMordidoVasconcelos22}.
\CMV is the variant of session types that is extended in \cite{casalMordidoVasconcelos22} with a mixed-choice-construct in order to obtain \CMVmix, \ie \CMV has traditional branching and selection but not their mix.
As it is the case for many variants of session types, \CMV can express separate choice but has no construct for mixed choice.
By analysing this encoding, we underpin our claim that mixed choice in \CMVmix is not more expressive than separate choice in the \piCal.

\begin{wrapfigure}{R}{0.35\textwidth}
	\centering
	\begin{tikzpicture}[bend angle=20]
		\node (pi) at (2, 2) {$ \pi $};
		\node (CMV) at (4, 0) {\CMV};
		\node (CMVmix) at (2, 0) {\CMVmix};
		\path[->, thick, color=darkgreen] (CMVmix) edge (CMV);
		\path[->, color=red, dashed] (pi) edge [bend right] node[left] {$ \mathsf{LE} $\;} node {$ \times $} (CMVmix);
		\path[->, color=red, dashed] (pi) edge [bend left] node[right] {\;\patternStar} node {$ \times $} (CMVmix);
	\end{tikzpicture}
\end{wrapfigure}

Our contributions are summarised in the picture on the right.
In \S~\ref{sec:separateMixedSessionsLeaderElection} we prove that there exists no good encoding from the \piCal (with mixed choice)
into \CMVmix, where we use the leader election problem by Palamidessi in \cite{palamidessi97} (\textcolor{red}{$ \mathsf{LE} $}) as distinguishing feature (the first \begin{tikzpicture}[baseline=-1mm] \path[->, dashed, color=red] (0, 0) edge node {$ \times $} (1, 0); \end{tikzpicture}).
In \S~\ref{sec:separateMixedSessionsFromPiSynchronisationPatterns} we reprove this result using the \emph{synchronisation pattern} \textcolor{red}{\patternStar} from \cite{petersNestmannGoltz13} instead as distinguishing feature (the second \begin{tikzpicture}[baseline=-1mm] \path[->, dashed, color=red] (0, 0) edge node {$ \times $} (1, 0); \end{tikzpicture}).
Then we prove soundness of the encoding presented in
\cite{casalMordidoVasconcelos22} closing their open problem in
\S~\ref{sec:encodeMixedSessions} (\begin{tikzpicture}[baseline=-1mm]
  \path[->, thick, color=darkgreen] (0, 0) edge (0.5, 0); \end{tikzpicture}).
By this encoding source terms in \CMVmix and their literal
translations in \CMV are related by \emph{coupled similarity} \cite{parrow1992}, \ie \CMVmix is encoded into \CMV up to coupled similarity.
From the separation results in \S~\ref{sec:separateMixedSessionsLeaderElection} and \S~\ref{sec:separateMixedSessionsFromPiSynchronisationPatterns} and the encoding into session types with separate choice in \S~\ref{sec:encodeMixedSessions} we conclude that \emph{mixed sessions in \cite{casalMordidoVasconcelos22} can express only separate choice}.

This technical report extends a paper presented at the workshop EXPRESS/SOS'22.
In particular, we present detailed proofs and some additional material on the considered languages such as their type systems.


\section{Technical Preliminaries: Mixed Sessions and Encodability Criteria}
\label{sec:technicalPreliminaries}

A \emph{process calculus} is a language $ \lang = \left( \proc, \step \right) $ that consists of a set of process terms $ \proc $ (its syntax) and a relation $ {\step} : \proc \times \proc $ on process terms (its reduction semantics), typically building upon some structural congruence ${\equiv}  : \proc \times \proc$. We often refer to process terms also simply as processes or terms and use upper case letters $ P, Q, R, \ldots, P', P_1, \ldots $ to range over them.
Typed languages often define their syntax by a grammar defining the untyped processes $ \proc^{\indexUntyped} $ and a set of typing rules that define the subset of well-typed processes $ \proc \subset \proc^{\indexUntyped} $ of the language.

Assume a countably-infinite set $ \names $, whose elements are called \emph{names}.
We use lower case letters such as $ a, b, c, \ldots, a', a_1, \ldots $ to range over names.
For the \piCal we additionally assume a set $ \Set{ \overline{y} \mid y \in \names } $ of \emph{co-names}.
Let $ \tau \notin \names \cup \Set{ \overline{y} \mid y \in \names } $.
The typed languages considered here intuitively distinguish names into \emph{session channels}, ranged over by $ x, y, \ldots $, and name variables, ranged over by $ z, \ldots $.
There is, however, no need to formally distinguish between different kinds of names.
We also assume a set of type variables, ranged over by $ t, t', \ldots $, and a set of process variables, ranged over by $ X, X', \ldots $.

The \emph{syntax} of a process calculus is usually defined by a context-free grammar defining operators, \ie functions $ \operatorname{op} : \names^n \times \proc^m \to \proc $. An operator of arity $ 0 $, \ie $ m = 0 $, is a \emph{constant}. The arguments that are again process terms are called \emph{subterms} of $ P $.

\begin{definition}[Subterms]
	\label{def:subterms}
	Let $ \left( \proc, \step \right) $ be a process calculus and $ P \in \proc $. The set of \emph{subterms} of $ P = \operatorname{op}\left( x_1, \ldots, x_n, P_1, \ldots, P_m \right) $ is defined recursively as:
	\begin{align*}
		\Set{ P } \cup \Set{ P' \mid \exists i \in \Set{ 1, \ldots, m } \logdot P' \text{ is a subterm of } P_i }
	\end{align*}
\end{definition}

\noindent
With Definition~\ref{def:subterms}, every term is a subterm of itself; constants have no further subterms.
Terms that appear as subterm underneath some (action) prefix are called \emph{guarded}, because the guarded subterm can not be executed before the guarding action has been performed.
Also conditionals, such as if-then-else-constructs, guard their respective subterms.

\emph{Expressions}, ranged over by $ e, e', \ldots $, are constructed from variables, unit, and standard boolean operators.
We assume an evaluation function $ \Eval{\cdot} $ that evaluates expressions to \emph{values}, ranged over by $ v, v', \ldots $:
\begin{align*}
	v & \deffTerms x \sepTerms \true \sepTerms \false \sepTerms \unit & \text{Values}
\end{align*}

A \emph{scope} defines an area in which a particular name is known and can be used. For several reasons, it can be useful to restrict the scope of a name. For instance to forbid interaction between two processes or with an unknown and, hence, potentially untrusted environment. Names whose scope is restricted such that they cannot be used beyond their scope are called \emph{bound names}. The remaining names are called \emph{free names}.
Let $ \FreeNames{P} $ denote the set of free names of $ P $.
In the case of bound names, their syntactical representation as lower case letters serves as a place holder for any fresh name, \ie any name that does not occur elsewhere in the term. To avoid confusion between free and bound names or different bound names, bound names can be replaced with fresh bound names by \emph{$ \alpha $-conversion} $ \equiv_{\alpha} $.

We assume that the \emph{semantics} is given as an \emph{operational semantics} consisting of inference rules defined on the operators of the language \cite{Plotkin04}. For many process calculi, the semantics is provided in two forms, as \emph{reduction semantics} and as \emph{labelled transition semantics}. We assume that at least the reduction semantics $ \step $ is given as part of the definition, because its treatment is easier in the context of encodings.
A single application of the reduction semantics is called a \emph{(reduction) step} and is written as $ P \step P' $. If $ P \step P' $, then $ P' $ is called \emph{derivative} of $ P $. Let $ P \step $ (or $ P \noStep $) denote the existence (absence) of a step from $ P $, and let $ \steps $ denote the reflexive and transitive closure of $ \step $. A sequence of reduction steps is called a \emph{reduction}.
We write $ P \step^{\omega} $ if $ P $ has an infinite sequence of steps.
We also use \emph{execution} to refer to a reduction starting from a particular term.
A process that cannot reduce is called \emph{stuck}.

A substitution $ \sigma $ is a finite mapping from names to names defined by a set of renamings of the form $ \Set{ \Subst{y_1}{x_1}, \ldots, \Subst{y_n}{x_n} } = \Set{ \Subst{y_1, \ldots, y_n}{x_1, \ldots, x_n}} $, where we assume that the $ x_1, \ldots, x_n $ are pairwise distinct.
The application $ P\Set{ \Subst{y_1}{x_1}, \ldots, \Subst{y_n}{x_n} } $ of a substitution on a term is defined as the result of simultaneously replacing all free occurrences of $ x_i $ by $ y_i $ for $ i \in \Set{ 1, \ldots, n } $, possibly applying $ \alpha $-conversion to avoid capture or name clashes.
For all names in $ \names \setminus \Set{ x_1, \ldots, x_n } $ the substitution behaves as the identity mapping.
We naturally extend substitution to the substitution of name variables by values and type variables by types.
In these cases we often denote substitution as the instantiation of the variable by the respective value or type.

To simplify the presentation, we sometimes tread sequences $ \tilde{x} = x_1, \ldots, x_n $ as sets and apply set operations (such as union) on sequences.

To reason about environments of terms, we use functions on process terms called contexts. More precisely, a \emph{context} $ \Context{}{}{\hole_1, \ldots, \hole_{n + m}} : \names^n \times \proc^m \to \proc $ with $ n + m $ holes is a function from $ n $ names and $ m $ terms into a term, \ie the term $ \Context{}{}{x_1, \ldots, x_n, P_1, \ldots, P_m} $ is the result of inserting $ x_1, \ldots, x_n, P_1, \ldots, P_m $ in the corresponding order into the holes of~$ \context $.

We use \emph{barbs} or observables to reason about and to compare the behaviour of processes.
We write $ P\Barb{\beta} $ if $ P $ emits the barb $ \beta $, where this definition is language specific, \ie implemented slightly differently in the considered languages.
In all considered languages $ P $ reaches a barb $ \beta $, denoted as $ P\WeakBarb{\beta} $, if there is some $ P' $ such that $ P \steps P' $ and $ P'\Barb{\beta} $.

Two terms of a language are usually compared using some kind of a behavioural simulation relation.
The most commonly known behavioural simulation relation is bisimulation.
A relation $ \mathcal{R} $ is a bisimulation if any two related processes mutually simulate their respective sequences of steps, such that the derivatives are again related.

\begin{definition}[Bisimulation]
	\label{def:bisimulation}
	$ \mathcal{R} $ is a (weak reduction, barbed) bisimulation if for each $ {\left( P, Q \right)} \in \mathcal{R} $:
	\begin{itemize}
		\item $ P \steps P' $ implies $ \exists Q'\logdot Q \steps Q' \wedge {\left( P', Q' \right)} \in \mathcal{R} $
		\item $ Q \steps Q' $ implies $ \exists P'\logdot P \steps P' \wedge {\left( P', Q' \right)} \in \mathcal{R} $
		\item $ P\WeakBarb{\beta} $ iff $ Q\WeakBarb{\beta} $ for all barbs $ \beta $
	\end{itemize}
	Two terms are bisimilar if there exists a bisimulation that relates them.
	For a language $ \lang $, let $ \approx_{\lang} $ denote bisimilarity on $ \lang $.
\end{definition}

Another interesting behavioural simulation relation is coupled similarity.
It was introduced in \cite{parrow1992} and discussed \eg in \cite{bispingNestmannPeters19}.
It is strictly weaker than bisimilarity.
As pointed out in \cite{parrow1992}, in contrast to bisimilarity it essentially allows for intermediate states (see \S~\ref{sec:encodeMixedSessions}).
Each symmetric coupled simulation is a bisimulation.

\begin{definition}[Coupled Simulation]
	\label{def:coupledSimulation}
	A relation $ \mathcal{R} $ is a (weak reduction, barbed) coupled simulation if for each $ {\left( P, Q \right)} \in \mathcal{R} $:
	\begin{itemize}
		\item $ P \steps P' $ implies $ \exists Q'\logdot Q \steps Q' \wedge {\left( P', Q' \right)} \in \mathcal{R} $
		\item $ P \steps P' $ also implies $ \exists Q'\logdot Q \steps Q' \wedge {\left( Q', P' \right)} \in
\mathcal{R} $ 
		\item $ P\WeakBarb{\beta} $ implies $ Q\WeakBarb{\beta} $ for all barbs $ \beta $
	\end{itemize}
	Two terms are coupled similar if they are related by a coupled simulation in both directions.
\end{definition}

\subsection{The Pi-Calculus with Mixed Choice}
\label{sec:piCalculus}

The \piCal was introduced by Milner, Parrow, and Walker in \cite{MilnerParrowWalker92} and is one of the most well-known process calculi.
Over the time a large number of variants and extensions of the \piCal emerged.
We are relying on the variant used in \cite{palamidessi97}, since we want to reuse some results and proof techniques of this paper.
Accordingly, we consider a variant of the \piCal with mixed guarded choice and replication but without matching.
This variant is often called the synchronous or full \piCal.
In the following we denote this calculus simply as the \piCal.

The set of processes $ \procPi $ of the \piCal, \ie its sytax, is given by:
\begin{align*}
	\alpha & \deffTerms \InpPi{y}{x} \sepTerms \OutPi{y}{z} \sepTerms \tau & \text{Prefixes}\\
	P & \deffTerms \sum_{i \in \indexSet} \alpha_i.P_i \sepTerms \ResPi{x}{P} \sepTerms P \mid P \sepTerms !P & \text{Processes}
\end{align*}

A choice $ \sum_{i \in \indexSet} \alpha_i.P_i $ offers for each $ i $ in the index set $ \indexSet $ a subterm guarded by some action prefix $ \alpha_i $.
An action prefix is either an input action $ \InpPi{y}{x} $, and output action $ \OutPi{y}{z} $, or an internal action denoted as $ \tau $.
We abbreviate the empty sum, \ie $ \sum_{i \in \indexSet} \alpha_i.P_i $ for $ \indexSet = \emptyset $, by the inactive process $ \inactPi $.
Moreover, we often write $ \alpha_1.P_1 + \ldots + \alpha_n.P_n $ for a choice $ \sum_{i \in \Set{1, \ldots, n}} \alpha_i.P_i $.
The remaining operators introduce restriction $ \ResPi{x}{P} $, parallel composition $ P \mid P $, and replication $ !P $.

The name $ x $ is bound in $ P $ by inputs $ \InpPi{y}{x}.P $ and restriction $ \ResPi{x}{P} $. All other names are free.
To simplify the presentation we often omit trailing $ \inactPi $.
Moreover, we sometimes omit the argument of action prefixes if it is irrelevant, \ie we write $ y.P $ instead of $ \InpPi{y}{x}.P $ if $ x \notin \FreeNames{P} $ and we write $ \overline{y}.P $ instead of $ \OutPi{y}{z}.P $ if for all matching receivers $ \InpPi{y}{x}.Q $ we have $ x \notin \FreeNames{Q} $.

\begin{figure}[tp]
	\begin{displaymath}\begin{array}{c}
		\ruleComPi \; \OutPi{y}{z}.P + M \mid \InpPi{y}{x}.Q + N \stepPi P \mid Q\Set{\Subst{z}{x}}
		\hspace{2em}
		\ruleTauPi \; \tau.P + M \stepPi P
		\vspace{0.5em}\\
		\ruleParPi \; \dfrac{P \stepPi P'}{P \mid Q \stepPi P' \mid Q}
		\hspace{2em}
		\ruleResPi \; \dfrac{P \stepPi P'}{\ResPi{x}{P} \stepPi \ResPi{x}{P'}}
		\vspace{0.5em}\\
		\ruleStructPi \; \dfrac{P \scPi Q \quad Q \stepPi Q' \quad Q' \scPi P'}{P \stepPi P'}
	\end{array}\end{displaymath}
	\caption{Reduction Semantics ($ \stepPi $) of the \PiCal.}
	\label{fig:semanticsPi}
\end{figure}

The semantics of the \piCal is given by the rules in
Figure~\ref{fig:semanticsPi}, where structural congruence $ \scPi $ is
the least congruence that contains $ \alpha $-conversion and satisfies
the rules:
\begin{displaymath}\begin{array}{c}
	\ResPi{x}{\inactPi} \scPi \inactPi
	\hspace{2em}
	\ResPi{x}{\ResPi{y}{P}} \scPi \ResPi{y}{\ResPi{x}{P}}
	\hspace{2em}
	P \mid \ResPi{x}{Q} \scPi \ResPi{x}{{\left( P \mid Q \right)}} \quad \text{if } x \notin \FreeNames{P}
	\vspace{0.5em}\\
	P \mid \inactPi \scPi P
	\hspace{2em}
	P \mid Q \scPi Q \mid P
	\hspace{2em}
	P \mid {\left( Q \mid R \right)} \scPi {\left( P \mid Q \right)} \mid R
	\hspace{2em}
	!P \scPi P \mid {!P}
\end{array}\end{displaymath}
Since choice is introduced as a set of summands, we naturally have commutativity and associativity of the summands.
We rely on the commutativity and associativity of summands then writing choices as $ \alpha_1.P_1 + \ldots + \alpha_n.P_n $.

Rule \ruleComPi defines communication as an interaction of an output and an input action that are composed in parallel choices.
As result of the communication step the respective alternative summands of the two choices that contained the output and the input as well as the action prefixes are removed, and in the subterm of the input the name $ x $ is substituted by the received name $ z $.
This step unguards the respective subterms of the output and input.

An internal step with rule \ruleTauPi reduces only a single choice.
As result again the subterm of the internal action is unguarded and the alternative summands are removed.

Rule \ruleParPi allows a process to reduce in the context of parallel processes and \ruleResPi allows the subterm of a restriction to reduce.
Finally, \ruleStructPi allows processes to reduce modulo structural congruence.

A process $ P $ emits an output barb $ \overline{y} $, denoted as $ P\Barb{\overline{y}} $, if $ P $ has an output $ \OutPi{y}{z}.P' $ as unguarded subterm and if $ y $ is free in $ P $, \ie $ y \in \FreeNames{P} $.
Similarly, $ P $ has an input barb $ y $, denoted as $ P\Barb{y} $, if $ P $ has an input $ \InpPi{y}{x}.P' $ as unguarded subterm with $ y \in \FreeNames{P} $.

\subsection{Mixed Sessions}
\label{sec:mixedSessions}

\emph{Mixed sessions} are variant of binary session types introduced by Casal, Mordido, and Vasconcelos in \cite{casalMordidoVasconcelos22} with a choice-construct that combines prefixes for sending and receiving.
We denote this language as \CMVmix.

A central idea of \CMVmix (and the language \CMV it is based on) is that channels are separated in two \emph{channel endpoints} and that interaction is by two processes acting on the respective different ends of such a channel.
The set of untyped processes $ \procCMVmixUntyped $ of \CMVmix is given as:
\begin{align*}
	P & \deffTerms \ChoiceCMVmix{q}{y}{\sum_{i \in \indexSet}M_i} \sepTerms P \mid P \sepTerms \ResCMVmix{y}{z}{P} \sepTerms \ConditionalCMVmix{v}{P}{P} \sepTerms \inactCMVmix & \text{Processes}\\
	M & \deffTerms \BranchCMVmix{\Label}{*}{v}{P} & \text{Branches}\\
	* & \deffTerms ! \sepTerms ? & \text{Polarities}\\
	q & \deffTerms \linCMVmix \sepTerms \unCMVmix & \text{Qualifiers}
\end{align*}

A choice $ \ChoiceCMVmix{q}{y}{\sum_{i \in \indexSet}M_i} $ is declared as either linear ($ \linCMVmix $) or unrestricted ($ \unCMVmix $) by the qualifier $ q $.
It proceeds on a single channel endpoint $ y $.
For every $ i $ in the index set $ \indexSet $ it offers a branch $ M_i $.
A branch $ \BranchCMVmix{\Label}{*}{v}{P} $ specifies a label $ \Label $, a polarity $ * $ ($ ! $ for sending or $ ? $ for receiving), a name $ v $ (a value in output actions or a variable for input actions), and a continuation $ P $.
We abbreviate the empty sum, \ie $ \ChoiceCMVmix{q}{y}{\sum_{i \in \indexSet}M_i} $ for $ \indexSet = \emptyset $, by $ \inactCMVmix $.
Moreover, we often write $ q \; x \left( M_1 + \ldots + M_n \right) $ for a choice $ \ChoiceCMVmix{q}{y}{\sum_{i \in \Set{ 1, \ldots, n }}M_i} $.
Restriction $ \ResCMVmix{y}{z}{P} $ binds the two channel endpoints $ y $ and $ z $ of a single channel to $ P $.
The remaining operators introduce parallel composition $ P \mid P $, conditionals $ \ConditionalCMVmix{v}{P}{P} $, and inaction $ \inactCMVmix $.
We sometimes abbreviate $ P_1 \mid \ldots \mid P_n $ by $ \prod_{i \in \Set{1, \ldots, n}} P_i $.

The variable $ x $ is bound in $ P $ by input branches $ \BranchCMVmix{\Label}{?}{x}{P} $ and the two endpoints of a channel $ x, y $ are bound in $ P $ by restriction $ \ResCMVmix{x}{y}{P} $.
All other names are free.

\begin{figure}[tp]
	\begin{displaymath}\begin{array}{c}
		\ruleRIfTCMVmix \; \ConditionalCMVmix{\true}{P}{Q} \stepCMVmix P
		\hspace{2em}
		\ruleRIfFCMVmix \; \ConditionalCMVmix{\false}{P}{Q} \stepCMVmix Q
		\vspace{0.5em}\\
		\ruleRLinLinCMVmix \; \begin{array}{l}
				\ResCMVmix{y}{z}{{\left( \ChoiceCMVmix{\linCMVmix}{y}{{\left( \OutCMVmix{\Label}{v}{P} + M \right)}} \mid \ChoiceCMVmix{\linCMVmix}{z}{{\left( \InpCMVmix{\Label}{x}{Q} + N \right)}} \mid R \right)}} \stepCMVmix\\
				\ResCMVmix{y}{z}{{\left( P \mid Q\Set{\Subst{v}{x}} \mid R \right)}}
			\end{array}
		\vspace{0.5em}\\
		\ruleRLinUnCMVmix \; \begin{array}{l}
				\ResCMVmix{y}{z}{{\left( \ChoiceCMVmix{\linCMVmix}{y}{{\left( \OutCMVmix{\Label}{v}{P} + M \right)}} \mid \ChoiceCMVmix{\unCMVmix}{z}{{\left( \InpCMVmix{\Label}{x}{Q} + N \right)}} \mid R \right)}} \stepCMVmix\\
				\ResCMVmix{y}{z}{{\left( P \mid Q\Set{\Subst{v}{x}} \mid \ChoiceCMVmix{\unCMVmix}{z}{{\left( \InpCMVmix{\Label}{x}{Q} + N \right)}} \mid R \right)}}
			\end{array}
		\vspace{0.5em}\\
		\ruleRUnLinCMVmix \; \begin{array}{l}
				\ResCMVmix{y}{z}{{\left( \ChoiceCMVmix{\unCMVmix}{y}{{\left( \OutCMVmix{\Label}{v}{P} + M \right)}} \mid \ChoiceCMVmix{\linCMVmix}{z}{{\left( \InpCMVmix{\Label}{x}{Q} + N \right)}} \mid R \right)}} \stepCMVmix\\
				\ResCMVmix{y}{z}{{\left( P \mid Q\Set{\Subst{v}{x}} \mid \ChoiceCMVmix{\unCMVmix}{y}{{\left( \OutCMVmix{\Label}{v}{P} + M \right)}} \mid R \right)}}
			\end{array}
		\vspace{0.5em}\\
		\ruleRUnUnCMVmix \; \begin{array}{l}
				\ResCMVmix{y}{z}{{\left( \ChoiceCMVmix{\unCMVmix}{y}{{\left( \OutCMVmix{\Label}{v}{P} + M \right)}} \mid \ChoiceCMVmix{\unCMVmix}{z}{{\left( \InpCMVmix{\Label}{x}{Q} + N \right)}} \mid R \right)}} \stepCMVmix\\
				\ResCMVmix{y}{z}{{\left( P \mid Q\Set{\Subst{v}{x}} \mid \ChoiceCMVmix{\unCMVmix}{y}{{\left( \OutCMVmix{\Label}{v}{P} + M \right)}} \mid \ChoiceCMVmix{\unCMVmix}{z}{{\left( \InpCMVmix{\Label}{x}{Q} + N \right)}} \mid R \right)}}
			\end{array}
		\vspace{0.5em}\\
		\ruleRParCMVmix \; \dfrac{P \stepCMVmix P'}{P \mid Q \stepCMVmix P' \mid Q}
		\hspace{2em}
		\ruleRResCMVmix \; \dfrac{P \stepCMVmix P'}{\ResCMVmix{y}{z}{P} \stepCMVmix \ResCMVmix{y}{z}{P'}}
		\vspace{0.5em}\\
		\ruleRStructCMVmix \; \dfrac{P \equiv Q \quad Q \stepCMVmix Q' \quad Q' \equiv P'}{P \stepCMVmix P'}
	\end{array}\end{displaymath}
	\caption{Reduction Rules ($ \stepCMVmix $) of \CMVmix.}
	\label{fig:semanticsCMVmix}
\end{figure}

The semantics of \CMVmix is given by the rules in Figure~\ref{fig:semanticsCMVmix}, where structural congruence $ \scCMVmix $ is the least congruence that contains $ \alpha $-conversion and satisfies the rules:
\begin{displaymath}\begin{array}{c}
	P \mid Q \scCMVmix Q \mid P
	\hspace{2em}
	{\left( P \mid Q \right)} \mid R \scCMVmix P \mid {\left( Q \mid R \right)}
	\hspace{2em}
	P \mid \inactCMVmix \scCMVmix P
	\hspace{2em}
	\ResCMVmix{y}{z}{\inactCMVmix} \scCMVmix \inactCMVmix
	\vspace{0.5em}\\
	P \mid \ResCMVmix{y}{z}{Q} \scCMVmix \ResCMVmix{y}{z}{{\left( P \mid Q \right)}} \quad \text{if } y, z \notin \FreeNames{P}
	\vspace{0.5em}\\
	\ResCMVmix{y}{z}{P} \scCMVmix \ResCMVmix{z}{y}{P}
	\hspace{2em}
	\ResCMVmix{w}{x}{\ResCMVmix{y}{z}{P}} \scCMVmix \ResCMVmix{y}{z}{\ResCMVmix{w}{x}{P}}
\end{array}\end{displaymath}
The commutativity and associativity of summands within choices again follows from choices being defined via a set of summands.

A conditional is reduced to its first subterm with Rule~\ruleRIfTCMVmix if its condition evaluates to $ \true $ and to its second subterm with Rule~\ruleRIfFCMVmix if its condition evaluates to $ \false $.
Communication is by one of the Rules~\ruleRLinLinCMVmix, \ruleRLinUnCMVmix, \ruleRUnLinCMVmix, or \ruleRUnUnCMVmix.
In all four cases, the continuation of the sender and der receiver are unguarded and in the receiver $ x $ is substituted by the received value $ v $.
These four rules differ \wrt the qualifiers of the involved choices.
Linear choices (qualifier $ \linCMVmix $) are removed in a reduction step, whereas unrestricted choices are persistent.
Rule \ruleRParCMVmix allows a process to reduce in the context of parallel processes and \ruleRResCMVmix allows the subterm of a restriction to reduce.
Finally, \ruleRStructCMVmix allows processes to reduce modulo structural congruence.

The process $ P $ emits the barb $ y $, denoted as $ P\Barb{y} $, if $ P $ has an unguarded choice $ \ChoiceCMVmix{q}{y}{\sum_{i \in \indexSet}M_i} $ on a free channel endpoint $ y \in \FreeNames{P} $.
We do not distinguish between output and input barbs here, but instead have barbs on different end points of a channel.

To obtain from $ \procCMVmixUntyped $ the set $ \procCMVmix $ of well-typed processes of \CMVmix, a type system is introduced in \cite{casalMordidoVasconcelos22}.
The syntax of types is given as:
\begin{align*}
	T & \deffTerms \ChoiceTCMVmix{q}{\#}{\Set{B_i}_{i \in \indexSet}} \sepTerms \finCMVmix \sepTerms \unitT \sepTerms \boolT \sepTerms \RecCMVmix{t}{T} \sepTerms t & \text{Types}\\
	B & \deffTerms \BranchCMVmix{\Label}{*}{T}{T} & \text{Branches}\\
	\# & \deffTerms \oplus \sepTerms \& & \text{Views}\\
	\Gamma & \deffTerms \cdot \sepTerms \Gamma, \At{x}{T} & \text{Contexts}
\end{align*}

A type of the form $ \ChoiceTCMVmix{q}{\#}{\Set{B_i}_{i \in \indexSet}} $ denotes a channel endpoint, where the view $ \# $ is either $ \oplus $ for internal choice or $ \& $ for external choice.
We often call it a choice type.
In a branch $ \BranchCMVmix{\Label}{*}{T_1}{T_2} $ the type $ T_1 $ specifies the communicated value whereas $ T_2 $ is the type of the continuation.
Besides channel endpoints there are types for inaction, the base types for unit and boolean, and types for recursion.

Following \cite{casalMordidoVasconcelos22}, we assume that the index sets $ \indexSet $ in types are not empty, that the label-priority-pairs $ \Label * $ are pairwise distinct in the branches of a choice type, and recursive types are contractive, \ie contain no subterm of the form $ \RecCMVmix{t_1}{\ldots\RecCMVmix{t_n}{t_1}} $ with $ n \geq 1 $.
A type variable $ t $ is bound in $ T $ by $ \RecCMVmix{t}{T} $.
All other type variables are free.

Type equivalence $ \simeq $ is coinductively defined by the rules:
\begin{displaymath}\begin{array}{c}
	\finCMVmix \simeq \finCMVmix
	\hspace{2em}
	\unitT \simeq \unitT
	\hspace{2em}
	\boolT \simeq \boolT
	\vspace{0.5em}\\
	\dfrac{T_i \simeq T_i' \quad U_i \simeq U_i' \quad {\left( \forall i \in \indexSet \right)}}{\ChoiceTCMVmix{q}{\#}{\Set{\BranchCMVmix{\Label}{*_i}{T_i}{U_i}}_{i \in \indexSet}} \simeq \ChoiceTCMVmix{q}{\#}{\Set{\BranchCMVmix{\Label}{*_i}{T_i'}{U_i'}}_{i \in \indexSet}}}
	\hspace{2em}
	\dfrac{T\Set{\Subst{\RecCMVmix{t}{T}}{t}} \simeq U}{\RecCMVmix{t}{T} \simeq U}
	\hspace{2em}
	\dfrac{T \simeq U\Set{\Subst{\RecCMVmix{t}{U}}{t}}}{T \simeq \RecCMVmix{t}{U}}
\end{array}\end{displaymath}

Two types are dual to each other if they describe well-coordinated behaviour of the two endpoints of a channel.
In particular, input is dual to output and internal choice is dual to external choice.
The operator $ \Dual{\cdot}{\cdot} $ for type duality is defined coinductively by the rules:
\begin{displaymath}\begin{array}{c}
	\Dual{!}{?}
	\hspace{2em}
	\Dual{?}{!}
	\hspace{2em}
	\Dual{{\oplus}}{\&}
	\hspace{2em}
	\Dual{\&}{\oplus}
	\hspace{2em}
	\Dual{\finCMVmix}{\finCMVmix}
	\vspace{0.5em}\\
	\dfrac{\Dual{\#}{\flat} \quad \Dual{{*_i}}{\bullet_i} \quad T_i \simeq T_i' \quad \Dual{U_i}{U_i'} \quad {\left( \forall i \in \indexSet \right)}}{\Dual{\ChoiceTCMVmix{q}{\#}{\Set{\BranchCMVmix{\Label}{*_i}{T_i}{U_i}}_{i \in \indexSet}}}{\ChoiceTCMVmix{q}{\flat}{\Set{\BranchCMVmix{\Label}{\bullet_i}{T_i'}{U_i'}}_{i \in \indexSet}}}}
	\hspace{2em}
	\dfrac{\Dual{T\Set{\Subst{\RecCMVmix{t}{T}}{t}}}{U}}{\Dual{\RecCMVmix{t}{T}}{U}}
	\hspace{2em}
	\dfrac{\Dual{T}{U\Set{\Subst{\RecCMVmix{t}{U}}{t}}}}{\Dual{T}{\RecCMVmix{t}{U}}}
\end{array}\end{displaymath}

Subtyping introduces more flexibility to the usage of types.
In external choices subtyping allows additional branches in the supertype; for internal choice we have the opposite.
The operator $ \Subtype{T_1}{T_2} $ ($ T_1 $ is a subtype of $ T_2 $) is defined coinductively by the rules:
\begin{displaymath}\begin{array}{c}
	\dfrac{\Subtype{T_2}{T_1} \quad \Subtype{U_1}{U_2}}{\Subtype{\OutCMVmix{\Label}{T_1}{U_1}}{\OutCMVmix{\Label}{T_2}{U_2}}}
	\hspace{2em}
	\dfrac{\Subtype{T_1}{T_2} \quad \Subtype{U_1}{U_2}}{\Subtype{\InpCMVmix{\Label}{T_1}{U_1}}{\InpCMVmix{\Label}{T_2}{U_2}}}
	\vspace{0.5em}\\
	\Subtype{\finCMVmix}{\finCMVmix}
	\hspace{2em}
	\Subtype{\unitT}{\unitT}
	\hspace{2em}
	\Subtype{\boolT}{\boolT}
	\vspace{0.5em}\\
	\dfrac{\indexSet[J] \subseteq \indexSet \quad \Subtype{B_j}{C_j} \quad {\left( \forall j \in \indexSet[J] \right)}}{\Subtype{\IntCMVmix{q}{\Set{B_i}_{i \in \indexSet}}}{\IntCMVmix{q}{\Set{C_j}_{j \in \indexSet[J]}}}}
	\hspace{2em}
	\dfrac{\indexSet \subseteq \indexSet[J] \quad \Subtype{B_i}{C_i} \quad {\left( \forall i \in \indexSet \right)}}{\Subtype{\ExtCMVmix{q}{\Set{B_i}_{i \in \indexSet}}}{\ExtCMVmix{q}{\Set{C_j}_{j \in \indexSet[J]}}}}
	\vspace{0.5em}\\
	\dfrac{\Subtype{T\Set{\Subst{\RecCMVmix{t}{T}}{t}}}{U}}{\Subtype{\RecCMVmix{t}{T}}{U}}
	\hspace{2em}
	\dfrac{\Subtype{T}{U\Set{\Subst{\RecCMVmix{t}{U}}{t}}}}{\Subtype{T}{\RecCMVmix{t}{U}}}
\end{array}\end{displaymath}

The predicate $ \UnT{\cdot} $ that is defined by the rules
\begin{displaymath}\begin{array}{c}
	\UnT{\finCMVmix}
	\hspace{2em}
	\UnT{\unitT}
	\hspace{2em}
	\UnT{\boolT}
	\hspace{2em}
	\UnT{\ChoiceTCMVmix{\unCMVmix}{\#}{\Set{B_i}_{i \in \indexSet}}}
	\hspace{2em}
	\dfrac{\UnT{T}}{\UnT{\RecCMVmix{t}{T}}}
\end{array}\end{displaymath}
identifies unrestricted types, \ie types without an unguarded linear choice type.

Typing contexts $ \Gamma $ collect assignments $ \At{x}{T} $ of names to their types.
We extend the predicate $ \UnT{\cdot} $ to a typing context $ \Gamma $, by requiring that for $ \UnT{\Gamma} $ all types in $ \Gamma $ are unrestricted.
In contrast, all typing contexts are linear, denoted as $ \Gamma \, \linCMVmix $.
The operation $ \cdot \circ \cdot $ allows to split a typing context into two typing contexts provided that all assignments with linear types are on distinct names.
Assignments with unrestricted types can be shared by the two parts.
\begin{displaymath}\begin{array}{c}
	\cdot = \cdot \circ \cdot
	\hspace{2em}
	\dfrac{\Gamma_1 \circ \Gamma_2 = \Gamma \quad \UnT{T}}{\Gamma, \At{x}{T} = {\left( \Gamma_1, \At{x}{T} \right)} \circ {\left( \Gamma_2, \At{x}{T} \right)}}
	\vspace{0.5em}\\
	\dfrac{\Gamma_1 \circ \Gamma_2 = \Gamma}{\Gamma, \At{x}{\linCMVmix \, p} = {\left( \Gamma_1, \At{x}{\linCMVmix \, p} \right)} \circ \Gamma_2}
	\hspace{2em}
	\dfrac{\Gamma_1 \circ \Gamma_2 = \Gamma}{\Gamma, \At{x}{\linCMVmix \, p} = \Gamma_1 \circ {\left( \Gamma_2, \At{x}{\linCMVmix \, p} \right)}}
\end{array}\end{displaymath}

The operation $ \cdot + \cdot $ adds a new assignment to a typing context, while ensuring that in a typing context all assignments are on pairwise distinct names and an assignment can be added to a typing context twice only if its type is unrestricted.
\begin{displaymath}\begin{array}{c}
	\dfrac{\At{x}{U} \notin \Gamma}{\Gamma + \At{x}{T} = \Gamma, \At{x}{T}}
	\hspace{2em}
	\dfrac{\UnT{T}}{{\left( \Gamma, \At{x}{T} \right)} + \At{x}{T} = \Gamma, \At{x}{T}}
\end{array}\end{displaymath}

\begin{figure}[tp]
	\begin{displaymath}\begin{array}{c}
		\ruleTUnitCMVmix \; \dfrac{\UnT{\Gamma}}{\Gamma \vdash \At{\unit}{\unitT}}
		\vspace{0.5em}\\
		\ruleTTrueCMVmix \; \dfrac{\UnT{\Gamma}}{\Gamma \vdash \At{\true}{\boolT}}
		\hspace{2em}
		\ruleTFalseCMVmix \; \dfrac{\UnT{\Gamma}}{\Gamma \vdash \At{\false}{\boolT}}
		\vspace{0.5em}\\
		\ruleTVarCMVmix \; \dfrac{\UnT{\Gamma_1, \Gamma_2}}{\Gamma_1, \At{x}{T}, \Gamma_2 \vdash \At{x}{T}}
		\hspace{2em}
		\ruleTSubCMVmix \; \dfrac{\Gamma \vdash \At{v}{T} \quad \Subtype{T}{U}}{\Gamma \vdash \At{v}{U}}
		\vspace{0.5em}\\
		\ruleTOutCMVmix \; \dfrac{\Gamma_1 \vdash \At{v}{T} \quad \Gamma_2 \vdash P}{\Gamma_1 \circ \Gamma_2 \vdash \At{\OutCMVmix{\Label}{v}{P}}{\OutCMVmix{\Label}{T}{U}}}
		\hspace{2em}
		\ruleTInCMVmix \; \dfrac{\Gamma, \At{x}{T} \vdash P}{\Gamma \vdash \At{\InpCMVmix{\Label}{x}{P}}{\InpCMVmix{\Label}{T}{U}}}
		\vspace{0.5em}\\
		\ruleTInactCMVmix \; \dfrac{\UnT{\Gamma}}{\Gamma \vdash \inactCMVmix}
		\hspace{2em}
		\ruleTParCMVmix \; \dfrac{\Gamma_1 \vdash P_1 \quad \Gamma_2 \vdash P_2}{\Gamma_1 \circ \Gamma_2 \vdash P_1 \mid P_2}
		\vspace{0.5em}\\
		\ruleTIfCMVmix \; \dfrac{\Gamma_1 \vdash \At{v}{\boolT} \quad \Gamma_2 \vdash P \quad \Gamma_2 \vdash Q}{\Gamma_1 \circ \Gamma_2 \vdash \ConditionalCMVmix{v}{P}{Q}}
		\hspace{1em}
		\ruleTResCMVmix \; \dfrac{\Gamma, \At{x}{T}, \At{y}{U} \vdash P \quad \Dual{T}{U}}{\Gamma \vdash \ResCMVmix{x}{y}{P}}
		\vspace{0.5em}\\
		\ruleTChoiceCMVmix \; \dfrac{\begin{array}{c} {\left( \Gamma_1 \circ \Gamma_2 \right)} \, q_1 \quad \Gamma_1 \vdash \At{x}{\ChoiceTCMVmix{q_2}{\#}{\Set{\BranchCMVmix{\Label}{*_i}{T_i}{U_i}}_{i \in \indexSet}}}\\ \Gamma_2 + \At{x}{U_j} \vdash \At{\BranchCMVmix{\Label}{*_j}{v_j}{P_j}}{\BranchCMVmix{\Label}{*_j}{T_j}{U_j}} \quad \Set{\Label *_j}_{j \in \indexSet[J]} = \Set{\Label *_i}_{i \in \indexSet} \quad {\left( \forall j \in \indexSet[J] \right)} \end{array}}{\Gamma_1 \circ \Gamma_2 \vdash \ChoiceCMVmix{q_1}{x}{\sum_{j \in \indexSet[J]} \BranchCMVmix{\Label}{*_j}{v_j}{P_j}}}
	\end{array}\end{displaymath}
	\caption{Typing Rules of \CMVmix.}
	\label{fig:typingRulesCMVmix}
\end{figure}

A process $ P $ is well-typed if there is some typing context $ \Gamma $ such that the \emph{type judgement} $ \Gamma \vdash P $ can be derived from the typing rules in Figure~\ref{fig:typingRulesCMVmix}.

The typing Rules~\ruleTUnitCMVmix, \ruleTTrueCMVmix, and \ruleTFalseCMVmix type constants.
Rule~\ruleTVarCMVmix checks the type of a variable against its type as stored in the typing context.
With Rule~\ruleTSubCMVmix we can use subtyping in type derivations.
The Rules~\ruleTOutCMVmix and \ruleTInCMVmix type output and input branches.
They check that the label and polarity are as described by the type and check the continuation of the branch against the continuation of the type.
Moreover, the type of the submitted value in output branches is checked, whereas for input branches we add a suitable assumption on the type of the variable to the typing context.
Rule~\ruleTInactCMVmix checks that the typing context for inactive processes is unrestricted.
Rule~\ruleTParCMVmix splits the typing context for checking the two parts of a parallel composition.
A conditional is well-typed if its condition is boolean and if its two subterms are well-typed as specified by Rule~\ruleTIfCMVmix.
To check the subterm of restriction with Rule~\ruleTResCMVmix, we have to add two assignments for the two endpoints of the restricted channel such that the respective types are dual.
Choices are checked with Rule~\ruleTChoiceCMVmix.
It requires that the typing environment is unrestricted ($ \unCMVmix $) if and only if the analysed choice is qualified as $ \unCMVmix $; else both need to be linear ($ \linCMVmix $).
Then the typing context needs to assign an external or internal choice type to the channel endpoint of this choice, where the qualifier in the type is $ \linCMVmix $ if the choice is qualified as $ \linCMVmix $.
Finally, Rule~\ruleTChoiceCMVmix checks all branches of the choice term against the branches of the choice type.

The language \CMVmix is composed of well-typed processes and the semantics in Figure~\ref{fig:semanticsCMVmix}, \ie $ \CMVmix = \Tuple{\procCMVmix, \stepCMVmix} $, where $ \procCMVmix $ is the well-typed fragment of $ \procCMVmixUntyped $.

The language \CMV is the fragment of \CMVmix with a standard branching construct instead of mixed choice (compare to \cite{casalMordidoVasconcelos22}).
The set of untyped processes $ \procCMVUntyped $ replaces the choice construct of $ \procCMVmixUntyped $ by the following four constructs
\begin{align*}
	\OutCMV{y}{v}{P} \sepTerms \InpCMV{q}{y}{x}{P} \sepTerms \SelCMV{x}{\Label}{P} \sepTerms \BranCMV{x}{\Set{\BranchCMV{\Label_i}{P_i}}_{i \in \indexSet}}
\end{align*}
and keeps the constructs for parallel composition, restriction, conditionals, and inaction.
Output is implemented by $ \OutCMV{y}{v}{P} $ and $ \InpCMV{q}{y}{x}{P} $ implements an input.
Selection $ \SelCMV{x}{\Label}{P} $ allows to select the branch with label $ \Label $ from a branching $ \BranCMV{x}{\Set{\BranchCMV{\Label_i}{P_i}}_{i \in \indexSet}} $ provided that $ \Label \in \Set{ \Label_i }_{i \in \indexSet} $.

\begin{figure}[tp]
	\begin{displaymath}\begin{array}{c}
		\ruleRLinComCMV \; \ResCMV{x}{y}{{\left( \OutCMV{x}{v}{P} \mid \InpCMV{\linCMV}{y}{z}{Q} \mid R \right)}} \stepCMV \ResCMV{x}{y}{{\left( P \mid Q\Set{\Subst{v}{z}} \mid R \right)}}
		\vspace{0.5em}\\
		\ruleRUnComCMV \; \ResCMV{x}{y}{{\left( \OutCMV{x}{v}{P} \mid \InpCMV{\unCMV}{y}{z}{Q} \mid R \right)}} \stepCMV \ResCMV{x}{y}{{\left( P \mid Q\Set{\Subst{v}{z}} \mid \InpCMV{\unCMV}{y}{z}{Q} \mid R \right)}}
		\vspace{0.5em}\\
		\ruleRCaseCMV \; \dfrac{j \in \indexSet}{\ResCMV{x}{y}{{\left( \SelCMV{x}{\Label_j}{P} \mid \BranCMV{y}{\Set{\BranchCMV{\Label_i}{Q_i}}_{i \in \indexSet}} \mid R \right)}} \stepCMV \ResCMV{x}{y}{{\left( P \mid Q_j \mid R \right)}}}
	\end{array}\end{displaymath}
	and the Rules \ruleRIfTCMVmix, \ruleRIfFCMVmix, \ruleRParCMVmix, \ruleRResCMVmix, and \ruleRStructCMVmix from Figure~\ref{fig:semanticsCMVmix}.
	\caption{Reduction Rules ($ \stepCMV $) of \CMV.}
	\label{fig:semanticsCMV}
\end{figure}

The reduction semantics of \CMV is given in Figure~\ref{fig:semanticsCMV}.
Instead of the four communication rules in \CMVmix, we have two communication rules---one for a linear input and one for an unrestricted input---and a rule for branching.
The remaining Rules~\ruleRIfTCMV, \ruleRIfFCMV, \ruleRParCMV, \ruleRResCMV, and \ruleRStructCMV as well as the rules of structural congruence are inherited from \CMVmix.
Also the notions of free names inherited from \CMVmix.
The definition of barbs has to be adapted.
The process $ P $ emits the barb $ y $, denoted as $ P\Barb{y} $, if $ P $ has an unguarded output $ \OutCMV{y}{v}{P} $ or an unguarded input $ \InpCMV{q}{y}{x}{P} $ or an unguarded selection $ \SelCMV{y}{\Label}{P} $ or an unguarded branching $ \BranCMV{y}{\Set{\BranchCMV{\Label_i}{P_i}}_{i \in \indexSet}} $ on a free channel endpoint $ y \in \FreeNames{P} $.

The set of types of \CMV replaces the choice construct in the definition of types of \CMVmix by the following two constructs
\begin{align*}
	\ComTCMV{q}{{*}}{T}{T} \sepTerms \ChoiceTCMV{q}{\#}{\Set{\BranchCMV{\Label_i}{T_i}}_{i \in \indexSet}}
\end{align*}
and keeps the constructs for inaction, base types, and recursion.

We adapt the typing rule that allows to compare types for choices to the simpler rule
\begin{displaymath}\begin{array}{c}
	\dfrac{T_i \simeq T_i'}{\ChoiceTCMV{q}{\#}{\Set{\BranchCMV{\Label_i}{T_i}}_{i \in \indexSet}} \simeq \ChoiceTCMV{q}{\#}{\Set{\BranchCMV{\Label_i}{T_i'}}_{i \in \indexSet}}}
\end{array}\end{displaymath}
and keep the remaining rules for inaction, base types, and recursion as well as the rules for type equivalence.

The following two rules replace the rule for choice in the definition of duality.
\begin{displaymath}\begin{array}{c}
	\dfrac{\Dual{\bullet}{{*}} \quad T_1 \simeq T_2 \quad \Dual{U_1}{U_2}}{\Dual{\ComTCMV{q}{\bullet}{T_1}{U_1}}{\ComTCMV{q}{{*}}{T_2}{U_2}}}
	\hspace{2em}
	\dfrac{\Dual{\#}{\flat} \quad \Dual{T_i}{U_i} \quad {\left( \forall i \in \indexSet \right)}}{\Dual{\ChoiceTCMV{q}{\#}{\Set{\BranchCMV{\Label_i}{T_i}}_{i \in \indexSet}}}{\ChoiceTCMV{q}{\flat}{\Set{\BranchCMV{\Label_i}{U_i}}_{i \in \indexSet}}}}
\end{array}\end{displaymath}
We keep the rules for polarities, views, inaction, and recursion.

The following four rules replace the subtyping rules for inputs, outputs, and choice.
\begin{displaymath}\begin{array}{c}
	\dfrac{\Subtype{U}{T} \quad \Subtype{T'}{U'}}{\Subtype{\ComTCMV{q}{!}{T}{T'}}{\ComTCMV{q}{!}{U}{U'}}}
	\hspace{2em}
	\dfrac{\Subtype{T}{U} \quad \Subtype{T'}{U'}}{\Subtype{\ComTCMV{q}{?}{T}{T'}}{\ComTCMV{q}{?}{U}{U'}}}
	\vspace{0.5em}\\
	\dfrac{\indexSet[J] \subseteq \indexSet \quad \Subtype{T_j}{U_j} \quad {\left( \forall j \in \indexSet[J] \right)}}{\Subtype{\ChoiceTCMV{q}{\oplus}{\Set{\BranchCMV{\Label_i}{T_i}}_{i \in \indexSet}}}{\ChoiceTCMV{q}{\oplus}{\Set{\BranchCMV{\Label_j}{U_j}}_{j \in \indexSet[J]}}}}
	\hspace{2em}
	\dfrac{\indexSet \subseteq \indexSet[J] \quad \Subtype{T_i}{U_i} \quad {\left( \forall i \in \indexSet \right)}}{\Subtype{\ChoiceTCMV{q}{\&}{\Set{\BranchCMV{\Label_i}{T_i}}_{i \in \indexSet}}}{\ChoiceTCMV{q}{\&}{\Set{\BranchCMV{\Label_j}{U_j}}_{j \in \indexSet[J]}}}}
\end{array}\end{displaymath}
We keep the subtyping rules for inaction, base types, and recursion.

The following two rules replace the rule for choice in the definition of the predicate $ \UnT{\cdot} $.
\begin{displaymath}\begin{array}{c}
	\UnT{\ComTCMV{\unCMV}{{*}}{T}{U}}
	\hspace{2em}
	\UnT{\ChoiceTCMV{\unCMV}{\#}{\Set{\BranchCMV{\Label_i}{T_i}}_{i \in \indexSet}}}
\end{array}\end{displaymath}
We keep the rules for inaction, base types, and recursion.

\begin{figure}[tp]
	\begin{displaymath}\begin{array}{c}
		\ruleTOutCMV \; \dfrac{\Gamma_1 \vdash \At{x}{\ComTCMV{q}{!}{T}{U}} \quad \Gamma_2 + \At{x}{U} = \Gamma_3 \circ \Gamma_4 \quad \Gamma_3 \vdash \At{v}{T} \quad \Gamma_4 \vdash P}{\Gamma_1 \circ \Gamma_2 \vdash \OutCMV{x}{v}{P}}
		\vspace{0.5em}\\
		\ruleTInCMV \; \dfrac{{\left( \Gamma_1 \circ \Gamma_2 \right)} \, q_1 \quad \Gamma_1 \vdash \At{x}{\ComTCMV{q_2}{?}{T}{U}} \quad {\left( \Gamma_2 + \At{x}{U} \right)}, \At{y}{T} \vdash P}{\Gamma_1 \circ \Gamma_2 \vdash \InpCMV{q_1}{x}{y}{P}}
		\vspace{0.5em}\\
		\ruleTBranchCMV \; \dfrac{\Gamma_1 \vdash \At{x}{\ChoiceTCMV{q}{\&}{\Set{\BranchCMV{\Label_i}{T_i}}_{i \in \indexSet}}} \quad \Gamma_2 + \At{x}{T_i} \vdash P_i \quad {\left( \forall i \in \indexSet \right)}}{\Gamma_1 \circ \Gamma_2 \vdash \BranCMV{x}{\Set{\BranchCMV{\Label_i}{P_i}}_{i \in \indexSet}}}
		\vspace{0.5em}\\
		\ruleTSelCMV \; \dfrac{\Gamma_1 \vdash \At{x}{\ChoiceTCMV{q}{\oplus}{\Set{\BranchCMV{\Label}{T}}}} \quad \Gamma_2 + \At{x}{T} \vdash P}{\Gamma_1 \circ \Gamma_2 \vdash \SelCMV{x}{\Label}{P}}
	\end{array}\end{displaymath}
	and the Rules \ruleTUnitCMVmix, \ruleTTrueCMVmix, \ruleTFalseCMVmix, \ruleTVarCMVmix, \ruleTIfCMVmix, \ruleTSubCMVmix, \ruleTInactCMVmix, \ruleTParCMVmix, and \ruleTResCMVmix.
	\caption{Typing Rules of \CMV.}
	\label{fig:typingRulesCMV}
\end{figure}

The typing rules of \CMV are depicted in Figure~\ref{fig:typingRulesCMV}.
The Rules~\ruleTOutCMVmix, \ruleTInCMVmix, and \ruleTChoiceCMVmix are replaced by the depicted rules.
We inherit the remaining Rules, \ie the Rules \ruleTUnitCMV, \ruleTTrueCMV, \ruleTFalseCMV, \ruleTVarCMV, \ruleTSubCMV, \ruleTInactCMV, \ruleTParCMV, \ruleTIfCMV, and \ruleTResCMV from \CMVmix.

Again, $ \CMV = \Tuple{\procCMV, \stepCMV} $, where $ \procCMV $ is the well-typed fragment of $ \procCMVUntyped $.

\subsection{Encodings, Quality Criteria, and Distributability}
\label{app:encodings}

Let $ \sourceLang = \ProcCal{\procSource}{\stepSource} $ and $ \targetLang = \ProcCal{\procTarget}{\stepTarget} $ be two (untyped or typed) process calculi, denoted as \emph{source} and \emph{target language}.
In the simplest case, an \emph{encoding} from $ \sourceLang $ into $ \targetLang $ is a function $ \arbitraryEncoding : \procSource \to \procTarget $ that translates source terms into target terms.
If the source and target language are typed then we assume an additional encoding function on types and allow the encoding function on terms to use informations about the type of source terms.
We often use $ S, S', S_1, \ldots $ to range over $ \procSource $ and $ T, T', T_1, \ldots $ to range over $ \procTarget $. Encodings often translate single source term steps into a sequence or pomset of target term steps. We call such a sequence or pomset an \emph{emulation} of the corresponding source term step.

Within a single calculus systems are usually compared up to some form of simulation relation that uses the observables of the language to compare the behaviour of the systems. Comparing systems of different languages is more difficult, because they might not share the same set of observables.
In order to provide a general framework, Gorla in \cite{gorla10} suggests five criteria well suited for language comparison, because they are language independent and as shown in \cite{petersGlabbeek15} induce some kind of simulation relation between a source term and its literal translation. They are divided into two structural and three semantic criteria. The structural criteria include
(1) \emph{compositionality} and
(2) \emph{name invariance}. The semantic criteria include
(3) \emph{operational correspondence},
(4) \emph{divergence reflection}, and
(5) \emph{success sensitiveness}.
These criteria are well suited for \emph{encodability} and \emph{separation} results.
An encodability result proves the existence of an encoding, where the criteria rule out trivial or meaningless encodings.
A separation result separates two languages by showing that no encoding that satisfies the criteria exists, where the criteria are minimal assumptions on reasonable encodings.

The combination of the semantic criteria ensures that source terms and their literal translation are coupled similar (see \cite{petersGlabbeek15}), where success sensitiveness (\ie a form of testing) is used instead of observables.
In this paper, we consider languages that do not have the same barbs but barbs that are similar enough to allow for comparisons, because all considered languages are based on the \piCal.
Because of that, we replace the criterion of success sensitiveness by the slightly stronger criterion of barb sensitiveness.
We claim that all separation results of this paper remain valid if we replace barb sensitiveness with success sensitiveness. In this case the counterexamples need to be adapted to the reachability of success.

Note that a behavioural equivalence $ \asymp $ on the target language is assumed for the definition of name invariance and operational correspondence. Its purpose is to describe the abstract behaviour of a target process, where abstract refers to the behaviour of the source term.
Moreover, let $ \varphi : \names \to \names^k $ be a \emph{renaming policy}, \ie a mapping from a (source term) name to a vector of (target term) names that can be used by encodings to split names and to reserve special names, such that no two different names are translated into overlapping vectors of names and reserved names are not confused with translated source term names.

Intuitively, an encoding is compositional if the translation of an operator is the same for all occurrences of that operator in a term. Hence, the translation of that operator can be captured by a context that is allowed in \cite{gorla10} to be parametrised on the free names of the respective source term.

\begin{definition}[Compositionality, \cite{gorla10}]
	\label{def:compositionality}
	The encoding $ \arbitraryEncoding $ is \emph{compositional} if, for every operator $ \mathbf{op} : \names^n \times \procSource^m \to \procSource $ of $ \sourceLang $ and for every subset of names $ N $, there exists a context $ \Context{N}{\mathbf{op}}{\hole_1, \ldots , \hole_{n' + m}} : \names^{n'} \times \procSource^m \to \procTarget $ and $ y_1, \ldots, y_{n'} \in \names $ such that, for all $ x_1, \ldots, x_n \in \names $ and all $ S_1, \ldots, S_m \in \procSource $ with $ \FreeNames{S_1} \cup \ldots \cup \FreeNames{S_m} = N $ and $ \Set{ y_1, \ldots, y_{n'} } \subseteq \varphi\!\left( x_1 \right) \cup \ldots \cup \varphi\!\left( x_n \right) $, it holds that: $$ \ArbitraryEncoding{\mathbf{op}\left( x_1, \ldots, x_n, S_1, \ldots, S_m \right)} = \Context{N}{\mathbf{op}}{y_1, \ldots, y_{n'}, \ArbitraryEncoding{S_1}, \ldots, \ArbitraryEncoding{S_m}} $$
\end{definition}

Name invariance ensures that encodings are independent of specific names in the source.
We use projection to obtain the respective elements of a translated name, \ie if $ \varphi(a) = \left( a_1, a_2, a_3 \right) $ then $ \varphi(a).2 = a_2 $.
Slightly abusing notation, we sometimes use the tuples that are generated by the renaming policy as sets, \ie we require \eg $ \varphi(a) \cap \varphi(b) = \emptyset $ whenever $ a \neq b $.
An encoding is name invariant if it preserves substitutions modulo the
relation $ \asymp $ on the target language.

\begin{definition}[Name Invariance, \cite{gorla10}]
	\label{def:nameInvariance}
	The encoding $ \arbitraryEncoding $ is \emph{name invariant} \wrt $ \asymp $ if, for every $ S \in \procSource $ and every substitution $ \sigma $, it holds that
	\begin{align*}
		\ArbitraryEncoding{S\sigma}
		\begin{cases}
			= \ArbitraryEncoding{S}\sigma' & \text{if } \sigma \text{ is injective}\\
			\asymp \ArbitraryEncoding{S}\sigma' & \text{otherwise}
		\end{cases}
	\end{align*}
	where $ \sigma' $ is such that $ \varphi(\sigma(a)) = \sigma'{\left( \varphi(a) \right)} $ for all $ a \in \names $.
\end{definition}

To simplify the presentation in the workshop paper that we presented at EXPRESS/SOS'22, we omit the renaming policy and instead assumed that the names reserved by the encoding function from \CMVmix into \CMV of \cite{casalMordidoVasconcelos22} are different from all source term names.
Under this assumption the encoding of \cite{casalMordidoVasconcelos22} satisfies the variant
\begin{center}
	For every $ S $ and every substitution $ \sigma $, it holds that $ \ArbitraryEncoding{S\sigma} \asymp \ArbitraryEncoding{S}\sigma $.
\end{center}
of name invariance, that we present as name invariance criterion in our workshop paper.
Indeed the purpose of the renaming policy is to allow to implement such an assumption.
When we prove the correctness of the encoding from \CMVmix into \CMV of \cite{casalMordidoVasconcelos22}, we consider both variants of name invariance.

The first semantic criterion is operational correspondence. It consists of a soundness and a completeness condition. \emph{Completeness} requires that every computation of a source term can be emulated by its translation. \emph{Soundness} requires that every computation of a target term corresponds to some computation of the corresponding source term.

\begin{definition}[Operational Correspondence, \cite{gorla10}]
	\label{def:operationalCorrespondence}
	The encoding $ \arbitraryEncoding $ satisfies \emph{operational correspondence} if it satisfies:
	\begin{center}
		{\tabcolsep3pt
		\begin{tabular}{ll}
			\emph{Completeness}: & For all $ S \stepsSource S' $, it holds $ \ArbitraryEncoding{S} \stepsTarget \asymp \ArbitraryEncoding{S'} $.\\
			\emph{Soundness}: & For all $ \ArbitraryEncoding{S} \stepsTarget T $, there exists an $ S' $ such that $ S \stepsSource S' $ and $ T \stepsTarget \asymp \ArbitraryEncoding{S'} $.
		\end{tabular}}
	\end{center}
\end{definition}

\noindent
The definition of operational correspondence relies on the equivalence $ \asymp $ to get rid of junk possibly left over within computations of target terms. Sometimes, we refer to the completeness criterion of operational correspondence as \emph{operational completeness} and, accordingly, for the soundness criterion as \emph{operational soundness}.

The next criterion concerns the role of infinite computations in encodings.

\begin{definition}[Divergence Reflection, \cite{gorla10}]
	\label{def:divergenceReflection}
	The encoding $ \arbitraryEncoding $ \emph{reflects divergence} if, for every source term $ S $, $ \ArbitraryEncoding{S} \stepTarget^{\omega} $ implies $ S \stepSource^{\omega} $.
\end{definition}

The last criterion links the behaviour of source terms to the behaviour of their encodings.

\begin{definition}[Barb Sensitiveness, \cite{petersGlabbeek15}]
	\label{def:barbSensitiveness}
	The encoding $ \arbitraryEncoding $ is \emph{barb-sensitive} if, for every source term $ S $ and every barb $ y $, $ S\WeakBarb{y} $ iff $ \ArbitraryEncoding{S}\WeakBarb{y} $.
\end{definition}

\noindent
This criterion only links the behaviours of source terms and their literal translations, but not of their derivatives. To do so, Gorla relates success sensitiveness and operational correspondence by requiring that the equivalence on the target language $ \asymp $ never relates two processes with different success behaviours.
Similarly, we require that $ \asymp $ respects barbs.

\begin{definition}[Barb Respecting]
	\label{def:asympBarbRespecting}
	$ \asymp $ is \emph{barb respecting} if, for every $ P $ and $ Q $ and every barb $ y $ with $ P\WeakBarb{y} $ and $ Q\NotWeakBarb{y} $, it holds that $ P \not\asymp Q $.
\end{definition}

\noindent
According to \cite{gorla10} a ``good'' equivalence $ \asymp $ is often defined in the form of a barbed equivalence (as described \eg in \cite{milnerSangiorgi92}) or can be derived directly from the reduction semantics and is often a congruence, at least with respect to parallel composition. For the separation results presented in this paper, we require only that $ \asymp $ is a barb respecting reduction bisimulation.

Since \cite{casalMordidoVasconcelos22} considers an encoding between two typed languages, they use an additional criterion, called \emph{type soundness}.
It requires that a source term that is well-typed \wrt some type environment $ \Gamma $ is translated into a target term that is well-typed \wrt the translation of $ \Gamma $.
We do not explicitly consider this criterion here, because it was already shown in \cite{casalMordidoVasconcelos22} that it is satisfied for the encoding $ \EncCMVmixCMV{\cdot} $ from \CMVmix into \CMV presented in \cite{casalMordidoVasconcelos22}.

Both of the papers \cite{casalMordidoVasconcelos22} and \cite{palamidessi97} require as additional criterion that the parallel operator is translated homomorphically.
As explained in \cite{palamidessi97} this criterion was mend to ensure that encodings preserve the degree of distribution in terms.
Indeed, \cite{petersNestmann12} presents an encoding of the \piCal with mixed choice into the asynchronous \piCal without choice that respects all of the above criteria.
Requiring that the degree of distribution is preserved is essential for the separation result in \cite{palamidessi97}.
Unfortunately, as explained in \cite{petersNestmann12, peters12} the homomorphic translation of the parallel operator is rather strict and rules out encodings that intuitively do preserve the degree of distribution.
Because of that, \cite{petersNestmann12, peters12, petersNestmannGoltz13} propose an alternative criterion for the preservation of the degree of distribution that we will use here to strengthen our separation results.
The encoding of \cite{casalMordidoVasconcelos22} that we discuss in \S~\ref{sec:encodeMixedSessions} translates the parallel operator homomorphically.

Intuitively, a distribution of a process means the extraction (or: separation) of its (sequential) components and their association to different locations.
Since all languages considered in this paper are based on the \piCal, we can rely on the intuition that the parallel operator splits locations.
Accordingly, a process $ P $ is distributable into $ P_1, \ldots, P_n $ if and only if we have $ P \equiv \ResPi{\tilde{y}}{\left( P_1 \mid \ldots \mid P_n \right)} $ for $ P \in \procPi $ or $ P \equiv \ResCMVmix{\tilde{y}}{\tilde{z}}{\left( P_1 \mid \ldots \mid P_n \right)} $ for $ P \in \procCMV $ or $ P \in \procCMVmix $.

Preservation of distributability  means that the target term is at least as distributable as the source~term.

\begin{definition}[Preservation of Distributability, \cite{petersNestmannGoltz13}]
	\label{def:distributabilityPreservation}
	An encoding $ \arbitraryEncoding : \procSource \to \procTarget $ \emph{preserves distributability} if for every $ S \in \procSource $ and for all terms $ S_1, \ldots, S_n \in \procSource $ that are distributable within $ S $ there are some $ T_1, \ldots, T_n \in \procTarget $ that are distributable within $ \ArbitraryEncoding{S} $ such that $ T_i \asymp \ArbitraryEncoding{S_i} $ for all $ 1 \leq i \leq n $.
\end{definition}

In essence, this requirement is a distributability-enhanced adaptation of operational completeness.
It respects both the intuition on distribution as separation on different locations---an encoded source term is at least as distributable as the source term itself---as well as the intuition on distribution as independence of processes and their executions---implemented by $ T_i \asymp \ArbitraryEncoding{S_i} $.

The preservation of distributability completes our set or criteria for encodings.

\begin{definition}[Good Encoding]
	\label{def:goodEncoding}
	We consider an encoding $ \arbitraryEncoding $ to be \emph{good} if it
	(1) is compositional,
	(2) is name invariant,
	(3) satisfies operational correspondence,
	(4) reflects divergence,
	(5) is barb-sensitive, and
	(6) preserves distributability.
	Moreover we require that the equivalence $ \asymp $ is a barb respecting (weak) reduction bisimulation.
\end{definition}

We inherit some of the machinery introduced in \cite{petersNestmannGoltz13} to work with distributability.
Note that in contrast to \cite{petersNestmannGoltz13} we do not need to distinguish between parallel and distributable processes or steps, because all considered languages in this paper are based on the \piCal.
If a single process can perform two different steps, then we call these steps alternative to each other.
Two alternative steps are in \emph{conflict}, if performing one step disables the other step. Otherwise they are \emph{distributable}.
For instance the reductions on the channels $ a $ and $ b $ are distributable in the term $ \overline{a} \mid \overline{b} \mid a \mid b $, but they are in conflict in $ \overline{a} \mid \overline{b} \mid a + b $, because the choice is reduced in both steps.
More precisely, two steps in the \piCal are in conflict if they reduce the same choice, two steps in \CMVmix are in conflict if they reduce the same choice or the same conditional, and two steps in \CMV are are in conflict if the reduce the same output, input, selection prefix, branching prefix, or conditional.
Note that reducing the same choice not necessarily means to reduce the same summand in this choice.
We lift the definition of conflict and distributable steps to executions, \ie sequences of steps.

\begin{definition}[Distributable Executions]
	\label{def:distributableSequences}
	Let $ \ProcCal{\proc}{\step} $ be a process calculus, $ P \in \proc $, and let $ A $ and $ B $ denote two executions of $ P $. $ A $ and $ B $ are in \emph{conflict}, if a step of $ A $ and a step of $ B $ are in conflict, else $ A $ and $ B $ are \emph{distributable}.
\end{definition}

As shown in \cite{petersNestmannGoltz13}, two executions of a term $ P $ are distributable iff $ P $ is distributable into two subterms such that each performs one of these executions.

\begin{lemma}[Distributable Executions, \cite{petersNestmannGoltz13}]
	\label{lem:distributabilityReductionsVsProcesses}
	Let $ \lang = \ProcCal{\proc}{\step} $ be a process calculus, $ P \in \proc $, and $ A_1, \ldots, A_n $ a set of executions of $ P $.
	The executions $ A_1, \ldots, A_n $ are pairwise distributable within $ P $ iff $ P $ is distributable into $ P_1, \ldots, P_n \in \proc $ such that, for all $ 1 \leq i \leq n $, $ A_i $ is an execution of $ P_i $, \ie during $ A_i $ only parts of $ P_i $ are reduced or removed.
\end{lemma}

Because of that, an operationally complete encoding is distributability-preserving only if it preserves the distributability of sequences of source term steps.

\begin{lemma}[Distributability-Preservation, \cite{petersNestmannGoltz13}]
	\label{lem:distributabilityPreservation}
	An operationally complete encoding $ \arbitraryEncoding : \procSource \to \procTarget $ that preserves distributability also preserves distributability of executions, \ie for all source terms $ S \in \procSource $ and all sets of pairwise distributable executions of $ S $, there exists an emulation of each execution in this set such that all these emulations are pairwise distributable in $ \ArbitraryEncoding{S} $.
\end{lemma}


\section{Separating Mixed Sessions and the Pi-Calculus via Leader Election}
\label{sec:separateMixedSessionsLeaderElection}

The first expressiveness result on the \piCal that focuses on mixed choice is the separation result by Palamidessi in \cite{palamidessi97, palamidessi03}.
This result uses the problem of leader election in symmetric networks as distinguishing feature.

Following \cite{palamidessi97} we assume that the set of names $ \names $ contains names that identify the processes of the network and that are never used as bound names within electoral systems.
For simplicity, we use natural numbers for this kind of names.
A leader is announced by unguarding an output on its id.
Then a network $ P = \ResPi{\tilde{x}}{{\left( P_1 \mid \ldots \mid P_k \right)}} $ in $ \procPi $ or $ P = \ResCMVmix{\tilde{x}}{\tilde{y}}{{\left( P_1 \mid \ldots \mid P_k \right)}} $ in $ \procCMVmix $ is an \emph{electoral system} if in every maximal execution exactly one leader is announced.
We adapt the definition of electoral systems of \cite{palamidessi97} to obtain electoral systems in the \piCal and in \CMVmix.

\begin{definition}[Electoral System]
	A network $ P = \ResPi{\tilde{x}}{{\left( P_1 \mid \ldots \mid P_k \right)}} $ in $ \procPi $ or $ P = \ResCMVmix{\tilde{x}}{\tilde{y}}{{\left( P_1 \mid \ldots \mid P_k \right)}} $ in $ \procCMVmix $ is an \emph{electoral system} if for every execution $ E: P \steps P' $ there exists an extension $ E': P \steps P' \steps P'' $ and some $ n \in \Set{ 1, \ldots, k } $ (the leader) such that $ P'''\Barb{n} $ for all $ P''' $ with $ P'' \steps P''' $, but $ P''\NotWeakBarb{m} $ for any $ m \in \Set{1, \ldots, k} $ with $ m \neq n $.
\end{definition}

Accordingly, an electoral system in the \piCal announces a leader by unguarding some output on $ n $ that cannot be reduced or removed, where $ n $ is the id of the leader.
In \CMVmix a leader is announced by unguarding a choice on the channel $ n $.
Since $ n $ is free this choice cannot be removed.
A network is an electoral system if in every maximal execution exactly one leader $ n $ is announced.

We adapt the definition of hypergraphs that are associated to a network of processes in the \piCal defined in \cite{palamidessi97} to networks in \CMVmix.
The hypergraph connects the nodes $ 1, \ldots, k $ of the network by edges representing the free channels that they share, where we ignore the outer restrictions of the network.

\begin{definition}[Hypergraph]
	Given a network $ P = \ResPi{\tilde{x}}{{\left( P_1 \mid \ldots \mid P_k \right)}} $ in $ \procPi $ or $ P = \ResCMVmix{\tilde{x}}{\tilde{y}}{{\left( P_1 \mid \ldots \mid P_k \right)}} $ in $ \procCMVmix $, the \emph{hypergraph} associated to $ P $ is $ \Hypergraph{P} = \Tuple{N, X, t} $ with $ N = \Set{ 1, \ldots, k } $, $ X = \FreeNames{P_1 \mid \ldots \mid P_n} \setminus N $, and $ t(x) = \Set{n \mid x \in \FreeNames{P_n}} $ for each $ x \in X $.
\end{definition}

Because we ignore the outer restrictions of the network in the above definition, the hypergraphs of two structural congruent networks may be different.
However, this is not crucial for our results.

Given a hypergraph $ H = \Tuple{N, X, t} $, an automorphism on $ H $ is a pair $ \sigma = \Tuple{\sigma_N, \sigma_X} $ such that $ \sigma_N: N \to N $ and $ \sigma_X: X \to X $ are permutations which preserve the type of arcs.
For simplicity, we usually do not distinguish between $ \sigma_N $ and $ \sigma_X $ and simply write $ \sigma $.
Moreover, since $ \sigma $ is a substitution, we allow to apply $ \sigma $ on terms $ P $, denoted as $ P\sigma $.
The orbit $ \Orbit{\sigma}{n} $ of $ n \in N $ generated by $ \sigma $ is defined as the set of nodes in which the various iterations of $ \sigma $ map $ n $, \ie $ \Orbit{\sigma}{n} = \Set{ n, \sigma(n), \ldots, \sigma^{h - 1}(n) } $, where $ \sigma^i $ represents the composition of $ \sigma $ with itself $ i $ times and $ \sigma^h = \id $.
We also adapt the notion of a symmetric system of \cite{palamidessi97} to obtain symmetric systems in the \piCal as well as in \CMVmix.

\begin{definition}[Symmetric System]
	Consider a network $ P = \ResPi{\tilde{x}}{{\left( P_1 \mid \ldots \mid P_k \right)}} $ in $ \procPi $ or a network $ P = \ResCMVmix{\tilde{x}}{\tilde{y}}{{\left( P_1 \mid \ldots \mid P_k \right)}} $ in $ \procCMVmix $, and let $ \sigma $ be an isomorphism on its associated hypergraph $ \Hypergraph{P} = \Tuple{N, X, t} $. $ P $ is \emph{symmetric \wrt $ \sigma $} iff $ P_{\sigma(i)} \approx_{\pi} P_i\sigma $ or $ P_{\sigma(i)} \approx_{\CMVmix} P_i\sigma $ for each node $ i \in N $. $ P $ is \emph{symmetric} if it is symmetric \wrt all the automorphisms of $ \Hypergraph{P} $.
\end{definition}

\noindent
In contrast to \cite{palamidessi97} we use bisimilarity---$ \approx_{\pi} $ and $ \approx_{\CMVmix} $---instead of alpha conversion in the definition of symmetry.
With this weaker notion of symmetry, we compensate for the weaker criterion on distributability that we use instead of the homomorphic translation of the parallel operator.
Accordingly, we also consider networks as symmetric if they behave in a symmetric way; they do not necessarily need to be structurally symmetric.

In the \piCal we find symmetric electoral systems for many kinds of hypergraphs.
We use such a solution of leader election in a network with five nodes as counterexample to separate \CMVmix from the \piCal.

\begin{example}[Leader Election in the \PiCal]
	\label{exa:leaderElectionPi}
	Consider the network
	\begin{align*}
		\LEPi &= \ResPi{a, b, c, d, e, v, w, x, y, z}{\left( S_1 \mid S_2 \mid S_3 \mid S_4 \mid S_5 \right)} 
	\end{align*}
	where $ S_1 = \overline{e} + a.{\left( \overline{x} + v.\overline{1} \right)} $, $ S_2 = \overline{a} + b.{\left( \overline{y} + w.\overline{2} \right)} $, $ S_3 = \overline{b} + c.{\left( \overline{z} + x.\overline{3} \right)} $, $ S_4 = \overline{c} + d.{\left( \overline{v} + y.\overline{4} \right)} $, and $ S_5 = \overline{d} + e.{\left( \overline{w} + z.\overline{5} \right)} $.
	\qed
\end{example}

\begin{wrapfigure}{R}{0.275\textwidth}
	\centering
	\scalebox{0.8}{
	\begin{tikzpicture}[bend angle=20]
		\foreach \v/\x/\y/\z in {2/1/a/v,1/2/b/w,5/3/c/x,4/4/d/y,3/5/e/z}
        {
            \path (360*\v/5-55:1.75) node[draw, inner sep=0pt, minimum size=3pt] (p\x) {$ \begin{array}{c} \x\\ \textcolor{blue}{\y} \;\; \textcolor{red}{\z} \end{array} $};
        }
        \draw[-latex, color=blue] (p1) edge [bend right] node[above] {$ \overline{e} $} (p5);
        \draw[-latex, color=blue] (p2) edge [bend right] node[above] {$ \overline{a} $} (p1);
        \draw[-latex, color=blue] (p3) edge [bend right] node[right] {$ \overline{b} $} (p2);
        \draw[-latex, color=blue] (p4) edge [bend right] node[below] {$ \overline{c} $} (p3);
        \draw[-latex, color=blue] (p5) edge [bend right] node[left] {$ \overline{d} $} (p4);
        \draw[-latex, color=red] (p1) edge node[pos=0.1, right] {$ \overline{x} $} (p3);
        \draw[-latex, color=red] (p2) edge node[pos=0.1, below] {$ \overline{y} $} (p4);
        \draw[-latex, color=red] (p3) edge node[pos=0.1, below] {$ \overline{z} $} (p5);
        \draw[-latex, color=red] (p4) edge node[pos=0.1, left] {$ \overline{v} $} (p1);
        \draw[-latex, color=red] (p5) edge node[pos=0.1, above] {$ \overline{w} $} (p2);
	\end{tikzpicture}
	}
\end{wrapfigure}

$ \LEPi $ is symmetric.
Consider \eg the permutation $ \sigma $ that permutes the channels as follows: $ a \rightarrow b \rightarrow c \rightarrow d \rightarrow e \rightarrow a $, $ v \rightarrow w \rightarrow x \rightarrow y \rightarrow z \rightarrow v $, and $ 1 \rightarrow 2 \rightarrow 3 \rightarrow 4 \rightarrow 5 \rightarrow 1 $.
Then $ S_{\sigma(i)} = S_i\sigma $ for all $ i \in \Set{ 1, \ldots, 5 } $.
The network elects a leader in two stages.
The first stage (depicted as \textcolor{blue}{blue circle}) uses mixed choices on the channels $ a, b, c, d, e $; in the second stage (depicted as a \textcolor{red}{red star}) we have mixed choices on the channels $ v, w, x, y, z $.
The picture on the right gives $ \Hypergraph{\LEPi} $ extended by arrow heads to visualise the direction of interactions and the respective action prefixes.
The senders in the two stages are losing the leader election game, \ie are not becoming the leader.
In the first stage two processes can be receivers and continue with the second stage.
The process that is neither sender nor receiver in the first stage is stuck and also loses.
The receiver of the second stage then becomes the leader by unguarding an output on its id.
The channels used by $ \LEPi $ in its two stages are summarised in the tabular:
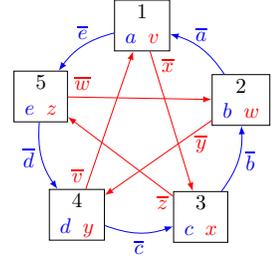
\begin{center}
	\begin{tabular}[b]{lccccc}
		Process ID & $ 1 $ & $ 2 $ & $ 3 $ & $ 4 $ & $ 5 $\\
		Input in First Stage & $ a $ & $ b $ & $ c $ & $ d $ & $ e $\\
		Input in Second Stage & $ v $ & $ w $ & $ x $ & $ y $ & $ z $
	\end{tabular}
	\hspace{4em}
\end{center}
Let $ \tilde{n} = a, b, c, d, e, v, w, x, y, z $.
The network $ \LEPi $ has 10 maximal executions (modulo structural congruence):
\begin{align*}
	\LEPi &\stepPi \ResPi{\tilde{n}}{{\left( \overline{x} + v.\overline{1} \mid S_3 \mid S_4 \mid S_5 \right)}} \stepPi \ResPi{\tilde{n}}{{\left( \overline{x} + v.\overline{1} \mid \overline{z} + x.\overline{3} \mid S_5 \right)}} \stepPi \overline{3} \mid \ResPi{\tilde{n}}{S_5} \noStep\\
	\LEPi &\stepPi \ResPi{\tilde{n}}{{\left( \overline{x} + v.\overline{1} \mid S_3 \mid S_4 \mid S_5 \right)}} \stepPi \ResPi{\tilde{n}}{{\left( \overline{x} + v.\overline{1} \mid S_3 \mid \overline{v} + y.\overline{4} \right)}} \stepPi \overline{1} \mid \ResPi{\tilde{n}}{S_3} \noStep\\
	\LEPi &\stepPi \ResPi{\tilde{n}}{{\left( S_1 \mid \overline{y} + w.\overline{2} \mid S_4 \mid S_5 \right)}} \stepPi \ResPi{\tilde{n}}{{\left( S_1 \mid \overline{y} + w.\overline{2} \mid \overline{v} + y.\overline{4} \right)}} \stepPi \overline{4} \mid \ResPi{\tilde{n}}{S_1} \noStep\\
	\LEPi &\stepPi \ResPi{\tilde{n}}{{\left( S_1 \mid \overline{y} + w.\overline{2} \mid S_4 \mid S_5 \right)}} \stepPi \ResPi{\tilde{n}}{{\left( \overline{y} + w.\overline{2} \mid S_4 \mid \overline{w} + z.\overline{5} \right)}} \stepPi \overline{2} \mid \ResPi{\tilde{n}}{S_4} \noStep\\
	\LEPi &\stepPi \ResPi{\tilde{n}}{{\left( S_1 \mid S_2 \mid \overline{z} + x.\overline{3} \mid S_5 \right)}} \stepPi \ResPi{\tilde{n}}{{\left( \overline{x} + v.\overline{1} \mid \overline{z} + x.\overline{3} \mid S_5 \right)}} \stepPi \overline{3} \mid \ResPi{\tilde{n}}{S_5} \noStep\\
	\LEPi &\stepPi \ResPi{\tilde{n}}{{\left( S_1 \mid S_2 \mid \overline{z} + x.\overline{3} \mid S_5 \right)}} \stepPi \ResPi{\tilde{n}}{{\left( S_2 \mid \overline{z} + x.\overline{3} \mid \overline{w} + z.\overline{5} \right)}} \stepPi \overline{5} \mid \ResPi{\tilde{n}}{S_2} \noStep\\
	\LEPi &\stepPi \ResPi{\tilde{n}}{{\left( S_1 \mid S_2 \mid S_3 \mid \overline{v} + y.\overline{4} \right)}} \stepPi \ResPi{\tilde{n}}{{\left( \overline{x} + v.\overline{1} \mid S_3 \mid \overline{v} + y.\overline{4} \right)}} \stepPi \overline{1} \mid \ResPi{\tilde{n}}{S_3} \noStep\\
	\LEPi &\stepPi \ResPi{\tilde{n}}{{\left( S_1 \mid S_2 \mid S_3 \mid \overline{v} + y.\overline{4} \right)}} \stepPi \ResPi{\tilde{n}}{{\left( S_1 \mid \overline{y} + w.\overline{2} \mid \overline{v} + y.\overline{4} \right)}} \stepPi \overline{4} \mid \ResPi{\tilde{n}}{S_1} \noStep\\
	\LEPi &\stepPi \ResPi{\tilde{n}}{{\left( S_2 \mid S_3 \mid S_4 \mid \overline{w} + z.\overline{5} \right)}} \stepPi \ResPi{\tilde{n}}{{\left( \overline{y} + w.\overline{2} \mid S_4 \mid \overline{w} + z.\overline{5} \right)}} \stepPi \overline{2} \mid \ResPi{\tilde{n}}{S_4} \noStep\\
	\LEPi &\stepPi \ResPi{\tilde{n}}{{\left( S_2 \mid S_3 \mid S_4 \mid \overline{w} + z.\overline{5} \right)}} \stepPi \ResPi{\tilde{n}}{{\left( S_2 \mid \overline{z} + x.\overline{3} \mid \overline{w} + z.\overline{5} \right)}} \stepPi \overline{5} \mid \ResPi{\tilde{n}}{S_2} \noStep
\end{align*}
These executions can be obtained form the first execution in the above list by symmetry on the first two steps.
In each maximal execution exactly one leader is elected.

We show that there exists no symmetric electoral system for networks of
size five in \CMVmix; or more generally no symmetric electoral system
for networks of odd size in \CMVmix.
A key ingredient to separate the \piCal with mixed choice from the asynchronous \piCal in \cite{palamidessi97} is a confluence lemma.
It states that in the asynchronous \piCal a step reducing an output and an alternative step reducing an input cannot be conflict to each other and thus can be executed in any order.
In the full \piCal this confluence lemma is not valid, because inputs and outputs can be combined within a single choice construct and can thus be in conflict.
For \CMVmix we observe that steps that reduce different endpoints can also not be in conflict to each other, because different channel endpoints cannot be combined in a single choice.

\begin{lemma}[Confluence]
	\label{lem:confluenceCMVmix}
	Let $ P, Q \in \procCMVmix $.
	Assume that $ A = \ResCMVmix{\tilde{x}}{\tilde{y}}{{\left( P \mid Q \right)}} $ can make two steps $ A \step \ResCMVmix{\widetilde{x_1}}{\widetilde{y_1}}{{\left( P_1 \mid Q_1 \right)}} = B $ and $ A \step \ResCMVmix{\widetilde{x_2}}{\widetilde{y_2}}{{\left( P_2 \mid Q_2 \right)}} = C $ such that $ P_1 $ is obtained modulo $ \scCMVmix $ from $ P $ by reducing a choice on channel endpoint $ a $ and $ P_2 $ is obtained modulo $ \scCMVmix $ from $ P $ by reducing a choice on channel endpoint $ b $ with $ a \neq b $.
	Then there exist $ P_3, Q_3 \in \procCMVmix $ and $ D = \ResCMVmix{\widetilde{x_3}}{\widetilde{y_3}}{{\left( P_3 \mid Q_3 \right)}} $ such that $ B \step D $ and $ C \step D $, where $ \widetilde{x_3} = \widetilde{x_1} \cup \widetilde{x_2} $ and $ \widetilde{y_3} = \widetilde{y_1} \cup \widetilde{y_2} $.
\end{lemma}

\begin{proof}
	Assume the two steps $ A \step B $ and $ A \step C $ as described above.
	By Figure~\ref{fig:semanticsCMVmix}, the steps $ A \step B $ and $ A \step C $ imply that $ P $ contains at least two choices, one on channel $ a $ and one on channel $ b $, that are modulo structural congruence combined in parallel (possibly surrounded by restrictions).
	Since the choice on $ a $ (or $ b $) is the only choice reduced in $ P $, another choice on the matching endpoint is reduced in $ Q $.
	Regardless of whether $ a $ and $ b $ are matching endpoints or not, we obtain with the same kind of reasoning that also $ Q $ contains at least two choices, one on the channel endpoint that matches $ a $ and one on the channel endpoint that matches $ b $, that are modulo structural congruence combined in parallel (possibly surrounded by restrictions).
	We conclude that the two steps of $ A = \ResCMVmix{\tilde{x}}{\tilde{y}}{{\left( P \mid Q \right)}} $ are distributable.
	This implies that these two steps can be executed in any order as required.
\end{proof}

\begin{wrapfigure}{R}{0.2\textwidth}
	\centering
	\scalebox{0.8}{
	\begin{tikzpicture}[bend angle=20]
		\node (a) at (0, 0.75) {$ A $};
		\node (b) at (1.5, 1.5) {$ B $};
		\node (c) at (1.5, 0) {$ C $};
		\node (d) at (3, 0.75) {$ D $};
		\path[|->] (a) edge (b);
		\path[|->] (a) edge (c);
		\path[|->] (b) edge (d);
		\path[|->] (c) edge (d);
	\end{tikzpicture}
	}
\end{wrapfigure}

The proof of this confluence lemma relies on the observation that the two steps of $ A $ to $ B $ and $ C $ have to reduce distributable parts of $ A $.
Then these two steps are distributable, which in turn allows us to perform them in any order.
Thus the expressive power of choice in \CMVmix is limited by the fact that syntactically the choice construct is fixed on a single channel endpoint.
With this alternative confluence lemma, we can show that there is no electoral system of odd degree in \CMVmix.

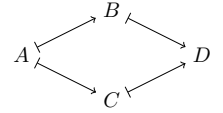
\begin{lemma}[No Electoral System]
	Consider a network $ P = \ResCMVmix{\tilde{x}}{\tilde{y}}{{\left( P_1 \mid \ldots \mid P_k \right)}} $ in \CMVmix with $ k > 1 $ being an odd number. Assume that the associated hypergraph $ \Hypergraph{P} $ admits an automorphism $ \sigma \neq \id $ with only one orbit, and that $ P $ is symmetric \wrt $ \sigma $. Then $ P $ cannot be an electoral system.
	\label{lem:noElectoralSystemCMVmix}
\end{lemma}

\begin{proof}
	Assume by contradiction that $ P $ is an electoral system.
	We will show that we can then construct an infinite execution $ E: P \steps P^0 \steps P^1 \steps \ldots $ such that, for each $ j $, $ E_j : P \steps P^j $ does not announce a unique leader and $ P^j $ is still symmetric \wrt $ \sigma_j $, where $ \sigma_j $ is the original automorphism enriched with associations on the new names possibly introduced by the communication actions.
	This is a contradiction, because the limit of this sequence is an infinite computation for $ P $ which does not announces exactly one leader.

	The proof is by induction on the number $ h $.
	Notice that the assumption of $ \sigma $ generating only one orbit implies that $ \Orbit{\sigma}{i} = \Set{ i, \sigma(i), \ldots, \sigma^{k - 1}(i) } = \Set{ 1, \ldots, k } $, for each $ i \in \Set{ 1, \ldots, k } $.
	Since $ \sigma_j $ is obtained from $ \sigma $ by adding substitutions on restricted names and since $ 1, \ldots, k $ are not used as bound names in electoral systems, the same holds for all the $ \sigma_j $.
	\begin{description}
		\item[Base Case ($ h = 0 $):] Define $ E_0 $ to be the empty execution, \ie $ E_0: P \steps P^0 $ with $ P^0 = P $.
		\item[Induction Step ($ h + 1 $):] Given $ E_h : P \steps P^h = \ResCMVmix{\widetilde{x_h}}{\widetilde{y_h}}{{\left( P_1^h \mid \ldots \mid P_k^h \right)}} $, we construct $ E_{h + 1} : P \steps P^{h + 1} $ as follows.

			If $ P^h $ announces a leader $ i $, then $ P^h\Barb{i} $ for some $ 1 \leq i \leq k $.
			By symmetry, then $ P^h\WeakBarb{\sigma(i)} $, \ie more than one leader is announced.
			This is a contradiction.

			Since $ P $ is an electoral system but $ P^h $ does not yet announces a leader, $ P^h $ has to be able to reduce, \ie there is some $ P' $ such that $ P^h \step P' $. This step was performed by one or two of the processes in the network, \ie either $ P_i^h \step P_i' $ and $ P' = \ResCMVmix{\widetilde{x_h}}{\widetilde{y_h}}{{\left( P_1^h \mid \ldots \mid P_i' \mid \ldots \mid P_k^h \right)}} $ or $ \ResCMVmix{\widetilde{x_h}}{\widetilde{y_h}}{{\left( P_i^h \mid P_j^h \right)}} \step \ResCMVmix{\widetilde{x_{h, 1}}}{\widetilde{y_{h, 1}}}{{\left( P_{i, 1} \mid P_{j, 1} \right)}} $ and $ P' = \ResCMVmix{\widetilde{x_{h, 1}}}{\widetilde{y_{h, 1}}}{{\left( P_1^h \mid \ldots \mid P_{i, 1} \mid \ldots \mid P_{j, 1} \mid \ldots \mid P_k^h \right)}} $ with $ i \neq j $.
			\begin{description}
				\item[$ P_i^h \step P_i' $:] Regardless of whether the step $ P_i^h \step P_i' $ is reducing a conditional or performing a communication within part $ i $ of the network, symmetry ensures that the other parts of the network can perform a sequence of steps that leads to state symmetric to $ P_i' $.
					We choose $ P_i^{h + 1} = P_i' $.
					By symmetry, $ P_{\sigma_h(i)}^h \steps P_{\sigma_h(i)}^{h + 1}, \ldots, P_{\sigma_h^{k - 1}(i)}^h \steps P_{\sigma_h^{k - 1}(i)}^{h + 1} $ with $ P_i^{h + 1}\sigma_h \approx_{\CMVmix} P_{\sigma_h(i)}^{h + 1}, \ldots, P_i^{h + 1}\sigma_h^{k - 1} \approx_{\CMVmix} P_{\sigma_h^{k - 1}(i)}^{h + 1} $.
					Since the steps of the different parts of the network are distributable, we obtain $ E_{h + 1} : P \steps P^h \steps P^{h + 1} $, where $ P^{h + 1} = \ResCMVmix{\widetilde{x_h}}{\widetilde{y_h}}{{\left( P_1^{h + 1} \mid \ldots \mid P_k^{h + 1} \right)}} $ and $ P^{h + 1} $ is still symmetric \wrt $ \sigma_{h + 1} = \sigma_h $.
				\item[$ \ResCMVmix{\widetilde{x_h}}{\widetilde{y_h}}{{\left( P_i^h \mid P_j^h \right)}} \step \ResCMVmix{\widetilde{x_{h, 1}}}{\widetilde{y_{h, 1}}}{{\left( P_{i, 1} \mid P_{j, 1} \right)}} $:] Let us denote this sequence of one step by $ S_1 $.
					A step performed by two processes of the network (in \CMVmix) is a communication.
					By Figure~\ref{fig:semanticsCMVmix}, $ S_1 $ reduces a choice on some endpoint $ a $ in $ P_i^h $ and a choice on some endpoint $ b $ in $ P_j^h $ such that $ a $ and $ b $ are matching endpoints of the same channel and thus $ a \neq b $.
					By symmetry,
					\begin{align*}
						S_2: \ResCMVmix{\widetilde{x_h}}{\widetilde{y_h}}{{\left( P_{\sigma_h(i)}^h \mid P_{\sigma_h(j)}^h \right)}} & \steps \ResCMVmix{\widetilde{x_{h, 2}}}{\widetilde{y_{h, 2}}}{{\left( P_{\sigma_h(i), 2} \mid P_{\sigma_h(j), 2} \right)}}\\
						& \vdots\\
						S_k: \ResCMVmix{\widetilde{x_h}}{\widetilde{y_h}}{{\left( P_{\sigma_h^{k - 1}(i)}^h \mid P_{\sigma_h^{k - 1}(j)}^h \right)}} & \steps \ResCMVmix{\widetilde{x_{h, k}}}{\widetilde{y_{h, k}}}{{\left( P_{\sigma_h^{k - 1}(i), k} \mid P_{\sigma_h^{k - 1}(j), k} \right)}},	
					\end{align*}
					where we apply $ \alpha $-conversion to ensure that the pairwise intersection of elements in $ \widetilde{x_{h, 1}}, \ldots, \widetilde{x_{h, k}} $ is always $ \widetilde{x_h} $ and similarly the pairwise intersection of elements in $ \widetilde{y_{h, 1}}, \ldots, \widetilde{y_{h, k}} $ is always $ \widetilde{y_h} $.
					In the sequences of steps $ S_1, \ldots, S_k $ each component of the network is used exactly twice to reduce a choice on endpoints $ \sigma_h^m(a) $ and $ \sigma_h^n(b) $ for some $ m, n \in \Set{ 0, \ldots, k - 1 } $ with $ m \neq n $.
					Since $ \sigma_h $ is an automorphism with only one orbit and since $ k $ is odd, $ \sigma_h^m(a) \neq \sigma_h^n(b) $ for all such cases.
					By repeatedly applying Lemma~\ref{lem:confluenceCMVmix}, then we can perform $ S_1, \ldots, S_k $ in sequence, \ie there are some $ P_1^{h + 1}, \ldots, P_k^{h + 1} $ such that $ E_{h + 1}: P \steps P^h \steps P^{h + 1} = \ResCMVmix{\widetilde{x_{h + 1}}}{\widetilde{y_{h + 1}}}{{\left( P_1^{h + 1} \mid \ldots \mid P_k^{h + 1} \right)}} $, where $ \widetilde{x_{h + 1}} $ is the union of $ \widetilde{x_{h, 1}}, \ldots, \widetilde{x_{h, k}} $, similarly $ \widetilde{y_{h + 1}} $ is the union of $ \widetilde{y_{h, 1}}, \ldots, \widetilde{y_{h, k}} $, the sequence $ P^h \steps P^{h + 1} $ is obtained from $ S_1, \ldots, S_k $, and we apply scope extrusion and the Rule~\textsc{(R-Struct)} to push restrictions to the outside.
					Let $ \sigma_{h + 1} $ be the automorphism obtained from $ \sigma_h $ by adding permutations for the names in $ \widetilde{x_{h + 1}} \setminus \widetilde{x_h} $ and $ \widetilde{y_{h + 1}} \setminus \widetilde{y_h} $.
					Finally, we observe that $ P^{h + 1} $ is still symmetric \wrt $ \sigma_{h + 1} $.
			\end{description}
	\end{description}
\end{proof}

In the proof we construct a potentially infinite sequence of steps such that the system constantly restores symmetry, \ie whenever a step destroys symmetry we can perform a sequence of steps that restores the symmetry.
Therefore we rely on the assumption of $ \sigma $ generating only one orbit.
This implies that $ \Orbit{\sigma}{i} = \Set{ i, \sigma(i), \ldots, \sigma^{k - 1}(i) } = \Set{ 1, \ldots, k } $, for each $ i \in \Set{ 1, \ldots, k } $.
Because of that, whenever part $ i $ performs a step that destroys symmetry or parts $ i $ and $ j $ together perform a step that destroys symmetry, the respective other parts of the originally symmetric network can perform symmetric steps to restore the symmetry of the network.
Because of the symmetry, the constructed sequence of steps does not elect a unique leader.
Accordingly, the existence of this sequence ensures that $ P $ is not an electoral system.

In contrast to \cite{palamidessi97}, the above lemma is for networks of odd degree.
This is necessary to ensure that in the last case of the proof the mentioned $ \sigma_h^m(a) $ and $ \sigma_h^n(b) $ reduced by a component of the network are distinct such that we can apply our confluence property of Lemma~\ref{lem:confluenceCMVmix}, which in turn ensures that we can always perform a sequence of steps to restore symmetry after the step that destroys the symmetry.

By the preservation of distributability, encodings preserve the
structure of networks; and 
by name invariance, they also preserve the symmetry of networks.
With operational correspondence and barb-sensitiveness, any good encoding of $ \LEPi $ is again a symmetric electoral system of size five, since the combination of these two criteria allows to distinguish between an electoral system and a system that does not elect exactly one leader in every maximal execution.
Since by Lemma~\ref{lem:noElectoralSystemCMVmix} this is not possible, 
we can separate \CMVmix from the \piCal by using $ \LEPi $ from Example~\ref{exa:leaderElectionPi} as counterexample.

\begin{theorem}[Separate \CMVmix from the \PiCal via Leader Election]
	\label{thm:separateCMVmixfromPiviaLeaderElection}
	$ $\\
	There is no good encoding from the \piCal into \CMVmix.
\end{theorem}

\begin{proof}
	Assume the contrary, \ie there is a good encoding $ \arbitraryEncoding $ from the \piCal into \CMVmix with the renaming policy $ \varphi $.
	Then this encoding translates $ \LEPi $ in Example~\ref{exa:leaderElectionPi}.
	By Definition~\ref{def:distributabilityPreservation},
	\begin{align*}
		\ArbitraryEncoding{\LEPi} \equiv \ResCMVmix{\tilde{y}}{\tilde{z}}{{\left( T_{\varphi(1)} \mid T_{\varphi(2)} \mid T_{\varphi(3)} \mid T_{\varphi(4)} \mid T_{\varphi(5)} \right)}}
	\end{align*}
	such that $ T_{\varphi(i)} \asymp \ArbitraryEncoding{S_i} $ for all $ i \in \Set{ 1, \ldots, 5 } $.
	Remember that $ \LEPi $ is symmetric.
	Below Example~\ref{exa:leaderElectionPi} we present an example for a permutation $ \sigma $, but here we consider all automorphisms of $ \Hypergraph{\LEPi} $.
	For all such automorphisms $ \sigma $ we have $ S_{\sigma(i)} = S_i\sigma $ for all $ i \in \Set{ 1, \ldots, 5 } $.
	Fix $ \sigma $, \ie let $ \sigma $ be an arbitrary such automorphism, and let $ \sigma' $ be such that $ \varphi(\sigma(a)) = \sigma'{\left( \varphi(a) \right)} $ for all $ a \in \names $.
	Then $ \sigma' $ is a permutation (on translated source term names).
	By Definition~\ref{def:nameInvariance}, then $ T_{\sigma'(\varphi(i))} = T_{\varphi(\sigma(i))} \asymp \ArbitraryEncoding{S_{\sigma(i)}} = \ArbitraryEncoding{S_i\sigma} \asymp \ArbitraryEncoding{S_i}\sigma' \asymp T_{\varphi(i)}\sigma' $ for all $ i \in \Set{ 1, \ldots, 5 } $.
	Since $ \asymp $ is a barb respecting weak reduction bisimulation (Definition~\ref{def:goodEncoding}), then $ T_{\sigma'(\varphi(i))} \approx_{\CMVmix} T_{\varphi(i)}\sigma' $ for all $ i \in \Set{ 1, \ldots, 5 } $ \ie $ \ArbitraryEncoding{\LEPi} $ is symmetric.
	By the combination of Definition~\ref{def:operationalCorrespondence} and Definition~\ref{def:barbSensitiveness}, $ \ArbitraryEncoding{\LEPi} $ is an electoral system, because every maximal execution has to be emulated with the same reachable barbs.
	Then $ \ArbitraryEncoding{\LEPi} $ is a symmetric electoral system of size five.
	This contradicts Lemma~\ref{lem:noElectoralSystemCMVmix}.
	We conclude that there is no good encoding from the \piCal into \CMVmix.
\end{proof}


\section{Separating Mixed Sessions and the Pi-Calculus via Synchronisation}
\label{sec:separateMixedSessionsFromPiSynchronisationPatterns}

\begin{wrapfigure}{R}{0.25\textwidth}
	\centering
	\scalebox{0.6}{
	\tikzstyle{place}=[circle,draw=black,thick,minimum size=5mm]
	\tikzstyle{transition}=[rectangle,draw=black,thick,minimum size=5mm]
	\begin{tikzpicture}
		\foreach \x/\xtext in {1/e,2/d,3/c,4/b,5/a}
        {
            \path (360*\x/5+125:0.8) node[transition] (\xtext) {$\xtext$};
            \path (360*\x/5-55:1.75) node[place,tokens=1] (p\x) {};
        }

        \draw[-latex] (p2) -- (a);
        \draw[-latex] (p2) -- (b);

        \draw[-latex] (p1) -- (b);
        \draw[-latex] (p1) -- (c);

        \draw[-latex] (p5) -- (c);
        \draw[-latex] (p5) -- (d);

        \draw[-latex] (p4) -- (d);
        \draw[-latex] (p4) -- (e);

        \draw[-latex] (p3) -- (e);
        \draw[-latex] (p3) -- (a);
	\end{tikzpicture}
	}
\end{wrapfigure}

In \cite{petersNestmannGoltz13} the technique used in \cite{palamidessi97} and its relation to synchronisation are analysed.
Two synchronisation patterns, the pattern \patternM and the pattern \patternStar, are identified that describe two different levels of synchronisation and allow to more clearly separate languages along their ability to express synchronisation.
These patterns are called \patternM and \patternStar, because their respective representations as a Petri net (see left and right picture) have these shapes.
The pattern \patternStar captures the power of synchronisation of the \piCal.
In particular it captures what is necessary to solve the leader election problem.

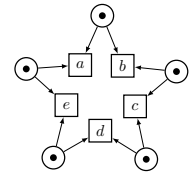
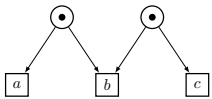
\begin{wrapfigure}{L}{0.25\textwidth}
	\centering
	\scalebox{0.6}{
	\tikzstyle{place}=[circle,draw=black,thick,minimum size=5mm]
	\tikzstyle{transition}=[rectangle,draw=black,thick,minimum size=5mm]
	\begin{tikzpicture}
		\node[place,tokens=1]	(p) at (1, 1.5) {};
		\node[place,tokens=1]	(q) at (3, 1.5) {};
		\node[transition]		(a) at (0, 0) {$ a $};
		\node[transition]		(b) at (2, 0) {$ b $};
		\node[transition]		(c) at (4, 0) {$ c $};

		\draw[-latex] (p) -- (a);
		\draw[-latex] (p) -- (b);
		\draw[-latex] (q) -- (b);
		\draw[-latex] (q) -- (c);
	\end{tikzpicture}
	}
\end{wrapfigure}

The pattern \patternM captures a very weak form of synchronisation, not enough to solve leader election but enough to make a fully distributed implementation of languages with this pattern difficult (see also \cite{petersNestmannIC20}).
This pattern was originally identified in \cite{Glabbeek2008} when studying the relevance of synchrony and distribution on Petri nets.
As shown in \cite{peters12, petersNestmannGoltz13}, the ability to express these different amounts of synchronisation in the \piCal lies in its different forms of choices: to express the pattern \patternStar the \piCal needs mixed choice, whereas separate choice allows to express the pattern \patternM.
Indeed we find the pattern \patternM in \CMVmix, but there are no \patternStar in \CMVmix.

We inherit the definition of the synchronisation pattern \patternM from \cite{petersNestmannGoltz13}, where we do not distinguish between local and non-local \patternM since in the \piCal there is no difference between parallel and distributable steps.

\begin{definition}[Synchronisation Pattern \patternM]
	\label{def:synchronisationPatternM}
	Let $ \ProcCal{\proc}{\step} $ be a process calculus and $ \PM \in \proc $ such that:
	\begin{enumerate}
		\item $ \PM $ can perform at least three alternative steps $ a{:}\; \PM \step P_a $, $ b{:}\; \PM \step P_b $, and $ c{:}\; \PM \step P_c $ such that $ P_a $, $ P_b $, and $ P_c $ are pairwise different.
		\item The steps $ a $ and $ c $ are parallel/distributable in $ \PM $.
		\item But $ b $ is in conflict with both $ a $ and $ c $.
	\end{enumerate}
	In this case, we denote the process $ \PM $ as \patternM.
\end{definition}

There are pattern $ \patternM $ in \CMVmix as for instance the next example.

\begin{example}[The \patternM in \CMVmix]
	\label{exa:CMVmixM}
	Consider the term $ \PMCMVmix $ and the types $ \Dual{T_1}{T_2} $ given as:
	\begin{align*}
		\PMCMVmix &= \ResCMVmix{x}{y}{(}\!\!\begin{array}[t]{l}
			\ChoiceCMVmix{\linCMVmix}{x}{{\left( \OutCMVmix{\Label}{\true}{P_1} + \InpCMVmix{\Label}{z}{P_2} \right)}} \mid \ChoiceCMVmix{\linCMVmix}{x}{{\left( \OutCMVmix{\Label}{\false}{P_3} + \InpCMVmix{\Label}{z}{P_4} \right)}} \mid\\
			\ChoiceCMVmix{\linCMVmix}{y}{{\left( \InpCMVmix{\Label}{z}{P_5} + \OutCMVmix{\Label}{\true}{P_6} \right)}} \mid \ChoiceCMVmix{\linCMVmix}{y}{{\left( \InpCMVmix{\Label}{z}{P_7} + \OutCMVmix{\Label}{\false}{P_8} \right)}} )
		\end{array}\\
		T_1 &= \IntCMVmix{\unCMVmix}{\Set{ \OutCMVmix{\Label}{\boolT}{T_{1, 1}}, \InpCMVmix{\Label}{\boolT}{T_{1, 2}} }}\\
		T_2 &= \ExtCMVmix{\unCMVmix}{\Set{ \InpCMVmix{\Label}{\boolT}{T_{2, 1}}, \OutCMVmix{\Label}{\boolT}{T_{2, 2}} }}
	\end{align*}
	The process $ \PMCMVmix $ with $ P_1 = \ldots = P_8 = \inactCMVmix $ and $ T_{1, 1} = T_{1, 2} = T_{2, 1} = T_{2, 2} = \finCMVmix $ is well-typed:
	\begin{align*}
		& D = \dfrac{\dfrac{D_1 \quad D_2 \quad D_3 \quad D_4}{\At{x}{T_1}, \At{y}{T_2} \vdash \ldots}\ruleTParCMVmix}{\vdash \PMCMVmix}\ruleTResCMVmix\\
		\\
		& D_1 = \dfrac{D_{1, 1} \quad D_{1, 2}}{\At{x}{T_1}, \At{y}{T_2} \vdash \ChoiceCMVmix{\linCMVmix}{x}{{\left( \OutCMVmix{\Label}{\true}{\inactCMVmix} + \InpCMVmix{\Label}{z}{\inactCMVmix} \right)}}}\ruleTChoiceCMVmix\\
		\\
		& D_{1, 1} = \dfrac{\dfrac{}{\At{x}{\finCMVmix}, \At{y}{T_2} \vdash \At{\true}{\boolT}}\ruleTTrueCMVmix \quad D_{1, 1, 1}}{\At{x}{\finCMVmix}, \At{y}{T_2} \vdash \At{\OutCMVmix{\Label}{\true}{\inactCMVmix}}{\OutCMVmix{\Label}{\boolT}{\finCMVmix}}}\ruleTOutCMVmix\\
		\\
		& D_{1, 1, 1} = \dfrac{}{\At{x}{\finCMVmix}, \At{y}{T_2} \vdash \inactCMVmix}\ruleTInactCMVmix\\
		\\
		& D_{1, 2} = \dfrac{\dfrac{}{\At{x}{\finCMVmix}, \At{y}{T_2}, \At{z}{\boolT} \vdash \inactCMVmix}\ruleTInactCMVmix}{\At{x}{\finCMVmix}, \At{y}{T_2} \vdash \At{\InpCMVmix{\Label}{z}{\inactCMVmix}}{\InpCMVmix{\Label}{\boolT}{\finCMVmix}}}\ruleTInCMVmix
	\end{align*}
	where the derivations of $ D_2 $, $ D_3 $, and $ D_4 $ are similar to the derivation of $ D_1 $:
	\begin{align*}
		D_2 &= \dfrac{D_{2, 1} \quad D_{1, 2}}{\At{x}{T_1}, \At{y}{T_2} \vdash \ChoiceCMVmix{\linCMVmix}{x}{{\left( \OutCMVmix{\Label}{\false}{\inactCMVmix} + \InpCMVmix{\Label}{z}{\inactCMVmix} \right)}}}\ruleTChoiceCMVmix\\
		\\
		D_{2, 1} &= \dfrac{\dfrac{}{\At{x}{\finCMVmix}, \At{y}{T_2} \vdash \At{\false}{\boolT}}\ruleTFalseCMVmix \quad D_{2, 1, 1}}{\At{x}{\finCMVmix}, \At{y}{T_2} \vdash \At{\OutCMVmix{\Label}{\false}{\inactCMVmix}}{\OutCMVmix{\Label}{\boolT}{\finCMVmix}}}\ruleTOutCMVmix\\
		\\
		D_{2, 1, 1} &= \dfrac{}{\At{x}{\finCMVmix}, \At{y}{T_2} \vdash \inactCMVmix}\ruleTInactCMVmix
	\end{align*}
	\begin{align*}
		D_3 &= \dfrac{D_{3, 1} \quad D_{3, 2}}{\At{x}{T_1}, \At{y}{T_2} \vdash \ChoiceCMVmix{\linCMVmix}{y}{{\left( \InpCMVmix{\Label}{z}{\inactCMVmix} + \OutCMVmix{\Label}{\true}{\inactCMVmix} \right)}}}\ruleTChoiceCMVmix\\
		\\
		D_{3, 1} &= \dfrac{\dfrac{}{\At{x}{T_1}, \At{y}{\finCMVmix}, \At{z}{\boolT} \vdash \inactCMVmix}\ruleTInactCMVmix}{\At{x}{T_1}, \At{y}{\finCMVmix} \vdash \At{\InpCMVmix{\Label}{z}{\inactCMVmix}}{\InpCMVmix{\Label}{\boolT}{\finCMVmix}}}\ruleTInCMVmix\\
		\\
		D_{3, 2} &= \dfrac{\dfrac{}{\At{x}{T_1}, \At{y}{\finCMVmix} \vdash \At{\true}{\boolT}}\ruleTTrueCMVmix \quad D_{3, 2, 1}}{\At{x}{T_1}, \At{y}{\finCMVmix} \vdash \At{\OutCMVmix{\Label}{\true}{\inactCMVmix}}{\OutCMVmix{\Label}{\boolT}{\finCMVmix}}}\ruleTOutCMVmix\\
		\\
		D_{3, 2, 1} &= \dfrac{}{\At{x}{T_1}, \At{y}{\finCMVmix} \vdash \inactCMVmix}\ruleTInactCMVmix
	\end{align*}
	\begin{align*}
		D_4 &= \dfrac{D_{3, 1} \quad D_{4, 2}}{\At{x}{T_1}, \At{y}{T_2} \vdash \ChoiceCMVmix{\linCMVmix}{y}{{\left( \InpCMVmix{\Label}{z}{\inactCMVmix} + \OutCMVmix{\Label}{\false}{\inactCMVmix} \right)}}}\ruleTChoiceCMVmix\\
		\\
		D_{4, 2} &= \dfrac{\dfrac{}{\At{x}{T_1}, \At{y}{\finCMVmix} \vdash \At{\false}{\boolT}}\ruleTFalseCMVmix \quad D_{3, 2, 1}}{\At{x}{T_1}, \At{y}{\finCMVmix} \vdash \At{\OutCMVmix{\Label}{\false}{\inactCMVmix}}{\OutCMVmix{\Label}{\boolT}{\finCMVmix}}}\ruleTOutCMVmix
	\end{align*}
	The process $ \PMCMVmix $ is a \patternM in \CMVmix:
	\begin{center}
		\begin{tikzpicture}[]
			\node (res) at (-0.4, 1) {$ \PMCMVmix = \ResCMVmix{x}{y}{(} $};
			\node[fill=blue!20, rounded corners=0.5cm] (l1) at (3, 1) {$ \begin{array}{c} \text{\begin{tiny} location 1 \end{tiny}}\\ \ChoiceCMVmix{\linCMVmix}{x}{{\left( \OutCMVmix{\Label}{\true}{P_1} + \InpCMVmix{\Label}{z}{P_2} \right)}}\\ \ChoiceCMVmix{\linCMVmix}{y}{{\left( \InpCMVmix{\Label}{z}{P_5} + \OutCMVmix{\Label}{\true}{P_6} \right)}} \end{array} $};
			\node (p1) at (5.25, 1) {$ \mid $};
			\node (p2) at (5.25, 0.55) {$ \mid $};
			\node[fill=green!20, rounded corners=0.5cm] (l2) at (7.5, 1) {$ \begin{array}{c} \text{\begin{tiny} location 2 \end{tiny}}\\ \ChoiceCMVmix{\linCMVmix}{x}{{\left( \OutCMVmix{\Label}{\false}{P_3} + \InpCMVmix{\Label}{z}{P_4} \right)}}\\ \ChoiceCMVmix{\linCMVmix}{y}{{\left( \InpCMVmix{\Label}{z}{P_7} + \OutCMVmix{\Label}{\false}{P_8} \right)}} \end{array} $};
			\node (p3) at (9.75, 1) {$ \mid $};
			\node (b) at (10, 0.55) {$ ) $};
		\end{tikzpicture}
	\end{center}
	For instance we can pick the steps $ a $, $ b $, and $ c $ as:
	\begin{description}
		\item[Step $ a $:] $ \PMCMVmix \step \ResCMVmix{x}{y}{\left( P_1 \mid \ChoiceCMVmix{\linCMVmix}{x}{{\left( \OutCMVmix{\Label}{\false}{P_3} + \InpCMVmix{\Label}{z}{P_4} \right)}} \mid P_5\Set{\Subst{\true}{z}} \mid \ChoiceCMVmix{\linCMVmix}{y}{{\left( \InpCMVmix{\Label}{z}{P_7} + \OutCMVmix{\Label}{\false}{P_8} \right)}} \right)} $
		\item[Step $ b $:] $ \PMCMVmix \step \ResCMVmix{x}{y}{\left( P_1 \mid \ChoiceCMVmix{\linCMVmix}{x}{{\left( \OutCMVmix{\Label}{\false}{P_3} + \InpCMVmix{\Label}{z}{P_4} \right)}} \mid \ChoiceCMVmix{\linCMVmix}{y}{{\left( \InpCMVmix{\Label}{z}{P_5} + \OutCMVmix{\Label}{\true}{P_6} \right)}} \mid P_7\Set{\Subst{\true}{z}} \right)} $
		\item[Step $ c $:] $ \PMCMVmix \step \ResCMVmix{x}{y}{\left( \ChoiceCMVmix{\linCMVmix}{x}{{\left( \OutCMVmix{\Label}{\true}{P_1} + \InpCMVmix{\Label}{z}{P_2} \right)}} \mid P_3 \mid \ChoiceCMVmix{\linCMVmix}{y}{{\left( \InpCMVmix{\Label}{z}{P_5} + \OutCMVmix{\Label}{\true}{P_6} \right)}} \mid P_7\Set{\Subst{\false}{z}} \right)} $
	\end{description}
	\qed
\end{example}

We use synchronisation patterns and the proof technique presented in \cite{petersNestmannGoltz13} to present an alternative way to prove Theorem~\ref{thm:separateCMVmixfromPiviaLeaderElection}.
By that we underpin our claim that the choice construct of \CMVmix is separate and not mixed, and we provide further intuition on why this choice construct is less expressive.

We inherit the definition of the synchronisation pattern \patternStar from \cite{petersNestmannGoltz13}, where we do not distinguish between local and non-local \patternStar since in the \piCal there is no difference between parallel and distributable steps.

\begin{definition}[Synchronisation Pattern \patternStar]
	\label{def:synchronisationPatternGreatM}
Let $ \ProcCal{\proc}{\step} $ be a process calculus and $ \PS \in \proc $ such that:
	\begin{itemize}
		\item $ \PS $ can perform at least five alternative reduction steps $ i : \PS \step P_i $ for $ i \in \Set{ a, b, c, d, e } $ such that the $ P_i $ are pairwise different;
		\item the steps $ a $, $ b $, $ c $, $ d $, and $ e $
                  form a circle such that $ a $ is in conflict with $
                  b $, $ b $ is in conflict with $ c $, $ c $ is in
                  conflict with $ d $, $ d $ is in conflict with $ e
                  $, and $ e $ is in conflict with $ a $; and 
		\item every pair of steps in $ \Set{ a, b, c, d, e } $ that is not in conflict due to the previous condition is distributable in $ \PS $.
	\end{itemize}
	In this case, we denote the process $ \PS $ as \patternStar.
\end{definition}

In contrast to \CMVmix we do find \patternStar in the \piCal. 

\begin{example}[The \patternStar in the \PiCal]
	\label{exa:piStar}
	Consider the following \patternStar in the \piCal:
	\begin{align*}
		\PSPi = \overline{a}{} + b.\overline{o_b} \mid \overline{b} + c.\overline{o_c} \mid \overline{c} + d.\overline{o_d} \mid \overline{d} + e.\overline{o_e} \mid \overline{e} + a.\overline{o_a} 
	\end{align*}
	The steps $ a, \ldots, e $ of Definition~\ref{def:synchronisationPatternGreatM} are the steps on the respective channels.
	\begin{description}
		\item[Step $ a $:] $ \PSPi \step S_a $ with $ S_a = \OutPi{b}{} + \InpPi{c}{}.\OutPi{o_c}{} \mid \OutPi{c}{} + \InpPi{d}{}.\OutPi{o_d}{} \mid \OutPi{d}{} + \InpPi{e}{}.\OutPi{o_e}{} \mid \OutPi{o_a}{} $,
		\item[Step $ b $:] $ \PSPi \step S_b $ with $ S_b = \OutPi{o_b}{} \mid \OutPi{c}{} + \InpPi{d}{}.\OutPi{o_d}{} \mid \OutPi{d}{} + \InpPi{e}{}.\OutPi{o_e}{} \mid \OutPi{e}{} + \InpPi{a}{}.\OutPi{o_a}{} $,
		\item[Step $ c $:] $ \PSPi \step S_c $ with $ S_c = \OutPi{a}{} + \InpPi{b}{}.\OutPi{o_b}{} \mid \OutPi{o_c}{} \mid \OutPi{d}{} + \InpPi{e}{}.\OutPi{o_e}{} \mid \OutPi{e}{} + \InpPi{a}{}.\OutPi{o_a}{} $,
		\item[Step $ d $:] $ \PSPi \step S_d $ with $ S_d = \OutPi{a}{} + \InpPi{b}{}.\OutPi{o_b}{} \mid \OutPi{b}{} + \InpPi{c}{}.\OutPi{o_c}{} \mid \OutPi{o_d}{} \mid \OutPi{e}{} + \InpPi{a}{}.\OutPi{o_a}{} $
		\item[Step $ e $:] $ \PSPi \step S_e $ with $ S_e = \OutPi{a}{} + \InpPi{b}{}.\OutPi{o_b}{} \mid \OutPi{b}{} + \InpPi{c}{}.\OutPi{o_c}{} \mid \OutPi{c}{} + \InpPi{d}{}.\OutPi{o_d}{} \mid \OutPi{o_e}{} $
	\end{description}
	The different outputs $ \OutPi{o_x} $ allow to distinguish between the different steps by their observables.\qed
\end{example}

We use the \patternStar $ \PSPi $ as counterexample to show that there is no good encoding from the \piCal into \CMVmix.
From Lemma~\ref{lem:noElectoralSystemCMVmix} we learned that \CMVmix cannot express certain electoral systems.
Accordingly, we are not surprised that \CMVmix cannot express the pattern \patternStar.

\begin{lemma}
	There are no \patternStar in \CMVmix.
	\label{lem:noStarCMVmix}
\end{lemma}

\begin{proof}
	Assume the contrary, \ie assume that there is a term $ \PSCMVmix $ in \CMVmix that is a \patternStar.
	Then $ \PSCMVmix $ can perform at least five alternative reduction steps $ a, b, c, d, e $ such that neighbouring steps in the sequence $ a, b, c, d, e, a $ are pairwise in conflict and non-neighbouring steps are distributable.
	Since steps reducing a conditional cannot be in conflict with any other step, none of the steps in $ \Set{a, b, c, d, e} $ reduces a conditional.
	Then all steps in $ \Set{a, b, c, d, e} $ are communication steps that reduce an output and an input that both are part of choices (with at least one summand).
	Because of the conflict between $ a $ and $ b $, these two steps reduce the same choice but this choice is not reduced in $ c $, because $ a $ and $ c $ are distributable.

	\vspace{0.25em}
	\noindent
	\begin{minipage}{\textwidth}
	\begin{wrapfigure}{R}{0.225\textwidth}
		\centering
		\scalebox{0.8}{
		\begin{tikzpicture}[]
			\foreach \x/\xlabel/\xtext/\ytext in {1/e/$ C_5 $/$ b $,2/d/$ C_4 $/$ a $,3/c/$ C_3 $/$ e $,4/b/$ C_2 $/$ d $,5/a/$ C_1 $/$ c $}
	        {
	            \path (360*\x/5+125:0.8) node (\xlabel) {\xtext};
	            \path (360*\x/5-55:1.75) node (p\x) {\ytext};
	        }

	        \draw[-latex] (p2) -- (a);
	        \draw[-latex] (p2) -- (b);

	        \draw[-latex] (p1) -- (b);
	        \draw[-latex] (p1) -- (c);

	        \draw[-latex] (p5) -- (c);
	        \draw[-latex] (p5) -- (d);

	        \draw[-latex] (p4) -- (d);
	        \draw[-latex] (p4) -- (e);

	        \draw[-latex] (p3) -- (e);
	        \draw[-latex] (p3) -- (a);
		\end{tikzpicture}
		}
	\end{wrapfigure}
	
	\hspace{1em}
	By repeating this argument, we conclude that in the steps $ a, b, c, d, e $ five choices $ C_1, \ldots, C_5 $ are reduced as depicted on the right,
	where \eg the step $ a $ reduces the choices $ C_1 $ and $ C_2 $.
	By the reduction semantics of \CMVmix, the two choices $ C_1 $ and $ C_2 $ that are reduced in step $ a $ need to use dual endpoints of the same channel.
	Without loss of generality, assume that $ C_1 $ is on channel endpoint $ x $ and $ C_2 $ is on channel endpoint $ y $.
	Then the choice $ C_3 $ needs to be on channel endpoint $ x $ again, because step $ b $ reduces $ C_2 $ (on $ y $) and $ C_3 $.
	By repeating this argument, then $ C_4 $ is on $ y $ and $ C_5 $ is on $ x $.
	But then step $ e $ reduces two choices $ C_1 $ and $ C_5 $ that are both on channel endpoint $ x $.
	Since the reduction semantics of \CMVmix does not allow such a step, this is a contradiction.
	\end{minipage}

	We conclude that there are no \patternStar in \CMVmix.
\end{proof}

The proof of the above lemma tells us more about why choice in \CMVmix is limited.
From the confluence property in \CMVmix we get the hint that the problem is the restriction of choice to a single channel endpoint.
A \patternStar is a circle of steps of odd degree, where neighbouring steps are in conflict.
More precisely, the star with five points in \patternStar is the smallest cycle of steps where neighbouring steps are in conflict and that contains non-neighbouring distributable steps.
The proof shows that the limitation of choice to a single channel endpoint and the requirement of the semantics that a channel endpoint can interact with exactly one other channel endpoint causes the problem.
This also explains why Lemma~\ref{lem:noElectoralSystemCMVmix} considers electoral systems of odd degree, because the odd degree does not allow to close the cycle as explained in the proof above.
Indeed, if we change the syntax to allow mixed choice with summands on more than one channel, we obtain the mixed-choice-construct of the \piCal.
Similarly, we invalidate our separation result in the Theorems~\ref{thm:separateCMVmixfromPiviaLeaderElection} and \ref{thm:separateCMVmixfromPiviaStar}, if we change the semantics to allow two choices to communicate even if they are on the same channel.
The latter may be more surprising, but indeed we do not need more than a single channel to solve leader election and build \patternStar, \eg $ \PSPi $ remains a star if we choose $ a = b = c = d = e $ (though we might want to pick different names $ o_a, \ldots, o_e $ to be able to distinguish the steps).

We use $ \PSPi $ in Example~\ref{exa:piStar} as counterexample to separate the \piCal from \CMVmix in Theorem~\ref{thm:separateCMVmixfromPiviaStar} below.
We prove first that the conflicts in the source term $ \PSPi $ have to be translated into conflicts of the corresponding emulations.

\begin{lemma}
	\label{lem:translateConflictsStar}
	Any good encoding $ \arbitraryEncoding $ from the \piCal into \CMVmix has to translate the conflicts in $ \PSPi $ given in Example~\ref{exa:piStar} into conflicts of the corresponding emulations.
\end{lemma}

\begin{proof}
	By operational completeness, all five steps of $ \PSPi $ have to be emulated in $ \ArbitraryEncoding{\PSPi} $, \ie there exist some $ T_a, T_b, T_c, T_d, T_e \in \procCMVmix $ such that $ \ArbitraryEncoding{\PSPi} \steps T_x \asymp \ArbitraryEncoding{S_x} $ for all $ x \in \Set{ a, b, c, d, e } $.
	Because $ \arbitraryEncoding $ preserves distributability, for each pair of steps $ x $ and $ y $ that are parallel in $ \PSPi $, the emulations $ X: \ArbitraryEncoding{\PSPi} \steps T_x $ and $ Y: \ArbitraryEncoding{\PSPi} \steps T_y $ such that $ T_x \asymp \ArbitraryEncoding{S_x} $ and $ T_y \asymp \ArbitraryEncoding{S_y} $ are distributable.
	Note that $ X $ and $ Y $ refer to the upper case variants of $ x $ and $ y $, respectively.

	Consider each triple of steps $ x, y, z \in \Set{ a, b, c, d, e } $ in $ \PSPi $ such that $ y $ is in conflict with $ x $ and $ z $ but $ x $ and $ z $ are parallel.
	Since $ \arbitraryEncoding $ as well as $ \asymp $ respect barbs, $ T_x\WeakBarb{\overline{o_x}} $, $ T_x\NotWeakBarb{\overline{o_y}} $, $ T_y\NotWeakBarb{\overline{o_x}} $, $ T_y\WeakBarb{\overline{o_y}} $, $ T_y\NotWeakBarb{\overline{o_z}} $, $ T_z\NotWeakBarb{\overline{o_y}} $, $ T_z\WeakBarb{\overline{o_z}} $, and thus $ T_x \not\asymp T_y \not\asymp T_z $.
	We conclude that, for all $ T_x, T_y, T_z \in \procCMVmix $ such that $ T_x \asymp \ArbitraryEncoding{S_x} $, $ T_y \asymp \ArbitraryEncoding{S_y} $, and $ T_z \asymp \ArbitraryEncoding{S_z} $ and for all sequences $ X : \ArbitraryEncoding{\PSPi} \steps T_x $, $ Y : \ArbitraryEncoding{\PSPi} \steps T_y $, and $ Z : \ArbitraryEncoding{\PSPi} \steps T_z $, there is a conflict between a step of $ X $ and a step of $ Y $, and there is a conflict between a step of $ Y $ and a step of $ Z $.
\end{proof}

Then we show that each good encoding of the counterexample $ \PSPi $ has to distribute one of its conflicts.

\begin{lemma}
	\label{lem:distributeMixedChoice}
	Any good encoding $ \arbitraryEncoding $ from the \piCal into \CMVmix has to split up a least one of the conflicts in $ \PSPi $ given by Example~\ref{exa:piStar} such that there exists a maximal execution in $ \ArbitraryEncoding{\PSPi} $ that emulates only one source term step.
\end{lemma}

\begin{proof}
	By operational completeness, all five steps of $ \PSPi $ have to be emulated in $ \ArbitraryEncoding{\PSPi} $, \ie there exist some $ T_a, T_b, T_c, T_d, T_e \in \procCMVmix $ such that $ X: \ArbitraryEncoding{\PSPi} \steps T_x \asymp \ArbitraryEncoding{S_x} $ for all $ x \in \Set{ a, b, c, d, e } $, where $ X $ is the upper case variant of $ x $.
	By Lemma~\ref{lem:translateConflictsStar}, for all $ T_a, T_b, T_c, T_d, T_e \in \procCMVmix $ and all $ x \in \Set{ a, b, c, d, e } $ such that $ T_x \asymp \ArbitraryEncoding{S_x} $, there is a conflict between a step of the following pairs of emulations: $ A $ and $ B $, $ B $ and $ C $, $ C $ and $ D $, $ D $ and $ E $, and $ E $ and $ A $.

	Since $ \arbitraryEncoding $ preserves distributability and by Lemma~\ref{lem:distributabilityPreservation}, each pair of distributable steps in $ \PSPi $ has to be translated into emulations that are distributable within $ \ArbitraryEncoding{\PSPi} $.
	Let $ X, Y, Z \in \Set{ A, B, C, D, E } $ be such that $ X $ and $ Z $ are distributable within $ \ArbitraryEncoding{\PSPi} $ but $ Y $ is in conflict with $ X $ as well as $ Z $.
	By Lemma~\ref{lem:distributabilityReductionsVsProcesses}, this implies that $ \ArbitraryEncoding{\PSPi} $ is distributable into $ T_1, T_2 \in \procCMVmix $ such that $ X $ is an execution of $ T_1 $ and $ Z $ is an execution of $ T_2 $. Since $ Y $ is in conflict with $ X $ and $ Z $ and because all three emulations are executions of $ \ArbitraryEncoding{\PSPi} $, there is one step of $ Y $ that is in conflict with one step of $ X $ and there is one (possibly the same) step of $ Y $ that is in conflict with one step of $ Z $.
	Moreover, since $ X $ and $ Z $ are distributable, if a single step of $ Y $ is in conflict with $ X $ as well as $ Z $ then this step is a communication between $ T_1 $ and $ T_2 $.

	Assume that for all such combinations $ X $, $ Y $, and $ Z $, the conflicts between $ Y $ and $ X $ or $ Z $ are ruled out by a single step of $ Y $, \ie both conflicts are ruled out by a communication step between some choice of $ X $ and some choice of $ Z $.
	Then this step reduces one endpoint in one of the executions $ X $ and $ Z $ and the respective other endpoint in the respective other execution, \ie $ X $ and $ Y $ compete for one endpoint and $ Y $ and $ Z $ compete for the respective other endpoint (compare to Lemma~\ref{lem:noStarCMVmix}).
	Without loss of generality let us assume that $ A $ and $ B $ compete for the channel endpoint $ x $ and, thus, $ B $ and $ C $ compete for the channel endpoint $ y $, $ C $ and $ D $ compete for $ x $, $ D $ and $ E $ compete for $ y $, $ E $ and $ A $ compete for $ x $, and $ A $ and $ B $ compete for $ y $. This is a contradiction, because $ A $ and $ B $ cannot compete for both channel endpoints $ x $ and $ y $.

	We conclude that there is at least one triple of emulations $ X $, $ Y $, and $ Z $ such that the conflict of $ Y $ with $ X $ and with $ Z $ results from two different steps in $ Y $.
	Because $ X $ and $ Z $ are distributable, the reduction steps of $ X $ that lead to the conflicting step with $ Y $ and the reduction steps of $ Z $ that lead to the conflicting step with $ Y $ are distributable.
	We conclude, that there is at least one emulation of $ y $, \ie one execution $ Y: \ArbitraryEncoding{\PSPi} \steps T_y \asymp \ArbitraryEncoding{S_y} $, starting with two distributable executions such that one is (in its last step) in conflict with the emulation of $ x $ in $ X: \ArbitraryEncoding{\PSPi} \steps T_x \asymp \ArbitraryEncoding{S_x} $ and the other one is in conflict with the emulation of $ z $ in $ Z: \ArbitraryEncoding{\PSPi} \steps T_z \asymp \ArbitraryEncoding{S_z} $. In particular this means that also the two steps of $ Y $ that are in conflict with a step in $ X $ and a step in $ Z $ are distributable.
	Hence, it is impossible to ensure that these two conflicts are decided consistently, \ie there is a maximal execution of $ \ArbitraryEncoding{\PSPi} $ that emulates $ X $ but neither $ Y $ nor $ Z $.

	In the set $ \Set{ A, B, C, D, E } $ there are---apart from $ X $, $ Y $, and $ Z $---two remaining executions.
	One of them, say $ X' $, is in conflict with $ X $ and the other one, say $ Z' $, is in conflict with $ Z $. Since $ X $ is emulated successfully, $ X' $ cannot be emulated.
	Moreover, note that $ Y $ and $ Z' $ are distributable.
	Thus, also $ Z' $ and the partial execution of $ Y $ that leads to the conflict with $ Z $ are distributable.
	Moreover, also the step of $ Y $ that already rules out $ Z $ cannot be in conflict with a step of $ Z' $.
	Thus, although the successful completion of $ Z $ is already ruled out by the conflict with $ Y $, there is some step of $ Z $ left, that is in conflict with one step in $ Z' $.
	Hence, the conflict between $ Z $ and $ Z' $ cannot be ruled out by the partial execution described so far that leads to the emulation of $ X $ but forbids to complete the emulations of $ X' $, $ Y $, and $ Z $.
	Thus, it cannot be avoided that $ Z $ wins this conflict, \ie that also $ Z' $ cannot be completed.
	We conclude that there is a maximal execution of $ \ArbitraryEncoding{\PSPi} $ such that only one of the five source term steps of $ \PSPi $ is emulated.
\end{proof}

Since each maximal execution of $ \PSPi $ given by Example~\ref{exa:piStar} consists of exactly two distributable steps, Lemma~\ref{lem:distributeMixedChoice} violates the requirements on a good encoding.

\begin{theorem}[Separate \CMVmix and the \PiCal via \patternStar]
	\label{thm:separateCMVmixfromPiviaStar}
	$ $\\
	There is no good and distributability preserving encoding from the \piCal into \CMVmix.
\end{theorem}

\begin{proof}[Proof of Theorem~\ref{thm:separateCMVmixfromPiviaStar}]
	Assume the opposite, \ie there is a good encoding $ \arbitraryEncoding $ from the \piCal into \CMVmix, and, thus, also of $ \PSPi $ given by Example~\ref{exa:piStar}.
	By Lemma~\ref{lem:distributeMixedChoice} there exists a maximal execution in $ \ArbitraryEncoding{\PSPi} $ in which only one source term step is emulated.
	Let us denote this step by $ x \in \Set{ a, b, c, d, e } $, \ie there is a maximal execution $ X: \ArbitraryEncoding{\PSPi} \steps T_x \steps \ldots $ with $ T_x \asymp \ArbitraryEncoding{S_x} $ in that only step $ x $ is emulated.
	Moreover, because $ \arbitraryEncoding $ is operationally corresponding and respects barbs and because no other source term step is emulated, $ T_x\Barb{\overline{o_x}} $ but $ T_x\NotWeakBarb{\overline{o_y}} $ for any $ y \in \Set{ a, b, c, d, e } $ with $ x \neq y $.
	Since for every $ S' $ with $ \PSPi \steps S' $ there are at least two $ i \in \Set{ a, b, c, d, e } $ such that $ S'\WeakBarb{\overline{o_i}} $, the execution $ X $ violates the combination of the criteria operational soundness and that $ \arbitraryEncoding $ respects barbs.
	We conclude that there cannot be such an encoding.
\end{proof}


\section{Encoding Mixed Sessions into Separate Choice}
\label{sec:encodeMixedSessions}

In \cite[\S~7]{casalMordidoVasconcelos22} an encoding of mixed sessions (\CMVmix) into the variant of this session type system \CMV with only separate choice (branching and selection) is presented.
The proof of soundness of this encoding is missing in \cite{casalMordidoVasconcelos22}.
They suggest to prove soundness modulo ``a weak form of bisimulation''.
As discussed below, the soundness criterion used in \cite{casalMordidoVasconcelos22} needs to be corrected first.
Prior to this discussion, we present the encoding $ \EncCMVmixCMV{\cdot} $ from \CMVmix into \CMV of \cite{casalMordidoVasconcelos22}.

To describe the encoding function $ \EncCMVmixCMV{\cdot} $ we reorder choices $ \ChoiceCMVmix{q}{y}{\sum_{h \in \indexSet[H]} M_h} $ into their respective send and receive actions for the same label
\begin{align*}
	\ChoiceCMVmix{q}{y}{\sum_{i \in \indexSet} {\left( \sum_{j \in \indexSet[J]_i} \BranchCMVmix{\Label_i}{!}{v_{i, j}}{P_{i, j}} + \sum_{k \in \indexSet[K]_i} \BranchCMVmix{\Label_i}{?}{x_{i, k}}{P'_{i, k}} \right)}}
\end{align*}
where $ i \in \indexSet $ is used to range over labels and for each label $ \Label_i $ the indices $ j \in \indexSet[J]_i $ iterate over send branches and $ k \in \indexSet[K]_i $ iterate over receive branches with this label.

The paper \cite{casalMordidoVasconcelos22} does not explicitly mention a renaming policy, but for the encoding to work properly, we need the names $ c, d $ and $ u, v $ to be fresh.
To increase readability, we omit the renaming policy and instead assume that $ c, d, u, v $ are different from all source term names.
A renaming policy can implement this freshness property.
Therefore, assume a renaming policy $ \RPCMVmixCMV{\cdot} $ that does not split names, \ie translates a source term name by a single target term name, but that reserves the names $ c, d, u, v $ such that $ \RPCMVmixCMV{y} \cap \Set{c, d, u, v} = \emptyset $ for all source term names $ y $.
Then replace all names $ n $ in target terms except $ c, d, u, v $ by $ \RPCMVmixCMV{n}.1 $.

\begin{figure}[tp]
	\begin{align*}
		& \EncCMVmixCMV{\Gamma \vdash \ChoiceCMVmix{\linCMVmix}{y}{\sum_{i \in \indexSet} {\left( \sum_{j \in \indexSet[J]_i} \BranchCMVmix{\Label_i}{!}{v_{i, j}}{P_{i, j}} + \sum_{k \in \indexSet[K]_i} \BranchCMVmix{\Label_i}{?}{x_{i, k}}{P'_{i, k}} \right)}}} =\\
		& \hspace{1em} \nDChoiceCMV\Bigg\{ \!\! \begin{array}[t]{l}
				\SelCMV{y}{\Label_{i, !}}{\nDChoiceCMV{\Set{\OutCMV{y}{v_{i, j}}{\EncCMVmixCMV{\Gamma_4 \vdash P_{i, j}}}}_{j \in \indexSet[J]_i}}},\\
				\SelCMV{y}{\Label_{i, ?}}{\nDChoiceCMV\Set{ \InpCMV{\linCMV}{y}{x_{i, k}}{\EncCMVmixCMV{{\left( \Gamma_2 + \At{y}{U_i'} \right)}, \At{x_{i, k}}{T_i'} \vdash P'_{i, k}}} }_{k \in \indexSet[K]_i}} \Bigg\}_{i \in \indexSet}
			\end{array}
	\end{align*}
	where $ \Gamma = \Gamma_1 \circ \Gamma_2 $ and $ \Gamma_1 \vdash \At{y}{\IntCMVmix{\linCMVmix}{\Set{\OutCMVmix{\Label_i}{T_i}{U_i}, \InpCMVmix{\Label_i}{T_i'}{U_i'}}_{i \in \indexSet}}} $ and $ \Gamma_2 + \At{y}{U_i} = \Gamma_3 \circ \Gamma_4 $ and $ \Gamma_3 \vdash \At{v_{i, j}}{T_i} $.
	\begin{align*}
		& \EncCMVmixCMV{\Gamma \vdash \ChoiceCMVmix{\linCMVmix}{y}{\sum_{i \in \indexSet} {\left( \sum_{j \in \indexSet[J]_i} \BranchCMVmix{\Label_i}{!}{v_{i, j}}{P_{i, j}} + \sum_{k \in \indexSet[K]_i} \BranchCMVmix{\Label_i}{?}{x_{i, k}}{P'_{i, k}} \right)}}} =\\
		& \hspace{1em} \BranCMV{y}{}\Bigg\{ \!\! \begin{array}[t]{l}
				\BranchCMV{\Label_{i, ?}}{\nDChoiceCMV\Set{\OutCMV{y}{v_{i, j}}{\EncCMVmixCMV{\Gamma_4 \vdash P_{i, j}}}}_{j \in \indexSet[J]_i}},\\
				\BranchCMV{\Label_{i, !}}{\nDChoiceCMV\Set{\InpCMV{\linCMV}{y}{x_{i, k}}{\EncCMVmixCMV{{\left( \Gamma_2 + \At{y}{U_i} \right)}, \At{x_{i, k}}{T_i} \vdash P_{i, k}'}}}_{k \in \indexSet[K]_i}} \Bigg\}_{i \in \indexSet}
			\end{array}
	\end{align*}
	where $ \Gamma = \Gamma_1 \circ \Gamma_2 $ and $ \Gamma_1 \vdash \At{y}{\ExtCMVmix{\linCMVmix}{\Set{\OutCMVmix{\Label_i}{T_i}{U_i}, \InpCMVmix{\Label_i}{T_i'}{U_i'}}_{i \in \indexSet}}} $ and $ \Gamma_2 + \At{y}{U_i'} = \Gamma_3 \circ \Gamma_4 $ and $ \Gamma_3 \vdash \At{v_{i, j}}{T_i'} $.
	\begin{align*}
		& \EncCMVmixCMV{\Gamma \vdash \ChoiceCMVmix{\linCMVmix}{y}{\sum_{i \in \indexSet} {\left( \sum_{j \in \indexSet[J]_i} \BranchCMVmix{\Label_i}{!}{v_{i, j}}{P_{i, j}} + \sum_{k \in \indexSet[K]_i} \BranchCMVmix{\Label_i}{?}{x_{i, k}}{P'_{i, k}} \right)}}} =\\
		& \hspace{1em} \nDChoiceCMV\Bigg\{ \!\! \begin{array}[t]{l}
				\ResCMV{c}{d}{{\left( \OutCMV{y}{c}{\SelCMV{d}{\Label_{i, !}}{\nDChoiceCMV{\Set{\OutCMV{d}{v_{i, j}}{\EncCMVmixCMV{\Gamma \vdash P_{i, j}}}}_{j \in \indexSet[J]_i}}}} \right)}},\\
				\ResCMV{c}{d}{{\left( \OutCMV{y}{c}{\SelCMV{d}{\Label_{i, ?}}{\nDChoiceCMV{\Set{\InpCMV{\linCMV}{d}{x_{i, k}}{\EncCMVmixCMV{\Gamma, \At{x_{i, k}}{T_i'} \vdash P_{i, k}'}}}_{k \in \indexSet[K]_i}}}} \right)}}  \Bigg\}_{i \in \indexSet}
			\end{array}
	\end{align*}
	where $ \UnT{\Gamma} $ and $ \Gamma \vdash \At{y}{\RecCMVmix{t}{\IntCMVmix{\unCMVmix}{\Set{\OutCMVmix{\Label_i}{T_i}{t}, \InpCMVmix{\Label_i}{T_i'}{t}}_{i \in \indexSet}}}} $ and $ \Gamma \vdash \At{v_{i, j}}{T_i} $.
	\begin{align*}
		& \EncCMVmixCMV{\Gamma \vdash \ChoiceCMVmix{\linCMVmix}{y}{\sum_{i \in \indexSet} {\left( \sum_{j \in \indexSet[J]_i} \BranchCMVmix{\Label_i}{!}{v_{i, j}}{P_{i, j}} + \sum_{k \in \indexSet[K]_i} \BranchCMVmix{\Label_i}{?}{x_{i, k}}{P'_{i, k}} \right)}}} =\\
		& \hspace{1em} \InpCMV{\linCMV}{y}{c}{}\BranCMV{c}{}\Bigg\{ \!\! \begin{array}[t]{l}
				\BranchCMV{\Label_{i, ?}}{\nDChoiceCMV\Set{\OutCMV{c}{v_{i, j}}{\EncCMVmixCMV{\Gamma \vdash P_{i, j}}}}_{j \in \indexSet[J]_i}},\\
				\BranchCMV{\Label_{i, !}}{\nDChoiceCMV\Set{\InpCMV{\linCMV}{c}{x_{i, k}}{\EncCMVmixCMV{\Gamma, \At{x_{i, k}}{T_i} \vdash P_{i, k}'}}}_{k \in \indexSet[K]_i}} \Bigg\}_{i \in \indexSet}
			\end{array}
	\end{align*}
	where $ \UnT{\Gamma} $ and $ \Gamma \vdash \At{y}{\RecCMVmix{t}{\ExtCMVmix{\unCMVmix}{\Set{\OutCMVmix{\Label_i}{T_i}{t}, \InpCMVmix{\Label_i}{T_i'}{t}}_{i \in \indexSet}}}} $ and $ \Gamma \vdash \At{v_{i, j}}{T_i'} $.
	\caption{The Encoding $ \EncCMVmixCMV{\cdot} $ from \CMVmix into \CMV from \cite{casalMordidoVasconcelos22} (Part I).}
	\label{fig:encCMVmixCMVA}
\end{figure}

\begin{figure}[tp]
	\begin{align*}
		& \EncCMVmixCMV{\Gamma \vdash \ChoiceCMVmix{\unCMVmix}{y}{\sum_{i \in \indexSet} {\left( \sum_{j \in \indexSet[J]_i} \BranchCMVmix{\Label_i}{!}{v_{i, j}}{P_{i, j}} + \sum_{k \in \indexSet[K]_i} \BranchCMVmix{\Label_i}{?}{x_{i, k}}{P'_{i, k}} \right)}}} = \\
		& \ResCMV{u}{v}{}\Bigg( \OutCMV{u}{\unit}{\inactCMV} \mid \InpCMV{\unCMV}{v}{\underline{\;}}{\nDChoiceCMV}\Bigg\{\\
		& \hspace{1em} \!\! \begin{array}[t]{l}
				\ResCMV{c}{d}{{\left( \OutCMV{y}{c}{\SelCMV{d}{\Label_{i, !}}{\nDChoiceCMV\Set{\OutCMV{d}{v_{i, j}}{\left( \OutCMV{u}{\unit}{\inactCMV} \mid \EncCMVmixCMV{\Gamma \vdash P_{i, j}} \right)}}_{j \in \indexSet[J]_i}}} \right)}},\\
				\ResCMV{c}{d}{{\left( \OutCMV{y}{c}{\SelCMV{d}{\Label_{i, ?}}{\nDChoiceCMV\Set{\InpCMV{\linCMV}{d}{x_{i, k}}{\left( \OutCMV{u}{\unit}{\inactCMV} \mid \EncCMVmixCMV{\Gamma, \At{x_{i, k}}{T_i'} \vdash P_{i, k}'} \right)}}_{k \in \indexSet[K]_i}}} \right)}} \Bigg\}_{i \in \indexSet} \Bigg)
			\end{array}
	\end{align*}
	where $ \UnT{\Gamma} $ and $ \Gamma \vdash \At{y}{\RecCMVmix{t}{\IntCMVmix{\unCMVmix}{\Set{\OutCMVmix{\Label_{i}}{T_i}{t}, \InpCMVmix{\Label_i}{T_i'}{t}}_{i \in \indexSet}}}} $ and $ \Gamma \vdash \At{v_{i, j}}{T_i} $.
	\begin{align*}
		& \EncCMVmixCMV{\Gamma \vdash \ChoiceCMVmix{\unCMVmix}{y}{\sum_{i \in \indexSet} {\left( \sum_{j \in \indexSet[J]_i} \BranchCMVmix{\Label_i}{!}{v_{i, j}}{P_{i, j}} + \sum_{k \in \indexSet[K]_i} \BranchCMVmix{\Label_i}{?}{x_{i, k}}{P'_{i, k}} \right)}}} = \\
		& \ResCMV{u}{v}{}\Bigg( \OutCMV{u}{\unit}{\inactCMV} \mid \InpCMV{\unCMV}{v}{\underline{\;}}{\InpCMV{\linCMV}{y}{c}{\BranCMV{c}{}}}\Bigg\{\\
		& \hspace{1em} \!\! \begin{array}[t]{l}
				\BranchCMV{\Label_{i, ?}}{\nDChoiceCMV\Set{\OutCMV{c}{v_{i, j}}{{\left( \OutCMV{u}{\unit}{\inactCMV} \mid \EncCMVmixCMV{\Gamma \vdash P_{i, j}} \right)}}}_{j \in \indexSet[J]_i}},\\
				\BranchCMV{\Label_{i, !}}{\nDChoiceCMV\Set{\InpCMV{\linCMV}{c}{x_{i, k}}{{\left( \OutCMV{u}{\unit}{\inactCMV} \mid \EncCMVmixCMV{\Gamma, \At{x_{i, k}}{T_i} \vdash P'_{i, k}} \right)}}}_{k \in \indexSet[K]_i}} \Bigg\}_{i \in \indexSet} \Bigg)
			\end{array}
	\end{align*}
	where $ \UnT{\Gamma} $ and $ \Gamma \vdash \At{y}{\RecCMVmix{t}{\ExtCMVmix{\unCMVmix}{\Set{\OutCMVmix{\Label_{i}}{T_i}{t}, \InpCMVmix{\Label_i}{T_i'}{t}}_{i \in \indexSet}}}} $ and $ \Gamma \vdash \At{v_{i, j}}{T_i'} $.
	\begin{align*}
		\EncCMVmixCMV{\Gamma_1 \circ \Gamma_2 \vdash P_1 \mid P_2} &= \EncCMVmixCMV{\Gamma_1 \vdash P_1} \mid \EncCMVmixCMV{\Gamma_2 \vdash P_2}\\
		\EncCMVmixCMV{\Gamma \vdash \ResCMVmix{y}{z}{P}} &= \ResCMV{y}{z}{\EncCMVmixCMV{\Gamma, \At{y}{T}, \At{z}{U} \vdash P}}\\
		\EncCMVmixCMV{\Gamma_1 \circ \Gamma_2 \vdash \ConditionalCMVmix{v}{P_1}{P_2}} &= \ConditionalCMV{v}{\EncCMVmixCMV{\Gamma_1 \vdash P_1}}{\EncCMVmixCMV{\Gamma_2 \vdash P_2}}\\
		\EncCMVmixCMV{\Gamma \vdash \inactCMVmix} &= \inactCMV
	\end{align*}
	where $ \Dual{T}{U} $ and $ \Gamma_1 \vdash \At{v}{\boolT} $.
	\caption{The Encoding $ \EncCMVmixCMV{\cdot} $ from \CMVmix into \CMV from \cite{casalMordidoVasconcelos22} (Part II).}
	\label{fig:encCMVmixCMVB}
\end{figure}

The encoding $ \EncCMVmixCMV{\cdot} $ of \cite{casalMordidoVasconcelos22} is then given by the Figures~\ref{fig:encCMVmixCMVA} and \ref{fig:encCMVmixCMVB}.
It relies on the predicate $ \nDChoiceCMV $, \ie a non-deterministic choice
\begin{align*}
	\nDChoiceCMV{\Set{P_i}_{i \in \indexSet}} = \ResCMV{s}{t}{{\left( \BranCMV{s}{\Set{\BranchCMV{\Label_i}{P_i}}_{i \in \indexSet}} \mid \prod_{i \in \indexSet} \SelCMV{t}{\Label_i}{\inactCMV} \right)}}
\end{align*}
introduced in \cite{casalMordidoVasconcelos22} for \CMV to non-deterministically choose one process from the set $ \Set{P_i}_{i \in \indexSet} $ in a single reduction step.
Let $ 1 \leq j \leq n $.
Then choosing option $ j $ we obtain
\begin{align*}
	\nDChoiceCMV{\Set{P_i}_{i \in \indexSet}} \step P_j \mid \ResCMV{s}{t}{{\left( \prod_{i \in \indexSet \setminus \Set{ j }} \SelCMV{t}{\Label_i}{\inactCMV} \right)}}
\end{align*}
where $ \ResCMV{s}{t}{{\left( \prod_{i \in \indexSet \setminus \Set{ j }} \SelCMV{t}{\Label_i}{\inactCMV} \right)}} $ remains as junk, \ie is stuck and does not emit barbs.
Then \cite{casalMordidoVasconcelos22} extend structural congruence $ \scCMV $ of \CMV to $ \scCMVGC $ by adding the rule
\begin{align*}
	\ResCMV{y}{z}{{\left( \prod_{i \in \indexSet} \SelCMV{y}{\Label_i}{\inactCMV} \right)}} \scCMVGC \inactCMV
\end{align*}
to garbage collect this kind of junk.
This relation is used in \cite{casalMordidoVasconcelos22} to prove completeness of the encoding $ \EncCMVmixCMV{\cdot} $.
For soundness we need something less restrictive, because the encoding allows the translation of an unrestricted choice to perform a step even if the original unrestricted choice in the source cannot be reduced (see \cite{casalMordidoVasconcelos22}).
As suggested we use $ \bisimCMV $ (as definition in Definition~\ref{def:bisimulation}), \ie a form of weak reduction barbed bisimilarity that we simply call bisimilarity in the following.

To prepare for the soundness proof, we show that steps reducing a non-deterministic choice always yield modulo bisimilarity one of its options.
Here we use bisimilarity to abstract from the junk produced by reducing non-deterministic choices and also show that a non-deterministic choice can do nothing but reduce to one of its options.

\begin{lemma}
	If $ \nDChoiceCMV\Set{P_i}_{i \in \indexSet} \stepCMV Q $ then there is some $ j \in \indexSet $ such that $ Q \bisimCMV P_j $.
	\label{lem:nDChoiceGarbageCollection}
\end{lemma}

\begin{proof}
	By the definition of $ \nDChoiceCMV $, there is some $ j \in \indexSet $ such that $ Q = P_j \mid J $ with $ J =\ResCMV{s}{t}{{\left( \prod_{i \in \indexSet \setminus \Set{ j }} \SelCMV{t}{\Label_i}{\inactCMV} \right)}} $.
	Since $ J $ is junk, \ie is stuck and does not emit barbs, $ Q \bisimCMV P_j $.
\end{proof}

The main idea of $ \EncCMVmixCMV{\cdot} $ is to encode the information about whether a summand is an output or an input into the label used in branching, where a label $ \Label_i $ used with polarity $ ! $ in a choice typed as internal becomes $ \Label_{i, !} $ and in a choice typed as external it becomes $ \Label_{i, ?} $.
The dual treatment of polarities \wrt the type ensures that the labels of matching communication partners are translated to the same label.

\begin{example}[Translation]
	\label{exa:translation}
	Consider for example the term $ S \in \procCMVmix $:
	\begin{align*}
		S &= \ResCMVmix{x}{y}{\left( \ChoiceCMVmix{\linCMVmix}{y}{\left( \textcolor{blue}{\OutCMVmix{\Label}{\false}{S_1}} + \textcolor{darkgreen}{\InpCMVmix{\Label}{z}{S_2}} \right)} \mid \ChoiceCMVmix{\linCMVmix}{x}{\left( \textcolor{orange}{\OutCMVmix{\Label}{\true}{\inactCMVmix}} + \textcolor{red}{\InpCMVmix{\Label}{z}{\inactCMVmix}} \right)} \mid \ChoiceCMVmix{\linCMVmix}{y}{\left( \OutCMVmix{\Label}{\false}{S_3} + \InpCMVmix{\Label}{z}{S_4} \right)} \right)}
	\end{align*}
	$ S $ is well-typed but the type system forces us to assign dual types to $ x $ and $ y $.
	Because of that, the choices on one channel need to be internal and on the other external.
	Let us assume that we have external choices on $ y $ and that the choice on $ x $ is internal.
	Moreover, we assume that both channels are marked as linear but typed as unrestricted.
	Then the translation\footnote{Note that \cite{casalMordidoVasconcelos22} introduces a typed encoding, thus $ \EncCMVmixCMV{P} $ actually means $ \EncCMVmixCMV{\Gamma \vdash P} $, where $ \Gamma \vdash P $ is the type statement ensuring that $ P $ is well-typed.} yields $ \EncCMVmixCMV{S} \steps T_1 $ with
	\begin{align*}
		T_1 = \ResCMV{x}{y}{\big(
			\!\!\begin{array}[t]{l}
				\InpCMV{\linCMV}{y}{c}{\BranCMV{c}{\Set{\BranchCMV{\textcolor{blue}{\Label_{?}}}{\left( \OutCMV{c}{\textcolor{blue}{\false}}{\textcolor{blue}{\EncCMVmixCMV{S_1}}} \mid J_1 \right)}, \quad \BranchCMV{\textcolor{darkgreen}{\Label_{!}}}{\left( \InpCMV{\linCMV}{c}{\textcolor{darkgreen}{z}}{\textcolor{darkgreen}{\EncCMVmixCMV{S_2}}} \mid J_2 \right)}}}}\\
				{}\mid \ResCMV{s}{t}{\big(
					\!\!\begin{array}[t]{l}
						\BranCMV{s}{\Set{ \BranchCMV{\Label_1}{\ResCMV{c}{d}{\left( \OutCMV{x}{c}{\SelCMV{d}{\textcolor{orange}{\Label_{!}}}{\left( \OutCMV{d}{\textcolor{orange}{\true}}{\textcolor{orange}{\inactCMV}} \mid J_3\right)}} \right)}}, \quad \BranchCMV{\Label_2}{\ResCMV{c}{d}{\left( \OutCMV{x}{c}{\SelCMV{d}{\textcolor{red}{\Label_{?}}}{\left( \InpCMV{\linCMV}{d}{\textcolor{red}{z}}{\textcolor{red}{\inactCMV}} \mid J_4\right)}} \right)}} }}\\
						{}\mid \SelCMV{t}{\Label_1}{\inactCMV} \mid \SelCMV{t}{\Label_2}{\inactCMV} \big)
					\end{array}}\\
				{}\mid \InpCMV{\linCMV}{y}{c}{\BranCMV{c}{\Set{\BranchCMV{\Label_{?}}{\left( \OutCMV{c}{\false}{\EncCMVmixCMV{S_3}} \mid J_5 \right)}, \quad \BranchCMV{\Label_{!}}{\left( \InpCMV{\linCMV}{c}{z}{\EncCMVmixCMV{S_4}} \mid J_6 \right)}}}} \big)
			\end{array}}
	\end{align*}
	where we already performed a few steps to hide some technical details of the encoding function $ \EncCMVmixCMV{\cdot} $ that are not relevant for this explanation and where the $ J_1, \ldots, J_6 $ remain as junk from performing these steps.
	We call terms junk if they are stuck and do not emit barbs, \ie we can ignore the junk.
	In particular, junk is invisible modulo $ \bisimCMV $.
	We observe, that in the translation of the first $ \ChoiceCMVmix{\linCMVmix}{y}{\left( \textcolor{blue}{\OutCMVmix{\Label}{\false}{S_1}} + \textcolor{darkgreen}{\InpCMVmix{\Label}{z}{S_2}} \right)} $ in the first line of $ T_1 $ the output with label $ \textcolor{blue}{\Label} $ is translated to the label $ \textcolor{blue}{\Label_{?}} $ and the input with label $ \textcolor{darkgreen}{\Label} $ is translated to the label $ \textcolor{darkgreen}{\Label_{!}} $, whereas in the translation of its dual $ \ChoiceCMVmix{\linCMVmix}{x}{\left( \textcolor{orange}{\OutCMVmix{\Label}{\true}{\inactCMVmix}} + \textcolor{red}{\InpCMVmix{\Label}{z}{\inactCMVmix}} \right)} $ in the second line of $ T_1 $ we obtain $ \textcolor{orange}{\Label_{!}} $ for the output and $ \textcolor{red}{\Label_{?}} $ for the input.
	To emulate the step $ S \step S_2' = \ResCMVmix{x}{y}{\left( S_2\Set{\Subst{\true}{z}} \mid \ChoiceCMVmix{\linCMVmix}{y}{\left( \OutCMVmix{\Label}{\false}{S_3} + \InpCMVmix{\Label}{z}{S_4} \right)} \right)} $ of $ S $ in that $ \textcolor{orange}{\true} $ is transmitted to $ \textcolor{darkgreen}{S_2} $, we start by picking the corresponding alternative, namely $ \Label_1 $ for sending, in the second and third line of $ T_1 $
	\begin{align*}
		T_1 \step T_2 = \ResCMV{x}{y}{\big(
			\!\!\begin{array}[t]{l}
				\InpCMV{\linCMV}{y}{c}{\BranCMV{c}{\Set{\BranchCMV{\textcolor{blue}{\Label_{?}}}{\left( \OutCMV{c}{\textcolor{blue}{\false}}{\textcolor{blue}{\EncCMVmixCMV{S_1}}} \mid J_1 \right)}, \quad \BranchCMV{\textcolor{darkgreen}{\Label_{!}}}{\left( \InpCMV{\linCMV}{c}{\textcolor{darkgreen}{z}}{\textcolor{darkgreen}{\EncCMVmixCMV{S_2}}} \mid J_2 \right)}}}}\\
				{}\mid \ResCMV{c}{d}{\left( \OutCMV{x}{c}{\SelCMV{d}{\textcolor{orange}{\Label_{!}}}{\left( \OutCMV{d}{\textcolor{orange}{\true}}{\textcolor{orange}{\inactCMV}} \mid J_3\right)}} \right)} \mid J_7\\
				{}\mid \InpCMV{\linCMV}{y}{c}{\BranCMV{c}{\Set{\BranchCMV{\Label_{?}}{\left( \OutCMV{c}{\false}{\EncCMVmixCMV{S_3}} \mid J_5 \right)}, \quad \BranchCMV{\Label_{!}}{\left( \InpCMV{\linCMV}{c}{z}{\EncCMVmixCMV{S_4}} \mid J_6 \right)}}}} \big)
			\end{array}}
	\end{align*}
	where $ J_7 $ again remains as junk.
	Then we perform a communication on $ xy $, where we chose the input on $ y $ in the first line:
	\begin{align*}
		T_2 \step T_3 = \ResCMV{x}{y}{\big(
			\!\!\begin{array}[t]{l}
				\ResCMV{c}{d}{\big(
					\!\!\begin{array}[t]{l}
						\BranCMV{c}{\Set{\BranchCMV{\textcolor{blue}{\Label_{?}}}{\left( \OutCMV{c}{\textcolor{blue}{\false}}{\textcolor{blue}{\EncCMVmixCMV{S_1}}} \mid J_1 \right)}, \quad \BranchCMV{\textcolor{darkgreen}{\Label_{!}}}{\left( \InpCMV{\linCMV}{c}{\textcolor{darkgreen}{z}}{\textcolor{darkgreen}{\EncCMVmixCMV{S_2}}} \mid J_2 \right)}}}\\
						{}\mid \SelCMV{d}{\textcolor{orange}{\Label_{!}}}{\left( \OutCMV{d}{\textcolor{orange}{\true}}{\textcolor{orange}{\inactCMV}} \mid J_3\right)} \big) \mid J_7
					\end{array}}\\
				{}\mid \InpCMV{\linCMV}{y}{c}{\BranCMV{c}{\Set{\BranchCMV{\Label_{?}}{\left( \OutCMV{c}{\false}{\EncCMVmixCMV{S_3}} \mid J_5 \right)}, \quad \BranchCMV{\Label_{!}}{\left( \InpCMV{\linCMV}{c}{z}{\EncCMVmixCMV{S_4}} \mid J_6 \right)}}}} \big)
			\end{array}}
	\end{align*}
	Finally, two more steps on $ cd $ resolve the branching and transmit $ \textcolor{orange}{\true} $:
	\begin{align*}
		T_3 \step\step T_4 = \ResCMV{x}{y}{\big(
			\!\!\begin{array}[t]{l}
				\textcolor{darkgreen}{\EncCMVmixCMV{S_2}}\Set{\Subst{\textcolor{orange}{\true}}{\textcolor{darkgreen}{z}}} \mid J_2 \mid J_3 \mid J_7 \mid J_8\\
				{}\mid \InpCMV{\linCMV}{y}{c}{\BranCMV{c}{\Set{\BranchCMV{\Label_{?}}{\left( \OutCMV{c}{\false}{\EncCMVmixCMV{S_3}} \mid J_5 \right)}, \quad \BranchCMV{\Label_{!}}{\left( \InpCMV{\linCMV}{c}{z}{\EncCMVmixCMV{S_4}} \mid J_6 \right)}}}} \big)
			\end{array}}
	\end{align*}
	This completes the emulation of $ S \step S_2' $, \ie the emulation of the single source term step $ S \step S_2' $ required a sequence of target term steps $ \EncCMVmixCMV{S} \steps T_1 \step T_2 \step T_3 \step \step T_4 $.
	\qed
\end{example}

The operational soundness is defined in \cite{casalMordidoVasconcelos22} as 
(adapting the notation):
\begin{equation}
	\label{eq:cmvsound}
	\text{If $ \ArbitraryEncoding{S} \stepTarget T $ then $ S \stepSource S' $ and $ T \stepsTarget \asymp \ArbitraryEncoding{S'} $.}
\end{equation}
As visualised above, the encoding translates a single source term step into a sequence of target term steps.
Unfortunately, for such encodings the statement in (\ref{eq:cmvsound}) is not strong enough: 
with (\ref{eq:cmvsound}), we check only that the first step on a literal translation does not introduce new behaviour.
The requirement $ T \stepsTarget \asymp \ArbitraryEncoding{S'} $ additionally checks that the emulation started with $ \ArbitraryEncoding{S} \stepTarget T $ can be completed, but not that there are no alternative steps introducing new behaviour.
Hence we prove a correct version of soundness as defined in \cite{gorla10} (see Definition~\ref{def:goodEncoding}).

We denote the steps that reduce the first non-deterministic choice of the translation of a choice typed as internal and steps reducing a conditional as \emph{starting-steps}.
The emulation of a step that reduces a conditional in the source is a single starting-step that also reduces a conditional in the target.
The emulation of a communication starts with a single starting-step followed by some other steps to complete the emulation (that might be interleaved with steps from other emulations).
Similarly, for branching we have again a single starting-step in the beginning.

For soundness we have to show that all steps of encoded terms belong modulo bisimilarity to the emulation of a source term step.

\begin{lemma}[Soundness, $ \EncCMVmixCMV{\cdot} $]
	The encoding $ \EncCMVmixCMV{\cdot} $ is operationally sound modulo $ \bisimCMV $, \ie $ \EncCMVmixCMV{S} \steps T $ implies $ S \steps S' $ and $ T \steps \bisimCMV \EncCMVmixCMV{S'} $.
	\label{lem:soundnessEncCMVmixCMV}
\end{lemma}

\begin{proof}[Proof of Lemma~\ref{lem:soundnessEncCMVmixCMV}]
	We have to prove that for all $ S \in \procCMVmix $ and all $ T \in \procCMV $ such that $ \EncCMVmixCMV{S} \steps T $ there is some $ S' \in \procCMVmix $ and some $ T' \in \procCMV $ such that $ S \steps S' $, $ T \steps T' $, and $ T' \bisimCMV \EncCMVmixCMV{S'} $.
	We further strengthen this goal by requiring that the sequence $ T \steps T' $ contains no starting steps.
	This ensures that the steps $ T \steps T' $ can only complete already started emulations instead of starting new emulations.
	We start with an induction on the number of steps in $ \EncCMVmixCMV{S} \steps T $.
	For the base case, \ie if $ T = \EncCMVmixCMV{S} $ is reached in zero steps, we choose $ S' = S $ and $ T' = T $ and obtain $ S \steps S' $, $ T \steps T' $, and $ T' \bisimCMV \EncCMVmixCMV{S'} $ as required.
	For the induction step we have $ \EncCMVmixCMV{S} \steps T_1 \step T $.
	By the induction hypothesis, there are $ S_2, T_2 $ such that $ S \steps S_2 $, $ T_1 \steps T_2 $, and $ T_2 \bisimCMV \EncCMVmixCMV{S_2} $, where $ T_1 \steps T_2 $ does not contain starting-steps.

	If the sequence $ T_1 \steps T_2 $ reduces in one step the same conditional or the same input and output or selection and branching constructs as reduced in $ T_1 \step T $ then we can reorder the sequence such that $ T_1 \step T \steps T_2 $.
	Then we can choose $ S' = S_2 $ and $ T' = T_2 $ such that $ S \steps S' $, $ T \steps T' $, and $ T' \bisimCMV \EncCMVmixCMV{S'} $ as required.
	
	Else, consider the case that $ T_1 \step T $ is not in conflict with any step in $ T_1 \steps T_2 $.
	If it is a part of an emulation but not a starting-step the corresponding emulation that was started before or by reaching $ T_1 $ was finished in $ T_1 \steps T_2 $ modulo $ \bisimCMV $ to ensure $ T_2 \bisimCMV \EncCMVmixCMV{S_2} $.
	Since $ T_1 \step T $ is not in conflict with any step in $ T_1 \step T_2 $, then either $ T_1 \bisimCMV T $ or the step $ T_1 \stepCMV T $ is a starting-step.
	Note that $ T_1 \bisimCMV T $ may result from a communication on the channel endpoints $ u, v $ introduced in the Cases~5 or 6 of the encoding function, but also \eg from a non-deterministic choice with a single option.
	If $ T_1 \bisimCMV T $ then $ T \steps T_2 $, \ie we can choose $ S' = S_2 $ and $ T' = T_2 $ such that $ S \steps S' $, $ T \steps T' $, and $ T' \bisimCMV \EncCMVmixCMV{S'} $ as required.
	Otherwise, if $ T_1 \step T $ is a starting step we complete this emulation with a sequence $ T_1 \step T \steps T'' $ as described in the completeness proof in \cite{casalMordidoVasconcelos22}.
	Since $ T_1 \step T $ is a starting-step, no step of the sequence $ T_1 \step T \steps T'' $ is in conflict with any step of $ T_1 \steps T_2 $.
	Then there is some $ T' $ such that $ T \steps T'' \steps T' $, where the sequence $ T'' \steps T' $ performs the steps of $ T_1 \steps T_2 $ starting in $ T'' $ instead of $ T_1 $.
	Moreover, there is some $ S' $ such that $ S \steps S_2 \step S' $, where the step $ S_2 \step S' $ is the step that is emulated in $ T_1 \step T \steps T'' $.
	Since $ T_2 \bisimCMV \EncCMVmixCMV{S_2} $ and by the construction of $ T' $ and $ S' $, we have $ T' \bisimCMV \EncCMVmixCMV{S'} $.

	Otherwise, there is exactly one step in the sequence $ T_1 \steps T_2 $ that is in conflict with the step $ T \step T' $.
	Every such conflict marks a decision in the emulation of one or another source term step.
	To conclude, we have to show that the sequences $ T_1 \steps T_2 $ and $ S \steps S_2 $ can be adapted to the alternative decisions in $ T_1 \step T $, \ie that all decisions of an encoded term lead to the emulation of a source term step.
	The procedure is similar for all decision points, we give a detailed proof for the first case. The other cases are similar or simpler.
	\begin{description}
		\item[Case 1:] The translation of a choice that is typed as linear and internal starts with a non-deterministic choice that has exactly one option for each summand of the source term choice, where summands with the same label and polarity are handled by the same option for the moment.
			The $ \nDChoiceCMV $ construct non-deterministically picks the translation of one of these summands.
			A conflict between $ T_1 \step T $ and one step in $ T_1 \steps T_2 $ then means that they both reduce this $ \nDChoiceCMV $ construct but pick different source term summands.
			Since $ S $ is well-typed, the source term choice can be reduced only with a communication partner that is typed as linear external choice and encoded by the second case and there is at most one such choice on the respective other channel endpoint.

			If there is no such choice on the other channel endpoint then the source term choice is stuck.
			Because the encoding does use source term channels only to encode a choice that is already on this source term channel, if the source term choice is stuck, so is its translation.
			In this case, the difference between $ T_1 \step T $ and its conflicting step in $ T_1 \steps T_2 $ cannot be observed modulo $ \bisimCMV $, since both translations of summands emit the same barb.
			Then we can simply replace the conflicting step in $ T_1 \step T_2 $ by the step $ T_1 \step T $ and reorder the sequence, \ie we have $ T_1 \step T \steps T' $, where $ T' $ is obtained from $ T_2 $ by exchanging the translations of the two summands.
			Then we choose $ S' = S_2 $ and have $ S \steps S' $, $ T \steps T' $, and $ T' \bisimCMV \EncCMVmixCMV{S'} $ as required.

			Otherwise, the type system ensures that for each combination of label and polarity requested by the internal choice the external choice offers a matching summand.
			The translation of choice typed as linear and internal introduces primitives for four consecutive steps: the outer non-deterministic choice, a selection construct, another non-deterministic choice, and an output or input.
			The translation of choice type as linear and external introduces a branching, then a non-deterministic choice, and an output or input.
			Since we reason modulo $ \bisimCMV $, the sequence $ T_1 \steps T_2 $ might not contain all of these steps.
			But, since we forbid for starting-steps in $ T_1 \steps T_2 $, there are no steps that rely on the emulation of this source term communication.
			Then we remove the conflicting step in $ T_1 \steps T_2 $ as well as all steps that also belong to this emulation attempt.
			Instead let $ T_1 \step T \steps T'' $ be the steps necessary to fully emulate a step with the summand picked in $ T_1 \step T $. That such a sequence of steps can be found was shown in the completeness theorem in \cite{casalMordidoVasconcelos22}.
			Then there is some $ T' $ such that $ T \steps T'' \steps T' $, where the sequence $ T'' \steps T' $ executes exactly the steps performed in $ T_1 \steps T_2 $ after removing the steps on the conflicting emulation.
			Then $ S \steps S' $ there $ S' $ is obtained from $ S_2 $ by exchanging the source term step whose emulation we removed by the source term step that is emulated in $ T \steps T'' $.
			Since $ T_2 \bisimCMV \EncCMVmixCMV{S_2} $ and by the construction of $ T' $ and $ S' $, then $ T' \bisimCMV \EncCMVmixCMV{S'} $.

			The second primitive introduced by Case~1 of the encoding is a selection construct.
			This step cannot be in conflict with any other step, because the matching branching construct in Case~2 provides exactly one branch for each combination of label and polarity.
			The third primitive is again a non-deterministic choice that allows to pick one value for transmission and matching continuation if there are several summands with the same label and polarity.
			The proof for this case is similar to the non-deterministic choice above.
			Finally, there is an output or input. Again this step cannot be in conflict with a step in $ T_1 \steps T_2 $, because it is not possible to unguard more than one input or output for a choice that is typed as linear.
		\item[Case 2:] This case is dual to the case above, but simpler since the first non-deterministic choice is missing.
		\item[Case 3:] The translation of a linear choice that is typed as unrestricted and internal starts again with a non-deterministic choice that has exactly one option for each summand of the source term choice, where summands with the same label and polarity are handled by the same option for the moment.
			A conflict between $ T_1 \step T $ and one step in $ T_1 \steps T_2 $ then means that they both reduce this $ \nDChoiceCMV $ construct but pick different source term summands.
			We proceed as with the non-deterministic choice in Case~1. Here, the translation of a linear choice typed as unrestricted and internal introduces primitives for five consecutive steps: the outer non-deterministic choice, an output, a selection construct, another non-deterministic choice, and an output or input.
			The translation of choice type as unrestricted and external introduces an output and matching input in Case~6 (but not Case~4), an input, a branching, then a non-deterministic choice, and an output or input.
			Accordingly, we might need to remove more steps from $ T_1 \steps T_2 $.

			The second primitive introduced by Case~3 is an output.
			Since in Case~3 a choice typed as unrestricted is translated that is matched by the type system with another choice typed as unrestricted, there can be several outputs on this channel endpoint or several inputs on the other channel endpoint.
			Since we forbid for starting-steps in $ T_1 \steps T_2 $, there are no steps that rely on the emulation of the source term communication.
			Again we remove the conflicting step in $ T_1 \steps T_2 $ as well as all steps that also belong to this emulation attempt.
			Instead let $ T_1 \step T \steps T'' $ be the steps necessary to fully emulate a step with the input and output picked in $ T_1 \step T $. That such a sequence of steps can be found was shown in the completeness theorem in \cite{casalMordidoVasconcelos22}.
			Then there is some $ T' $ such that $ T \steps T'' \steps T' $, where the sequence $ T'' \steps T' $ executes exactly the steps performed in $ T_1 \steps T_2 $ after removing the steps on the conflicting emulation.
			Then $ S \steps S' $ there $ S' $ is obtained from $ S_2 $ by exchanging the source term step whose emulation we removed by the source term step that is emulated in $ T \steps T'' $.
			Since $ T_2 \bisimCMV \EncCMVmixCMV{S_2} $ and by the construction of $ T' $ and $ S' $, then $ T' \bisimCMV \EncCMVmixCMV{S'} $.

			Next there is a selection construct matched by a branching construct in the Cases~4 or 6.
			The restriction on the channel endpoints $ c $ and $ d $ in Case~3 ensures that this step cannot be in conflict with any other step.
			The fourth primitive is again a non-deterministic choice that allows to pick one value for transmission and matching continuation if there are several summands with the same label and polarity.
			The proof for this case is similar to the non-deterministic choice in Case~1.
			Finally, there is an output or input. Again the restriction on the channel endpoints $ c $ and $ d $ in Case~3 ensures that this step cannot be in conflict with any other step.
		\item[Case 4:] This case is dual to the case above, but simpler since the first non-deterministic choice is missing.
		\item[Case 5:] In comparison to Case~3 there is only one additional internal step on the restricted channels $ u, v $.
			Because of the restriction, this step cannot be in conflict with any other step.
			The proof is then as in Case~3.
		\item[Case 6:] This case is dual to the case above, but simpler since the first non-deterministic choice is missing.
		\item[Case 7:] The translation of parallel composition does not introduce any steps, \ie there are no conflicts to be considered in this case.
		\item[Case 8:] The translation of restriction does not introduce any steps, \ie there are no conflicts to be considered in this case.
		\item[Case 9:] The translation of a conditional in \CMVmix yields a conditional in \CMV.
			Since steps reducing a conditional in \CMVmix (as well as \CMV) cannot be conflict to any other steps, \ie there are no conflicts to be considered in this case.
		\item[Case 10:] The translation of $ \inactCMVmix $ cannot perform steps, \ie there are no conflicts to be considered in this case.
	\end{description}
\end{proof}

In the proof we analyse the sequence of steps $ \EncCMVmixCMV{S} \steps T $ and identify all source term steps $ S \steps S' $ whose emulation is started within $ \EncCMVmixCMV{S} \steps T $ and the target term steps $ T \steps \bisimCMV \EncCMVmixCMV{S'} $ that are necessary to complete all started emulations modulo bisimulation.
Therefore, we use an induction on the number of steps in the sequence $ \EncCMVmixCMV{S} \steps T $ and analyse the encoding function in order to distinguish between different kinds of target term steps and the emulations of source term steps to that they belong.
Note that, as it is typical for many encodability results, the proof of operational soundness is more elaborate than the proof of operational completeness presented in \cite{casalMordidoVasconcelos22}.

In Example~\ref{exa:translation} we have $ T_4 \bisimCMV \EncCMVmixCMV{S_2'} $, because all differences between $ T_4 $ and $ \EncCMVmixCMV{S_2'} $ are due to junk that cannot be observed modulo $ \bisimCMV $.
In fact, we have already $ T_3 \bisimCMV \EncCMVmixCMV{S_2'} $, since we consider a weak form of bisimulation here.

In the above variant of soundness $ T $ can catch up with the source term $ S' $ by the steps $ T \steps \bisimCMV \EncCMVmixCMV{S'} $.
This allows for so-called \emph{intermediate states}: target terms that are strictly in between the translation of two source terms, \ie $ T $ such that $ S \step S' $, $ \EncCMVmixCMV{S} \steps T \steps \bisimCMV \EncCMVmixCMV{S'} $, but neither $ \EncCMVmixCMV{S} \bisimCMV T $ nor $ \EncCMVmixCMV{S'} \bisimCMV T $ (see \cite{parrow1992, petersNestmannGoltz13}).
In $ \EncCMVmixCMV{\cdot} $ such intermediate states are caused by mapping the task of finding matching communication partners of a single source term step onto several steps in the target.
Consider the term $ T_2 $ in the above emulation of $ S \step S_2' $.
By picking the branch with label $ \Label_1 $, we discarded the branch with label $ \Label_2 $.
Because of that, the emulation starting with $ \EncCMVmixCMV{S} \steps T_2 $ can no longer emulate source term steps of $ S $ that use channel $ x $ for receiving, \ie $ T_2 \not\bisimCMV \EncCMVmixCMV{S} $.
But, since we have not yet decided whether we emulate a communication with the first or second choice on $ y $, we also have $ T_2 \not\bisimCMV \EncCMVmixCMV{S_2'} $ whenever $ S_2 \not\bisimCMVmix S_4 $.
Indeed, if we assume that $ S_1, S_2, S_3, S_4 $ are pairwise not bisimilar, then $ T_2 \not\bisimCMV \EncCMVmixCMV{S'} $ for all $ S \step S' $, \ie $ T_2 $ is an intermediate state.

The existence of intermediate states prevents us from using stronger versions of soundness, \ie with $ T \asymp \ArbitraryEncoding{S'} $ instead of the requirement $ T \stepsTarget \asymp \ArbitraryEncoding{S'} $ in soundness.
The encoding $ \EncCMVmixCMV{\cdot} $ needs the steps in $ T \steps \bisimCMV \EncCMVmixCMV{S'} $ to complete the emulation of source term steps started in $ \EncCMVmixCMV{S} \steps T $.
With the soundness result we can complete the proof of \cite{casalMordidoVasconcelos22} that $ \EncCMVmixCMV{\cdot} $ presented in
\cite[\S~7]{casalMordidoVasconcelos22} is good.

\begin{theorem}[Encoding from \CMVmix into \CMV]
	\label{thm:encodeCMVmixintoCMV}
	The encoding $ \EncCMVmixCMV{\cdot} $ from \CMVmix into \CMV presented in \cite{casalMordidoVasconcelos22} is good. By this encoding source terms in \CMVmix and their literal translations in \CMV are related by coupled similarity.
\end{theorem}

\begin{proof}[Proof of Theorem~\ref{thm:encodeCMVmixintoCMV}]
	Compositionality follows from the encoding function in the Figures~\ref{fig:encCMVmixCMVA} and \ref{fig:encCMVmixCMVB}.
	Under the assumption that $ c, d, u, v $ are different from all source term names, we can translate source term names by themselves. This ensures name invariance.
	If instead of this assumption the renaming policy $ \RPCMVmixCMV{\cdot} $ is used, name invariance follows from the consequent use of this renaming policy.
	Operational completeness was shown in \cite{casalMordidoVasconcelos22} \wrt $ \scCMVGC $.
	Since $ \scCMVGC $ is contained in $ \bisimCMV $, we can inherit this completeness result.
	Operational soundness follows from Lemma~\ref{lem:soundnessEncCMVmixCMV}.
	By the Figures~\ref{fig:encCMVmixCMVA} and \ref{fig:encCMVmixCMVB}, all literal translations of source terms have the same barbs as the respective source term and the encoding does not introduce free names.
	Barb sensitiveness then follows from operational correspondence, since $ \bisimCMV $ respects barbs.
	Divergence reflection follows from operational correspondence, since every sequence of target term steps eventually emulates a source term step and since all emulations of a single source term step are finite.
	Distributability preservation follows from the homomorphic translation of the parallel operator.
	We conclude that the encoding $ \EncCMVmixCMV{\cdot} $ is good.

	As shown in \cite{petersGlabbeek15} the combination of operational correspondence, divergence reflection, and barb sensitiveness induces a (weak reduction, barbed) coupled similarity that relates all source terms and their literal translations.
\end{proof}

Note that the translation of choice and in particular the non-deterministic choices distribute the decision made by a single source term step into several smaller decisions on the target: first a label and polarity on the internal choice is chosen, then a matching summand in the external choice, \ldots.
As explained in \cite{petersNestmann12, peters12, bispingNestmannPeters19}, splitting decisions leads to intermediate states and prevents from a tighter connection between source and target, \ie this encoding relates source terms and their literal translations by coupled similarity and not bisimilarity as shown in \cite{petersGlabbeek15}.
To obtain a tighter connection such as the bisimilarity, we would need the stronger version of soundness with $ T \asymp \ArbitraryEncoding{S'} $ instead of $ T \stepsTarget \asymp \ArbitraryEncoding{S'} $ (see \cite{petersGlabbeek15}).

As mentioned, a key feature of the encoding is to translate the nature of its summands, \ie whether they are send or receive actions, into the label used by the target term.
That this is possible, \ie that the prefixes for send and receive in a choice of \CMVmix can be translated to labels in a separate choice of \CMV such that the difference is not observable modulo the criteria in Definition~\ref{def:goodEncoding}, gives us the last piece of evidence that we need.
\CMVmix does not allow to solve problems such as leader election (Theorem~\ref{thm:separateCMVmixfromPiviaLeaderElection}) that are standard problems for mixed choice; \CMVmix cannot express the synchronisation pattern \patternStar either that we associate with mixed choice (Theorem~\ref{thm:separateCMVmixfromPiviaStar}). 
Yet, \CMVmix can express the pattern \patternM which is associated with \emph{separate choice}, and is encoded by a language with only separate choice (Theorem~\ref{thm:encodeCMVmixintoCMV}).
We conclude that choice in \CMVmix is semantically rather a separate choice.

\begin{corollary}
	\label{col:separateChoiceCMVmix}
	$ $\\
	The extension of \CMV given by \CMVmix introduces a form of separate choice rather than mixed choice.
\end{corollary}


\section{Related Work and Outlook}
\label{sec:conclusions}

We conclude by discussing related work, summing up our results, and briefly discussing our next steps.

\subsection{Related Work}

Encodings or the proof of their absence are the main way to compare process calculi \cite{boerPalamidessi1991, parrow08, gorla10, fuLu10, peters12, vG12, parrow2014abstraction, fu16, glabbeek18}.
See \cite{peters19} for an overview and discussion on encodings.
We used this methodology to compare different variants of choice in
session types.

The relevance of mixed choice for the expressive power of the \piCal was extensively studied.
An important encodability result on choices 
is the existence of a good encoding from the choice-free synchronous \piCal into its asynchronous variant \cite{boudol92, honda.tokoro:objectcalculus}, since it proves the relevance of choice.
As for the separation result, 
\cite{palamidessi03, gorla10, breakingSymmetries16} have shown that there is no good encoding from the full \piCal, \ie the synchronous \piCal including mixed choice, into its asynchronous variant if an encoding should preserve the distribution of systems.
Palamidessi in \cite{palamidessi97} was the first to point out that mixed choice strictly raises the expressive power of the \piCal.
Later work studies the criteria under that this separation result
holds and alternative ways to prove this result:
\cite{nestmann00} studies the relevance of divergence reflection for this result and considers separate choice.
\cite{gorla10, parrow08} discuss how to reprove this result if the rather strict criterion on the homomorphic translation of the parallel operator is replaced by compositionality.
\cite{peters12, petersNestmann12} show that compositionality itself is not strong enough to replace the homomorphic translation of the parallel operator by presenting an encoding and then propose the preservation of distributability as criterion to regain the result of Palamidessi.
\cite{breakingSymmetries16} uses the more fundamental problem of breaking symmetries instead of leader election.
\cite{petersNestmannGoltz13} further simplifies this separation result by introducing synchronisation patterns to distinguish the languages.
\cite{peters16} shows that instead of the preservation of distributability or the homomorphic translation of the parallel operator also the preservation of causality can be used as criterion.

While there are a vast amount of
theories~\cite{DBLP:journals/csur/HuttelLVCCDMPRT16}, 
programming languages 
~\cite{DBLP:journals/ftpl/AnconaBB0CDGGGH16}, 
and tools~\cite{BETTYTOOLBOOK} of session types, 
as far as we know, the \CMVmix-calculus  
is the only session $\pi$-calculus which extends external and internal choices to 
their mixtures with full constructs, 
i.e.~delegation, shared (or unlimited) name passing, 
value passing, and recursion in its process syntax,   
proposes its typing system and proves type-safety. 
In the context of \emph{multiparty session types} \cite{HYC2016}, 
there are several works that extend 
the original form of global types where choice is fixed (from one sender to one receiver) with more flexible forms of choices:
Recent work in \cite{DBLP:conf/concur/MajumdarMSZ21} \eg allows the global type to specify a choice of one sender to transmit to one of several receivers.
In \cite{JY-ESOP20} flexible choices are discussed but their well-formedness (which ensures deadlock-freedom of local types) needs to be checked by bisimuluation.
These works focus on gaining expressiveness of behaviours of a set of local types (or a simple form of CCS-like processes which are equivalent to local types \cite{DBLP:conf/concur/MajumdarMSZ21}) which correspond to \emph{a single multiparty session}, without delegations, interleaved sessions, restrictions nor name passing.

More recently, \cite{DBLP:conf/esop/Glabbeek22} compares the expressive power of a variant of the \piCal (with implicit matching) and the variant of CCS where the result of a synchronisation of two actions is itself an action subject to relabelling or restriction.
Because of the connection between CCS-like languages and local types, it may be interesting to compare the expressiveness results in \cite{DBLP:conf/esop/Glabbeek22} with (variants of) multiparty session types.

\subsection{Summary and Outlook}

We proved that \CMVmix is strictly less expressive than the \piCal in two different ways: by showing that \CMVmix cannot solve leader election in symmetric networks of odd degree and that \CMVmix cannot express the synchronisation pattern \patternStar.
Then we provide the missing soundness proof for the encoding presented in \cite{casalMordidoVasconcelos22}.
From these results and the insights on the reasons of these results, we conclude that the choice primitive added to \CMV in \cite{casalMordidoVasconcelos22} is rather a separate choice and not a mixed choice at least with respect to its expressive power.

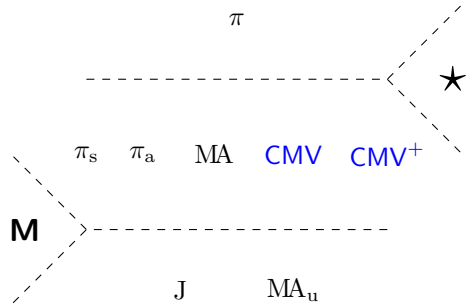
\begin{figure}
	\centering
	\begin{tikzpicture}[node distance=3cm, auto]
		\node (mix)		at (0, 1.8)		{\piMix};
		\node (sep)		at (-2, 0)	{\piSep};
		\node (asyn)	at (-1.25, 0)		{\piAsyn};
		\node (ma)		at (-0.3, 0)		{\MA};
		\node[color=blue] (CMV)		at (0.75, 0)		{\CMV};
		\node[color=blue] (CMVmix)		at (2, 0)		{\CMVmix};
		\node (join)	at (-0.75, -1.8)	{\join};
		\node (mau)		at (0.75, -1.8)	{\MAu};

		\node[scale=2] (star) at (2.75, 1) {\ \patternStar};
		\draw[dashed] (-2, 1) -- (2,1) -- (3,2);
		\draw[dashed] (2,1) -- (3,0);

		\node[scale=1.2] (M) at (-2.75, -1) {\patternM\quad};
		\draw[dashed] (2,-1) -- (-2,-1) -- (-3,-2);
		\draw[dashed] (-2,-1) -- (-3,0);
	\end{tikzpicture}
	\caption{Hierarchy of Pi-like Calculi.}
	\label{fig:hierarchy}
\end{figure}

With these results we can extend the hierarchy of pi-like calculi obtained in \cite{petersNestmannGoltz13, petersNestmannIC20} by two more languages as depicted in Figure~\ref{fig:hierarchy}.
This hierarchy orders languages according to their ability to express certain synchronisation patterns.
At the top we have the \piCal ($ \piMix $), because it can express the synchronisation pattern \patternStar.
In the middle are languages that can express \patternM but not \patternStar: the \piCal with separate choice ($ \piSep $) \cite{nestmann00}, the asynchronous \piCal without choice ($ \piAsyn $) \cite{honda.tokoro:objectcalculus, boudol92}, Mobile Ambients ($ \MA $) \cite{cardelliGordon00}, \CMV, and \CMVmix.
In the bottom we have the join-calculus ($ \join $) \cite{fournet.gonthier:join} and Mobile Ambients with unique Ambient names ($ \MAu $) \cite{petersNestmannIC20}, \ie the languages that cannot express \patternStar or \patternM.
That $ \piMix, \piSep, \piAsyn, \MA, \join $, and $ \MAu $ can or cannot express the respective pattern was shown in \cite{petersNestmannGoltz13, petersNestmannIC20}.

Linearity as enforced by the type system of \CMV/\CMVmix restricts the possible structures of communication protocols.
In particular, the type system ensures that it is impossible to unguard two competing inputs or outputs on the same linear channel at the same time.
Accordingly, it is not surprising that adding choice, even mixed choice, towards communication primitives under a type discipline that enforces linearity does not significantly increase the expressive power of the respective language (though it still might increase flexibility).
However, that adding mixed choice between unrestricted communication primitives does not significantly increase the expressive power of the language, did surprise us.
Unrestricted channels allow to have several in- or outputs on these channels in parallel, because the type system only ensures the absence of certain communication mismatches as \eg that the sort of a transmitted value is as expected by the receiver; but not linearity (compare also to shared channels as \eg in \cite{hondaVasconcelosKubo98}).
So, there is no obvious reason why the type system should limit the expressive power of unrestricted channels within a mixed choice.
Indeed, it turns out that the problem lies not in the type system.
In both ways to prove the separation result in \S~\ref{sec:separateMixedSessionsLeaderElection} and \S~\ref{sec:separateMixedSessionsFromPiSynchronisationPatterns} we completely ignore the type system and carry out the proof on the untyped version of the language, \ie it is already the untyped version of \CMVmix that cannot express mixed choice despite a mixed-choice-like primitive.
This limitation of the language definition, \ie in its syntax and semantics, is not obvious and indeed it was very hard to spot the problem.

We expect that adding mixed choice to the non-linear parts of other session type systems will instead significantly increase the expressive power. Accordingly, as the next step, we want to add a primitive for mixed choice between shared channels in session types such as described \eg in \cite{hondaVasconcelosKubo98, yoshidaVasconcelos06} and analyse the expressiveness of the resulting language.

\paragraph{Acknowledgements.}
The work is partially supported by EPSRC (EP/T006544/1, EP/K011715/1, \linebreak EP/K034413/1, EP/L00058X/1, EP/N027833/1, EP/N028201/1, EP/T006544/1, EP/T014709/1, \linebreak EP/V000462/1 and EP/X015955/1) and NCSS/EPSRC VeTSS.


\bibliographystyle{plainurl}
\bibliography{references}

\end{document}